\documentclass[prx,epsf,showpacs,twocolumn,superscriptaddress,longbibliography]{revtex4-2}
\usepackage[pdftex]{graphicx}
\usepackage[position=top,caption=false,singlelinecheck=off, justification=raggedright]{subfig}
\usepackage{dcolumn}
\usepackage{bm}
\usepackage{epsfig}
\usepackage{latexsym}
\usepackage{amsmath}
\usepackage{amsfonts}
\usepackage{color}
\usepackage{array}
\usepackage{braket}
\usepackage{bbold}
\usepackage{dsfont}
\usepackage{mathrsfs}
\usepackage{tikz}
\usetikzlibrary{quantikz}
\usepackage{hyperref}
\hypersetup{
    colorlinks,
    citecolor=blue,
    filecolor=red,
    linkcolor=magenta,
    urlcolor=red
}

\newcommand{\tr}{\textrm{tr}}

\makeatletter
\DeclareFontFamily{OMX}{MnSymbolE}{}
\DeclareSymbolFont{MnLargeSymbols}{OMX}{MnSymbolE}{m}{n}
\SetSymbolFont{MnLargeSymbols}{bold}{OMX}{MnSymbolE}{b}{n}
\DeclareFontShape{OMX}{MnSymbolE}{m}{n}{
    <-6>  MnSymbolE5
   <6-7>  MnSymbolE6
   <7-8>  MnSymbolE7
   <8-9>  MnSymbolE8
   <9-10> MnSymbolE9
  <10-12> MnSymbolE10
  <12->   MnSymbolE12
}{}
\DeclareFontShape{OMX}{MnSymbolE}{b}{n}{
    <-6>  MnSymbolE-Bold5
   <6-7>  MnSymbolE-Bold6
   <7-8>  MnSymbolE-Bold7
   <8-9>  MnSymbolE-Bold8
   <9-10> MnSymbolE-Bold9
  <10-12> MnSymbolE-Bold10
  <12->   MnSymbolE-Bold12
}{}

\let\llangle\@undefined
\let\rrangle\@undefined
\DeclareMathDelimiter{\llangle}{\mathopen}%
                     {MnLargeSymbols}{'164}{MnLargeSymbols}{'164}
\DeclareMathDelimiter{\rrangle}{\mathclose}%
                     {MnLargeSymbols}{'171}{MnLargeSymbols}{'171}

\newcommand{\eref}[1]{Eq.\,(\ref{#1})}

\begin{document}

\title{Robustness of near-thermal dynamics on digital quantum computers}

\author{Eli Chertkov}
\email{eli.chertkov@quantinuum.com}
\affiliation{Quantinuum, 303 S. Technology Ct., Broomfield, Colorado 80021, USA}
\author{Yi-Hsiang Chen}
\affiliation{Quantinuum, 303 S. Technology Ct., Broomfield, Colorado 80021, USA}
\author{Michael Lubasch}
\affiliation{Quantinuum, Partnership House, Carlisle Place, London SW1P 1BX, United Kingdom}
\author{David Hayes}
\affiliation{Quantinuum, 303 S. Technology Ct., Broomfield, Colorado 80021, USA}
\author{Michael Foss-Feig}
\affiliation{Quantinuum, 303 S. Technology Ct., Broomfield, Colorado 80021, USA}

\begin{abstract}
Understanding the impact of gate errors on quantum circuits is crucial to determining the potential applications of quantum computers, especially in the absence of large-scale error-corrected hardware. We put forward analytical arguments, corroborated by extensive numerical and experimental evidence, that Trotterized quantum circuits simulating the time-evolution of systems near thermal equilibrium are substantially more robust to both quantum gate errors and Trotter (discretization) errors than is widely assumed. In Quantinuum's trapped-ion computers, the weakly entangling gates that appear in Trotterized circuits can be implemented natively, and their error rate is smaller when they generate less entanglement; from benchmarking, we know that the error for a gate $\exp[-i (Z\otimes Z) \tau]$ decreases roughly linearly with $\tau$, up to a small offset at $\tau = 0$. We provide extensive evidence that this scaling, together with the robustness of near-thermal dynamics to both gate and discretization errors, facilitates substantial improvements in the achievable accuracy of Trotterized dynamics on near-term quantum computers.
We make heavy use of a new theoretical tool --- a statistical ensemble of random product states that approximates a thermal state, which can be efficiently prepared with low noise on quantum computers. We outline how the random product state ensemble can be used to predict, optimize, and design Hamiltonian simulation experiments on near-thermal quantum systems.
\end{abstract}
\maketitle

\section{Introduction}

Quantum computers have the potential to significantly outperform classical computers in simulating the dynamics of many-body quantum systems, with applications to materials science, chemistry, nuclear, and high-energy physics. Despite decades of effort, classical simulations of such dynamics is computationally prohibitive except for small problems, short-time evolution, weakly-entangled states, or systems with fine-tuned structure. Quantum computers are expected to allow access to generic, classically difficult simulation regimes, thereby expanding the range of scientific exploration of quantum dynamics.

In time-independent many-body quantum systems without symmetries, the only conserved quantity is the Hamiltonian operator (or energy), and quantum dynamics essentially randomizes the quantum state subject to the constraint of energy conservation, leading to thermalization. 
In this work, we examine the Hamiltonian simulation algorithm known as Trotterization or the product formula \cite{trotter1959,suzuki1976}. Specifically, we use a second-order Trotterization to approximate the continuous time-evolution of a one-dimensional mixed-field Ising spin chain on a digital quantum computer. The quantum circuit contains layers of $\exp[-i (Z\otimes Z) \tau]$ gates with angle $\tau$ that controls the discretization, or Trotter, error.

In current devices, the largest source of error is typically the two-qubit gate. A standard approach to estimating whether a quantum circuit can or cannot produce reliable outputs is to count the two-qubit gates and from the failure rate of individual gates determine the likelihood that none of them have failed.
In this work, we explore ways---some well-known and some not---in which the accuracy of local observables during quantum dynamics can be dramatically better than gate counting would suggest. Specifically, we show that for systems in or near thermal equilibrium, the values of local observables are dramatically less affected by gate errors than those of systems far from equilibrium. We demonstrate this distinction between thermal and non-thermal dynamics by comparing a thermalizing system to one having a quantum many-body scar state, an anomalous energy eigenstate with atypical non-thermal properties \cite{Moudgalya2021}. We find that for thermalizing dynamics, errors on local observables scale (1) linearly with time at short times, and (2) independently of system size at late times. In contrast, errors on local observables in non-thermal systems can grow in proportion to the circuit volume causally connected to them, which grows quadratically at short time and depends linearly on the system size at late times. These observations are corroborated with numerical simulations and experimental results on the H1-1 quantum computer \cite{H11,H11specsheet}. In addition, we show that if gate error rates scale linearly with the angle of rotation they generate (e.g., with $\tau$ for Quantinuum's native $\exp[-i (Z\otimes Z) \tau]$ gate), not only do simulations become proportionally better, but Trotter errors can be made arbitrarily small. We demonstrate that Quantinuum's H-Series devices have a nearly-linear angle-dependent error model with a constant offset, which enables high-accuracy Hamiltonian simulation with low Trotterization error \cite{moses2023}. Our work indicates that decreasing any residual error offset as the gate angle is taken to zero is an important technical goal for digital quantum computers that will facilitate high-accuracy large-scale Hamiltonian simulation.

While we mainly study the impact of gate errors, we also examine the impact of Trotter errors on thermalizing dynamics, a topic recently explored in Ref.~\onlinecite{heyl2019}. We observe that for short to intermediate time scales, the Trotter error remains essentially constant in time, with linear-in-time growth observed at late times. We find that the crossover to linear growth is determined by how close the initial state is to a thermal state (or ``diagonal ensemble'' \cite{rigol2008}). We show that the combined effects of gate and Trotter errors can be described by a simple heuristic model that agrees well with both numerical and experimental data. 

The numerical and experimental data presented in this work utilize a newly developed tool that we call the \emph{random product state ensemble} (RPE), a statistical ensemble of random unentangled states with fixed total energy for a quantum Hamiltonian. The RPE has a number of desirable properties that facilitate the study of thermalizing quantum dynamics. First, RPE product states can be efficiently prepared on digital quantum computers using a single layer of single-qubit gates. Second, the RPE provides a reasonable approximation to quantum thermal states. The RPE can be interpreted as a microcanonical ensemble for a classical version of the quantum Hamiltonian. We introduce an efficient classical Markov chain Monte Carlo algorithm for sampling states from the RPE. Empirically, we show that the RPE mixed state generated by averaging product states drawn from the RPE is close to the thermal diagonal ensemble.  Therefore, by initializing time-evolution experiments with the RPE mixed state, we can start much closer to thermal equilibrium than using only a single product state. We demonstrate how the RPE, when combined with our simple heuristic gate and Trotter error models, can be used to predict the errors of observables in thermalizing Trotterized dynamics experiments. From this model, one can efficiently estimate the optimal parameters (such as Trotter step $\tau$) to use for experiments on real hardware.

The paper is organized as follows: Section~\ref{sec:gate_error_predictions} presents arguments for how gate errors impact near-thermal Trotterized dynamics. 
Section \ref{sec:model} introduces the specific Hamiltonian and quantum circuit we study throughout this work. Section~\ref{sec:trotter_error_predictions} outlines how we expect Trotter errors to impact near-thermal dynamics. Section \ref{sec:experiment} presents experimental data from Quantinuum's H1-1 quantum computer, validating our gate and Trotter error predictions. Section \ref{sec:rpe} introduces the random product state ensemble.
Sections \ref{sec:single_error}~and~\ref{sec:many_errors} discuss numerical studies on how gate errors impact dynamics. Section~\ref{sec:trotter_errors} provides numerical simulations focused on Trotter errors. Section~\ref{sec:tools} describes heuristic ways to estimate the performance of Trotterized dynamics circuits on noisy quantum computers. Section~\ref{sec:discussion} presents an overview of this work and discusses potential future directions.

\section{Predicted impact of gate errors under different assumptions} \label{sec:gate_error_predictions}

Quantum computers are limited by the noisiest quantum operation, which is often a two-qubit (2Q) gate. A simple and often employed strategy to assess the expected performance of a quantum algorithm  is to assume that a single 2Q gate failure will lead (with high probability) to a random output state, with no correlation to the intended output.  If the output of the algorithm can be formulated as the expectation value of a (potentially multi-qubit) Pauli operator $\mathcal{O}$, whose value should be exponentially small in the Hilbert-space dimension for a random state, it is natural to estimate the output as

\begin{align}
\langle\mathcal{O}\rangle_{\rm error}\approx \langle \mathcal{O}\rangle_{\rm ideal}\times (1-p)^{N_{2Q}},
\label{eq:gc_1}
\end{align}
or that the observable error is
\begin{align}
|\langle\mathcal{O}\rangle_{\rm error} - \langle\mathcal{O}\rangle_{\rm ideal}|/|\langle\mathcal{O}\rangle_{\rm ideal}| \approx  p N_{2Q} \label{eq:obs_error_gc_1}
\end{align}
for $pN_{2Q} \ll 1$, where $p$ is the 2Q gate failure probability and $N_{2Q}$ is the number of 2Q gates. The suppression factor $(1-p)^{N_{2Q}}$ is the probability of no 2Q gate having failed, and so \eref{eq:gc_1} says the output will be entirely corrupted by noise if a single 2Q gate has failed. There are many reasons why \eref{eq:gc_1} might not properly account for the impact of gate errors on the output of a quantum algorithm.

In Hamiltonian simulation, a quantum computer approximately evolves a quantum state $\rho$ under a Hamiltonian $H$. A typical application of Hamiltonian simulation would be to prepare some meaningful many-body state (e.g., a ground state or thermal state) and then extract local information about that state (e.g., an order parameter, correlation functions, etc.) by measuring the expectation value $\langle \mathcal{O} \rangle$ of a few-body intensive \footnote{Intensive quantities are those that do not scale with system size, such as single-site observables or spatially-averaged observables.} observable $ \mathcal{O}$. Suppose that the Hamiltonian simulation is carried out using a Trotter decomposition, so that the system of size $N$ is evolved for time $t$ using $D$ Trotter steps of size $\tau=t/D$.  We are concerned with the question: 
How does the noisy expectation value of a local observable evolving under a generic local Hamiltonian behave? In particular, how does its error scale with $N$, $t$, and $\tau$?

Below we discuss how under different assumptions we can obtain different heuristic predictions for the scaling of observable error with $N,t,$ and $\tau$. Throughout this work, we assume that 2Q gate errors are incoherent. In practice, coherent errors can often be converted to incoherent ones using, for example, randomized compiling \footnote{While randomized compiling is generally formulated for randomizing coherent errors on Clifford two-qubit gates, many types of coherent errors on $e^{-i\tau ZZ}$ can also be made incoherent given physical access to the gate with $\tau\rightarrow -\tau$. For example, by assuming $e^{\pm i\tau ZZ}$ have the same error channel, one can Pauli twirl its error into a stochastic Pauli channel by implementing the $-\tau$ angle whenever a non-commuting Pauli (with $ZZ$) is sampled.} (or Pauli twirling) techniques \cite{hashim2021}.

\subsection{Causal constraints on error propagation}

\begin{figure}
\centering
\includegraphics[width=0.4\textwidth]{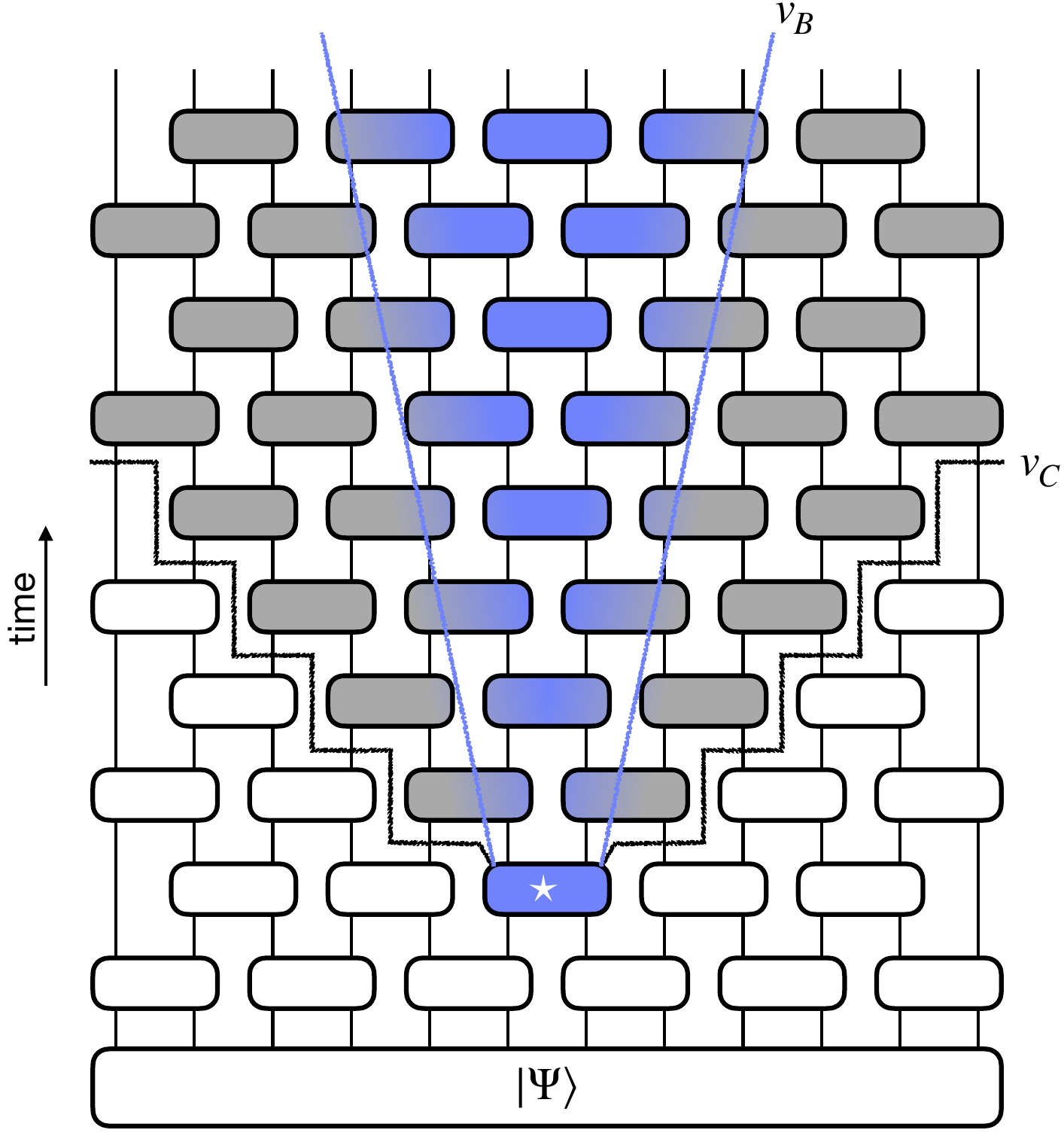}
\caption{Insertion of an error (here occurring on the gate marked with a ``$\star$'') can only ever impact an observable with support that overlaps the forward ``circuit causal cone'' (here the gates above the black lines spreading at the circuit velocity $v_{C}$). In Trotterized Hamiltonian simulation, where each gate is only weakly entangling, there is a ``physical circuit cone'' that spreads at the parametrically (in the step size $\tau$) smaller butterfly velocity (blue lines) of the physical model being simulated. The impact of an error outside of this cone will not be identically zero, but decays exponentially in the distance outside the cone.}
\label{fig:causality}
\end{figure}

Suppose that two-qubit gate errors propagate at the maximal velocity allowed in the quantum circuit, and that each error that evolves to have support on the region of space containing the observable has an \emph{equal effect on the observable}. Under these assumptions, the local observable error is determined only by the gates in the circuit-velocity causal cone of the local observable. Counting up these gates in $d$ spatial dimensions, we find that there are effectively
    \begin{align}
        N_{2Q,\textrm{eff}}=A_d D^{d+1} = A_d t^{d+1}/\tau^{d+1} \sim t^{d+1}/\tau^{d+1} \label{eq:N2q1}
    \end{align}
    gates at early times $t \leq t_1 \equiv B_d N^{1/d} \tau$ before the light cone has reached the boundaries and 
    \begin{align}
        N_{2Q,\textrm{eff}}&= C_d N^{1+1/d} + \alpha N D \nonumber \\
        &= C_d N^{1+1/d} + \alpha N t/\tau \sim Nt/\tau \label{eq:N2q2}
    \end{align}
    gates at late times $t > t_1$ when the light cone covers the entire system, where $A_d,B_d,C_d,\alpha$ are dimension- and geometry-dependent constants. 
    
    It is well known that in Hamiltonian simulation circuits, information propagates not at the circuit velocity, but at the butterfly velocity $v_B$, which is $O(1)$ in units of the energy scale of the Hamiltonian \cite{roberts2016}. 
    Suppose now that we assume that two-qubit gate errors propagate at the butterfly velocity but still \emph{equally affect the final observable} (see Fig.~\ref{fig:causality}).
    In a Trotterized circuit, the spatial distance quantum information travels in time $t$ is $L=v_B t$. Equivalently, it travels a distance $L= \tilde{v}_B D$, where $\tilde{v}_B=\tau v_B$, after $D$ Trotter layers of circuit evolution. Since $v_B$ is an $O(1)$ constant, this indicates that the butterfly velocity per Trotter layer $\tilde{v}_B$ is $\times \tau$ smaller than the circuit velocity.
    Counting up the number of 2Q gates in the butterfly causal cone, we find that the effective number of gates is
    \begin{align}
        N_{2Q,\textrm{eff}}&=A_d (v_{B}t)^{d} D = A_d v_{B}^{d} \tau^{d} D^{d+1} \nonumber \\
        &= A_d v_{B}^d t^{d+1}/\tau \sim t^{d+1}/\tau \label{eq:N2qbutterfly1}
    \end{align}
    at early times $t \leq t_2\equiv \frac{B_d}{v_B} N^{1/d}$ and 
    \begin{align}
        N_{2Q,\textrm{eff}}&=\frac{C_d}{v_B \tau} N^{1+1/d} + \alpha N D \nonumber \\
        &=\frac{C_d}{v_B \tau} N^{1+1/d} + \alpha N t/\tau \sim N t/\tau \label{eq:N2qbutterfly2}
    \end{align}
    at late times $t > t_2$.

    From gate counting and causal constraints, one would expect from Eqs.~(\ref{eq:obs_error_gc_1}),~(\ref{eq:N2qbutterfly1}),~and~(\ref{eq:N2qbutterfly2}) that observable errors would scale as 
    \begin{align}
    |\langle\mathcal{O}\rangle_{\rm error} - \langle\mathcal{O}\rangle_{\rm ideal}| \sim pt^{d+1}/\tau \label{eq:obs_error_gc_early_time}
    \end{align}
    for short times before the butterfly causal cone hits the boundaries and as
    \begin{align}
    |\langle\mathcal{O}\rangle_{\rm error} - \langle\mathcal{O}\rangle_{\rm ideal}| \sim pN t/\tau \label{eq:obs_error_gc_late_time}
    \end{align}
    for later times after the causal cone has covered the entire system.
     
\subsection{Errors heat up a thermalizing system}  

\begin{figure}
\centering
\includegraphics[width=0.45\textwidth]{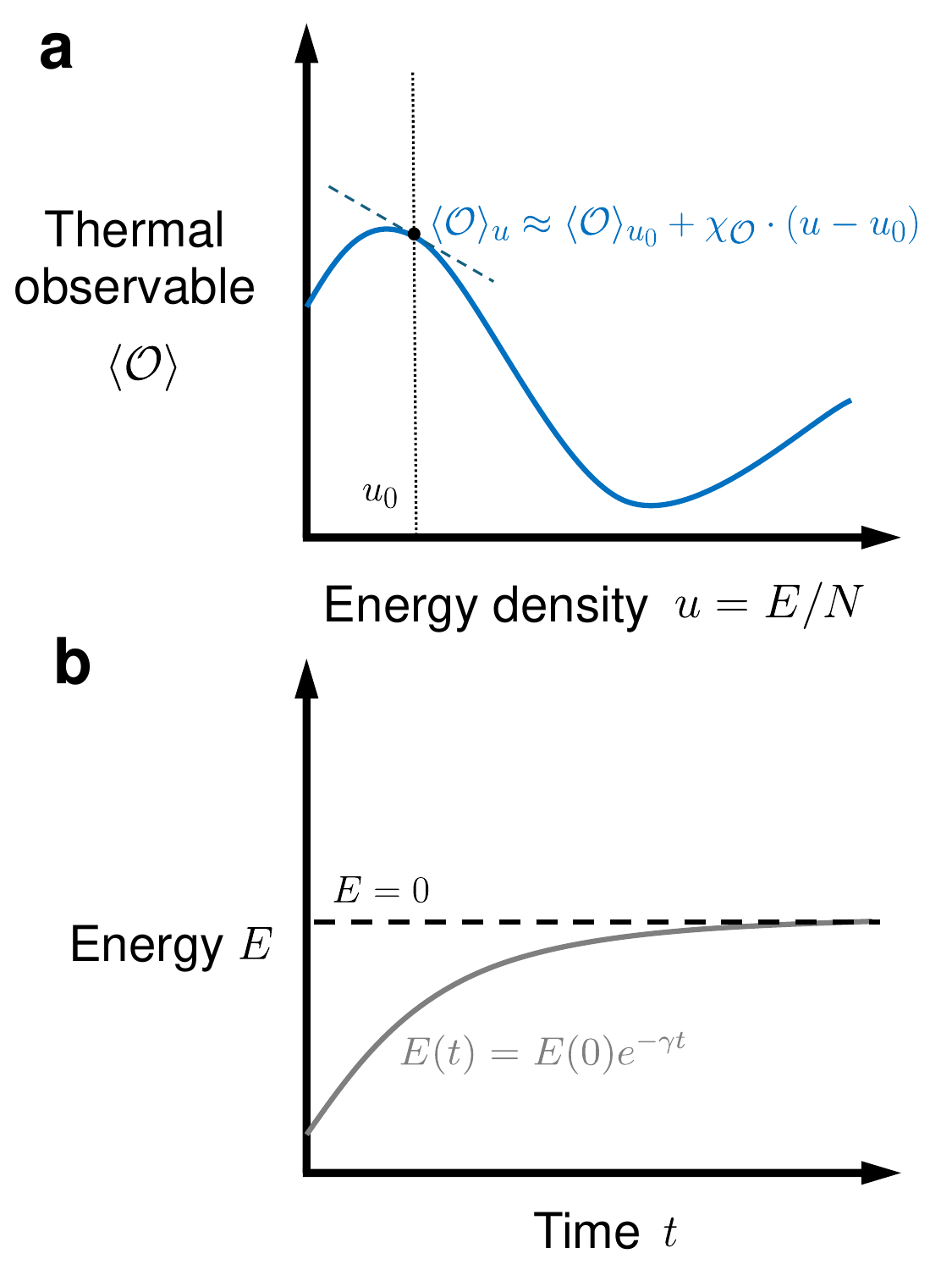}
\caption{\textbf{a} According to the eigenstate thermalization hypothesis, for energy eigenstates intensive local observables are smooth functions of the energy density. For a small change in the energy density, the change in the observable is approximately linear. \textbf{b} Gate errors in a Trotterized Hamiltonian simulation circuit heat up the system, causing the energy to change exponentially with time at a rate set by the gate errors.}
\label{fig:eth_schematic}
\end{figure}

The previous section concerns kinematic constraints on how an error can propagate, which dictate whether it can or cannot appreciably impact a particular local observable. While these considerations require few assumptions, in the case when an error \emph{can} impact a local observable they offer no information regarding \emph{how large} that impact will be. Recently, several authors have explored how errors impact local observables in ergodic many-body dynamics \cite{Yang2023,schiffer2024,granet2024}. An important conclusion of these works is that---contrary to the gate counting argument supplemented with locality constraints---causally relevant gate errors generally cannot simultaneously impact many local observables. Instead, as also seen in Ref.~\onlinecite{granet2024}, the effect of a gate error is \emph{diluted} as it spreads, with observables in the future less affected by the error than observables immediately following it.

To understand the above results, consider that at small Trotter step or short times, energy is approximately conserved and does not change with time (in fact, even for reasonably large Trotter steps energy can still be conserved over very long time scales when measured with respect to a quasi-local Floquet Hamiltonian \cite{Kuwahara2016,Yang2023}). If the Hamiltonian considered satisfies the eigenstate thermalization hypothesis (ETH), then after a short thermalization time any local intensive observable $\langle \mathcal{O}(t)\rangle$ should remain close to its thermal value at a temperature set by the energy density $u=E/N$ of the initial state $\rho$ (see Fig.~\ref{fig:eth_schematic}\textbf{a}) \cite{srednicki1994,nandkishore2015,dalessio2016,mori2018,deutsch2018,altman2018,abanin2019}. 

In general, we expect that gate errors during the evolution will lead to an exponential decay of energy with time (see Fig.~\ref{fig:eth_schematic}\textbf{b}), with the system eventually heating up to infinite temperature at $E=0$ (for a trace-less Hamiltonian). For a local Hamiltonian, an incoherent gate error will change the energy of the system by a constant amount. Using this fact, Ref.~\onlinecite{Yang2023} devised a heuristic stochastic model that predicted exponential energy decay (see Appendix~C of that work). Similarly, we show in Appendix~\ref{sec:xymodel} that the Trotterized evolution of the XY model in any geometry undergoing 2Q depolarizing error exhibits essentially perfect exponential decay of energy $E(t)=E(0)e^{-\gamma t}$ at a rate that scales as $\gamma \propto p/\tau$. This leads us to believe that exponential decay of energy likely holds to a good approximation in other models.

The behavior of thermal observables is more complicated. Assuming the system remains at all times well described by ETH at the instantaneous energy density, the observable expectation value is a non-trivial function of energy $\langle \mathcal{O}\rangle_u = f(u)$ and time $\langle \mathcal{O}\rangle \approx f(u_0e^{-\gamma t})$. However, at sufficiently short times, the observable change is approximately linear 
\begin{align*}
\langle \mathcal{O}\rangle_u \approx \langle \mathcal{O}\rangle_{u_0} + \chi_{\mathcal{O}} \cdot (u - u_0)
\end{align*}
where $\chi_{\mathcal{O}}\equiv\left.\frac{\partial \langle \mathcal{O}\rangle}{\partial u}\right|_{u=u_0}$ is the susceptibility of the observable to a change in energy density. This leads us to predict that for times short enough that the change in energy density is small, the error on an observable due to gate errors is
\begin{align}
    |\langle \mathcal{O}\rangle_{\textrm{error}}-\langle \mathcal{O}\rangle_{\textrm{ideal}}| &= pSD=pS t/\tau \label{eq:error_eth}
\end{align}
where $S \propto \chi_{\mathcal{O}}|u_0|$ is a model, initial-state and observable dependent constant that quantifies how much the thermal observable changes due to heating per Trotter step. When $\chi_{{\mathcal{O}}}$ is close to zero the observable is particularly robust to gate errors and when $\chi_{{\mathcal{O}}}$ diverges, as can happen at a thermal phase transition, the influence of errors is amplified. Note that \eref{eq:error_eth} (and the notion of ``short time'' justifying it) can be valid both before and after the butterfly cone contains the entire system, exhibiting linear dependence on time and independence of system size $N$ in both cases. The $N$-independence in the latter case is a nontrivial consequence of thermalization; a given observable will be causally connected to more errors in a larger system, but the energy input by each error is also diluted amongst that larger system, making the response independent of system size. 

\begin{figure}[!t]
\centering
\includegraphics[width=0.45\textwidth]{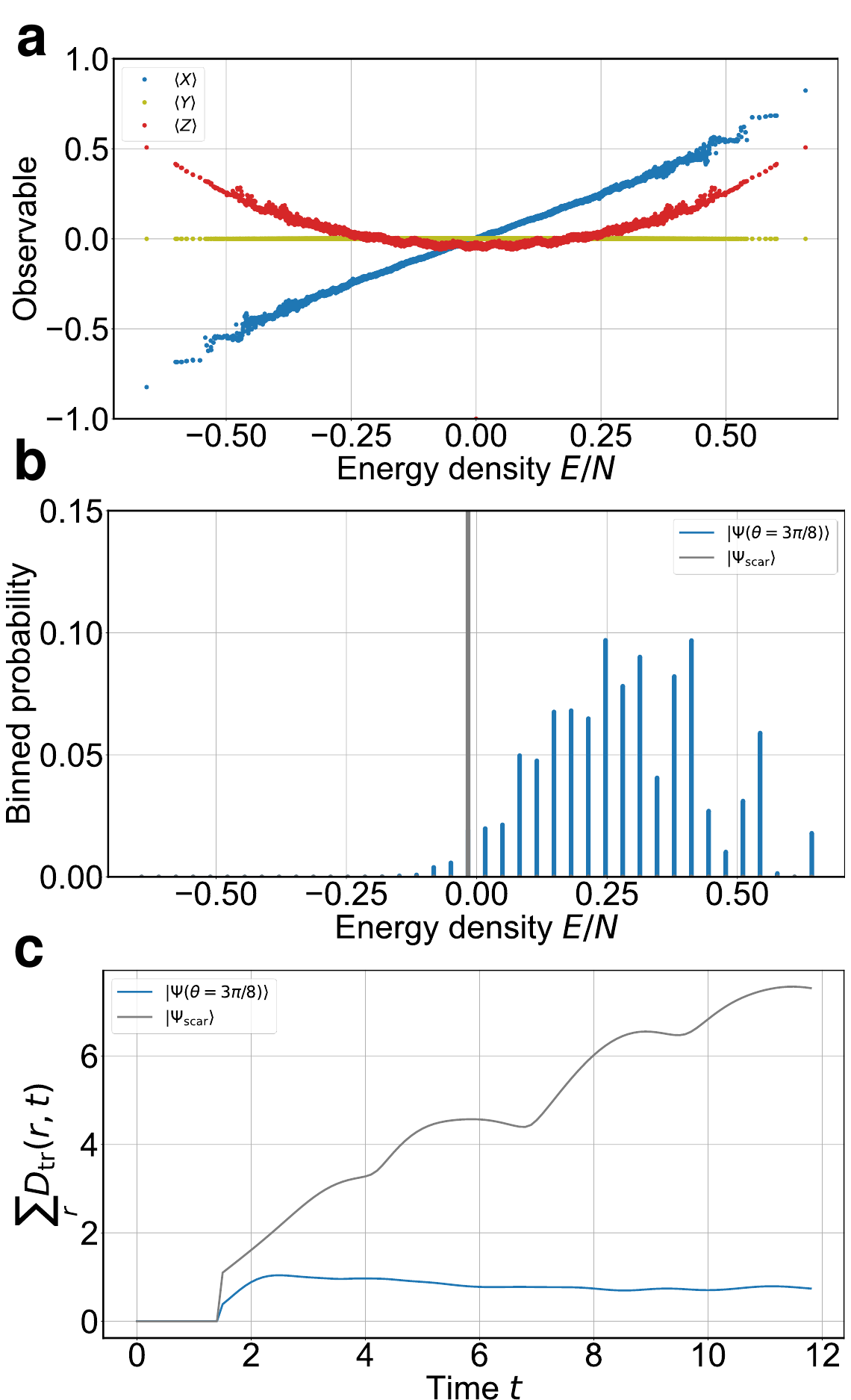}
\caption{\textbf{a} Local observables (average $X,Y,Z$ magnetizations) in eigenstates of $H_{\rm QE}$ as a function of their energy in a 14-site chain. Note the isolated quantum scar $\ket{\Psi_{\rm scar}}=\ket{0}^{\otimes N}$ at zero energy.  \textbf{b} Energy distributions for the scar state (grey; off the scale, but unity probability at zero energy) and the product state $\ket{\Psi(\theta = 3\pi/8)}$ (green). \textbf{c} Dynamics starting from either $\ket{\Psi_{\rm scar}}$ (black) or $\ket{\Psi(\theta = 3\pi/8)}$ (green) in a 24-site chain, showing that single errors generally have a conserved $O(1)$ total impact on local observables for thermalizing dynamics but more generally (absent thermalization) can be much more impactful. For this simulation we used a Trotter step of $\tau=0.1/J$.} \label{fig:qeast}
\end{figure}

A helpful way to think about why a single error does not evolve to cause substantial changes to all (causally connected) local observables in thermalizing dynamics is to consider the dynamics in the basis of energy eigenstates.  Local errors can induce transitions between nearby eigenstates, and in general a single error \emph{will}, via such transitions, lead to a new state with a substantially reduced fidelity. After several errors, the fidelity of the state with respect to what it should have been will be nearly zero. The robustness of local observables to errors in thermalizing systems stems from the fact that, according to ETH, all eigenstates close to each other in energy have similar values for local observables. Thus while the fidelity of the state may become low after a small number of errors, it is due to scattering into states of the Hilbert space \emph{that all look similar locally}. This picture makes it clear that constraints on the impact of errors imposed by thermalization are less robust than the kinematic constraints of the previous sections; they rely on the typicality of eigenstates supposed by ETH, which can be violated in a variety of settings.

A simple counter-example to these thermalization-based arguments can be found by embedding even a \emph{single} atypical state into an otherwise chaotic/ETH spectrum. Consider, as an example, a variant of the quantum-East model \cite{PhysRevB.92.100305} with the Hamiltonian
\begin{align}
H_{\rm QE}=\frac{J}{2}\sum_{j=1}^{N}(1-Z_j)X_{j+1}.
\label{eq:H_qeast}
\end{align}
Here we assume periodic boundary conditions (site $N+1$ being identified with site $1$ in the above summation).  Weakly randomizing the couplings $J\rightarrow J(1+\eta_j)$ (with $\eta_j\in [-0.1,0.1]$ a uniform random variable) to break translational symmetry, the spectral gaps of a small system can readily be seen to obey level repulsion in the manner expected for GOE statistics by the metric of Ref.~\onlinecite{PhysRevB.75.155111}, indicating that the model is ergodic. As seen in Fig.~\ref{fig:qeast}\textbf{a}, observables depend smoothly on energy throughout the spectrum with the notable exception of the zero-energy state $\ket{\Psi_{\rm scar}}=\ket{0}^{\otimes N}$; by construction of the Hamiltonian, $\ket{\Psi_{\rm scar}}$ is an eigenstate with eigenvalue zero, as it is annihilated by all of the projectors $(1-Z_j)$. Consider initiating time evolution under $H$ from either $\ket{\Psi_{\rm scar}}$ or a product state $\ket{\Psi(\theta)}=\Big(\prod_j \exp(-i Y_j\theta)\Big)\ket{\Psi_{\rm scar}}$ that has little overlap on $\ket{\Psi_{\rm scar}}$ (the energy distribution of such a state for $\theta=3\pi/8$ is shown in Fig.~\ref{fig:qeast}\textbf{b}).  By the thermalization arguments put forth above, insertion of a single Pauli error should lead to changes in local observables that are diluted as the error spreads.  We can quantify this expectation by computing the trace distance 
\begin{align}
D_{\textrm{tr}}(r,t) \equiv \frac{1}{2}||\rho_{\textrm{error}}(r,t)-\rho_{\textrm{ideal}}(r,t)||_1, \label{eq:tr_dist}
\end{align}
a direct measure of local observable error, where $||A||_1\equiv \tr \sqrt{A^\dagger A}$ and $\rho_{\textrm{error}}(r,t)$ and $\rho_{\textrm{ideal}}(r,t)$ are single-site reduced density matrices at site $r$ and time $t$ for the dynamics with and without the error, respectively. Summed over all sites, the trace distance should be roughly \emph{constant} in time if thermalization holds, which is indeed the behavior observed in Fig.~\ref{fig:qeast}\textbf{c} for the initial state $\ket{\Psi(3\pi/8)}$. To the contrary, a single error applied to $\ket{\Psi_{\rm scar}}$ leads to a (spatially summed) trace-distance error that grows roughly \emph{linearly} in time, which is as bad as kinematic (light-cone) constraints allow for. Physically, the linear growth is a consequence of the scar state transitioning (under local perturbations) into nearby energy states with dramatically different local observables. An error in any causally connected location with the measured local observable leads to strong changes in that observable. In the case of many errors occurring after each gate in the circuit, the $\ket{\Psi_{\rm scar}}$ scar state would show a quadratic-in-time growth of errors, consistent with Eq.~(\ref{eq:obs_error_gc_early_time}), while the $\ket{\Psi(3\pi/8)}$ thermalizing state would show a linear-in-time growth of errors, consistent with Eq.~(\ref{eq:error_eth}).

Note that Eq.~(\ref{eq:error_eth}) captures the general scaling with respect to $N,t,$ and $\tau$ that we expect for how local observable errors in thermalizing Hamiltonian simulation are affected by 2Q gate errors. We emphasize that accounting for the physics of thermalization leads to significantly more favorable scaling for observable accuracy. For example for $d=1$, at early times observables errors scale as $\sim t^2$ without thermalization (Eq.~(\ref{eq:obs_error_gc_early_time})) and $\sim t$ with thermalization (Eq.~(\ref{eq:error_eth})); at late times they scale as $\sim N t$ without thermalization (Eq.~(\ref{eq:obs_error_gc_late_time})) and $\sim t$ with thermalization (Eq.~(\ref{eq:error_eth})). However, there are potentially important details that this rough scaling does not capture, such as temporal oscillations of observables around their thermal values.

\subsection{The importance of angle-dependent two-qubit gate errors} \label{sec:angle_dep_gate_errors}

\begin{figure}[htbp!]
    \centering
    \includegraphics[width=0.45\textwidth]{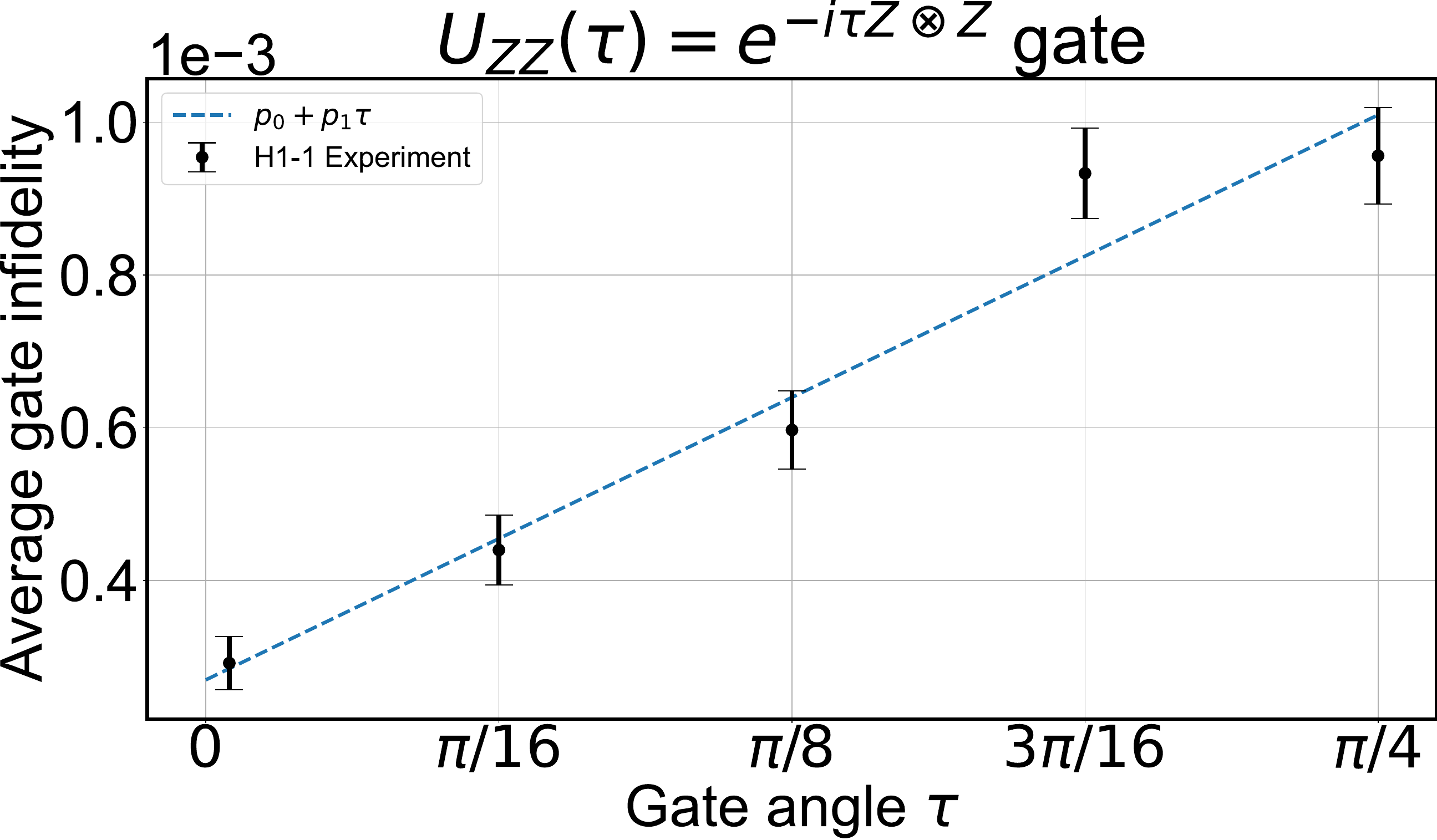}
    \caption{Experimentally measured average gate infidelity for the native arbitrary angle $U_{ZZ}(\tau)=e^{-i\tau ZZ}$ gate on the H1-1 quantum computer, along with a linear fit.}
    \label{fig:linear_angle_scaling}
\end{figure}

Some quantum computing architectures, such as trapped-ion devices \cite{pino2020,moses2023}, are able to natively implement 2Q gates with arbitrary angle rotations, such as $U_{ZZ}(\tau)=e^{-i\tau ZZ}$. In Quantinuum's H-Series trapped-ion quantum computers, it has been observed that to a good approximation the error rate for a $U_{ZZ}(\tau)$ gate is linear in the gate angle
\begin{align}
p(\tau) = p_0 + p_1 |\tau| \label{eq:linear_angle_gate_error}
\end{align}
with a slope of $p_1$ (per radian) and an error floor of $p_0$ at $\tau=0$ \cite{moses2023}. The average gate infidelity as measured on the H1-1 quantum computer using 2Q parameterized randomized benchmarking \cite{moses2023} is shown in Fig.~\ref{fig:linear_angle_scaling}, with a linear fit with parameters $p_0= 2.7\times 10^{-4} \pm 4\times 10^{-5}$ and $p_1= 9.4\times 10^{-4} \pm 1.0\times 10^{-4}$. Importantly, the small-angle error $p(\tau\approx 0) = 2.7 \times 10^{-4}$ is about $3 \times$ smaller than the maximum-angle gate error $p(\tau=\pi/4) \approx 1 \times 10^{-3}$, highlighting that for the H-Series devices there is a significant benefit to working at smaller Trotter steps (gate angles) $\tau$. In numerical simulations in this work, we use similar numbers based on older measurements, with parameters $p_0=3.5\times 10^{-4}$ and $p_1=9.5 \times 10^{-4}$.

Consider the effect of an angle-dependent error model of the form Eq.~(\ref{eq:linear_angle_gate_error}) on the accuracy of observable estimation for a Trotterized Hamiltonian simulation near thermal equilibrium. For a small gate error $p(\tau) \ll 1$, using Eq.~(\ref{eq:error_eth}) we predict that the observable error scales as
\begin{align}
|\langle \mathcal{O} \rangle_{\textrm{error}} - \langle \mathcal{O} \rangle_{\textrm{ideal}}| \approx p_0 S t/\tau + p_1 S t. \label{eq:gate_error_vs_tau_model}
\end{align}
For a constant error ($p=p_0,p_1=0$), we recover the previous prediction. However, for a perfectly linear-in-angle error ($p=p_1\tau, p_0=0)$, we find a surprising result: that the observable error is independent of Trotter step $\tau$ (or equivalently, independent of $D$). This indicates that for a perfectly linear 2Q gate error model, for a fixed total evolution time $t$, the Trotter step size $\tau=t/D$ can be decreased arbitrarily low without increasing the effect of 2Q gate errors. For such an error model, there is no penalty for choosing smaller Trotter step sizes. Another way to understand this result is that if gate errors scale linearly with $\tau$, then in the $\tau \rightarrow 0$ limit the circuit approaches a continuous-time Linbladian evolution with jump operator decay rates set by $p_1$. Of course, the scaling in Eq.~(\ref{eq:gate_error_vs_tau_model}) is only valid when $p S t/\tau \ll 1$, which holds when time or error rates are small enough (and $\tau$ is large enough if $p_0>0$).

\section{Model} \label{sec:model}

We study a one-dimensional mixed-field Ising model
\begin{align}
H = \underbrace{-\sum_{j=1}^{N-1} X_j X_{j+1}}_{H_2} \underbrace{- g\sum_{j=1}^N Z_j - h \sum_{j=1}^N X_j}_{H_1}, \label{eq:H}
\end{align}
with $h=0.9045,g=1.4$ and open boundary conditions, a thermalizing Hamiltonian studied recently in Ref.~\onlinecite{Thomas2023}. To time-evolve up to a time $t$, we apply $D$ layers of a 2nd order Trotter decomposition
\begin{align}
U_{\textrm{Trotter}}(\tau) &\equiv U_1(\tau/2) U_2(\tau) U_1(\tau/2) \nonumber \\
U_1(\tau/2) &\equiv e^{-i\tau H_1/2}=\prod_{j=1}^N e^{i\tau (g Z_j + h X_j)/2} \nonumber \\
U_2(\tau) &\equiv e^{-i\tau H_2} = \prod_{j=1}^{N-1} e^{i\tau X_j X_{j+1}} \label{eq:UTrotter}
\end{align}
with Trotter step size $\tau = t/D$. Our total evolution is given by 
\begin{align}
U_F(t)=U_{\textrm{Trotter}}^D \label{eq:U}
\end{align}
which approximates the continuous-time evolution operator $e^{-itH}$ when $\tau$ is small. Each $e^{i\tau X_j X_{j+1}}$ unitary in Eq.~(\ref{eq:UTrotter}) can be implemented with a single native arbitrary-angle $U_{ZZ}(\tau)$ 2Q gate (surrounded by 1Q gates) on the H-Series hardware. The total number of layers of such 2Q gates in the unitary circuit $U_F(t)$ is $2D$. The mixed-field Ising model is one of the simplest thermalizing models, essentially possessing no local conservation laws besides energy conservation, making it a good test-bed for this work (see Appendix~\ref{sec:mfi_symmetries} for additional details). 

The evolution operator $U_F(t)$ in Eq.~(\ref{eq:U}) is a Floquet unitary that repeats $U_{\textrm{Trotter}}$ periodically in time and defines a Floquet Hamiltonian $H_F$ through the relation $U_{\textrm{Trotter}} = e^{-i\tau H_F}$ (or equivalently, $U_F(t)=e^{-it H_F}$). For small $\tau$, the Floquet Hamiltonian can be computed perturbatively to lowest order in $\tau$ using the Magnus expansion \cite{Kuwahara2016}. For the second-order Trotter decomposition in Eq.~(\ref{eq:UTrotter}) and using the Baker–Campbell–Hausdorff formula, we find that the Floquet Hamiltonian takes the form
\begin{align}
H_F = H + \frac{\tau^2}{24}[H_1+2H_2, [H_1, H_2]] + O(\tau^4). \label{eq:HF}
\end{align}
Unlike the first-order Trotter result which is accurate to $O(\tau)$, this Floquet Hamiltonian is accurate to $O(\tau^2)$. Note also that the third-order term vanishes. 

\section{Predicted impact of Trotter errors on local observables} \label{sec:trotter_error_predictions}

In practice, since quantum computers have significant gate errors and smaller Trotter steps $\tau$ require deeper circuits with more gates, Trotterized circuits are usually implemented with non-trivially large $\tau$. In this regime, errors from Trotterization become non-negligible and potentially competitive with errors from noisy gates.

As discussed in Ref.~\onlinecite{heyl2019}, one can estimate the effects of Trotter errors for a thermalizing Hamiltonian by using time-dependent perturbation theory and assuming the system obeys ETH. By treating $H_F-H = O(\tau^2)$ in Eq.~(\ref{eq:HF}) as a perturbation, one can compute in the Heisenberg picture the time-evolution of a local observable $\mathcal{O}$ and compare it with the unperturbed result. To lowest order in $\tau$, this leads to the following heuristic prediction
\begin{align}
|\langle \mathcal{O}\rangle_{\textrm{error}} - \langle \mathcal{O}\rangle_{\textrm{ideal}}| \lesssim C\tau^2 + O(\tau^4) \label{eq:trotter_error_scaling}
\end{align}
for an observable, Hamiltonian, and initial-state-dependent constant $C$. Equation~(\ref{eq:trotter_error_scaling}) should be interpreted to mean that the Trotter error on an intensive local observable is \emph{independent of time $t$ and system size $N$} and only depends on the Trotter step $\tau$, up to potential temporal oscillations that can be bounded by a constant.
The system-size independence of $C$ for thermalizing observables was observed empirically in Ref.~\onlinecite{heyl2019}.
In Sec.~\ref{sec:trotter_scaling_with_time} and Appendix~\ref{sec:tdpt}, we present time-dependent perturbation theory calculations that provide further details about the Trotter error. In particular, we find that the time-independence of Eq.~(\ref{eq:trotter_error_scaling}) is only valid for states in the diagonal ensemble or for short times and that in general a linear-in-time growth term dominates at late times, a behavior also observed in Ref.~\onlinecite{chen2024}.

Intuitively, Eq.~(\ref{eq:trotter_error_scaling}) is capturing the idea that after a system has thermalized, all local observables take on their thermal values and so are essentially constant in time up to temporal oscillations. In Trotterized simulations, the values of local observables are effectively thermalizing to the thermal values of the Floquet Hamiltonian instead of the true Hamiltonian, causing this $O(\tau^2)$ discrepancy. As Ref.~\onlinecite{heyl2019} and others \cite{sieberer2019,chinni2022} have observed, errors in observables as well as properties of the Hamiltonian's spectrum dramatically change at a large enough critical Trotter step $\tau_C$, a phenomenon referred to as a Trotter transition.

Assuming that the effects of gate errors (Eq.~(\ref{eq:gate_error_vs_tau_model})) and Trotter errors (Eq.~(\ref{eq:trotter_error_scaling})) are independent and can be simply added together, we propose a simple form for how local observable errors behave:
\begin{align}
|\langle \mathcal{O}\rangle_{\textrm{error}} - \langle \mathcal{O}\rangle_{\textrm{ideal}}| \approx Sp_0 t/\tau + S p_1 t + C \tau^2. \label{eq:error_model}
\end{align}
To summarize: this model describes deviations from thermal observable values due to gate and Trotter errors and does not capture transient behaviors; it states that intensive observable errors are independent of system size $N$; and it only holds for times and Trotter steps that satisfy $S (p_0t/\tau + p_1 t) \ll 1$ and $\tau < \tau_C$ (i.e., generally holds for small gate error rates, short times, and small, but not too small, Trotter steps).

\section{Experimental results} \label{sec:experiment}

In our experiments, we execute the Trotterized quantum circuit in Eq.~(\ref{eq:U}) using different times $t$, system sizes $N$, and Trotter steps $\tau$. At the end of a circuit, we measure all qubits in either the $X$, $Y$, or $Z$ basis and gather many repetitions (shots) of measurement outcomes to obtain statistical estimates of the site-averaged $\langle X\rangle $, $\langle Y\rangle$, $\langle Z\rangle$ magnetizations. From these measurements, we obtain the site-averaged one-body reduced density matrix (1-RDM), $\rho^{(avg)}=\frac{1}{2}\left( I + \langle X\rangle X + \langle Y\rangle Y + \langle Z\rangle Z\right)$, which we compare with an ideal 1-RDM $\rho^{(avg)}_{\textrm{ideal}}$ using the trace distance error $D_{\textrm{tr}}^{(avg)}=\frac{1}{2}||\rho^{(avg)} - \rho^{(avg)}_{\textrm{ideal}}||_1$. This quantity provides a measure of local observable accuracy. We look at the trace distance of the \emph{site-averaged} 1-RDM rather than the site-averaged trace distance of \emph{single-site} 1-RDMs because the former could be computed more precisely with fewer shots on quantum hardware. We find numerically that the two alternative trace distance errors match very closely. Experimental error bars presented in the figures are 68\% confidence intervals obtained by bootstrap resampling with 1000 resamples.

Each circuit is initialized in a random product state drawn from the random product state ensemble described in Sec.~\ref{sec:rpe} using a classical compute environment available in the H-Series devices, which utilizes WebAssembly (Wasm) \cite{ryan2022implementing,moses2023}. We write a look-up table that stores the information needed to prepare each product state. At the beginning of each quantum circuit, the look-up table is queried to apply conditional 1Q gates to the qubits to create the product state. Ultimately, this procedure requires only a single circuit compilation and produces different initial states in each shot, resulting in efficient compilation and execution.

\textbf{Gate errors cause local observable errors to grow linearly in time.} At early times, we expect observable errors during thermalizing dynamics to grow linearly in time $O(t)$ for a fixed Trotter step (see Eq.~(\ref{eq:error_eth})). By comparison, for a non-thermalizing circuit, we would expect quadratic $O(t^2)$ growth of errors. 

\begin{figure}[htbp!]
    \centering
    \includegraphics[width=0.45\textwidth]{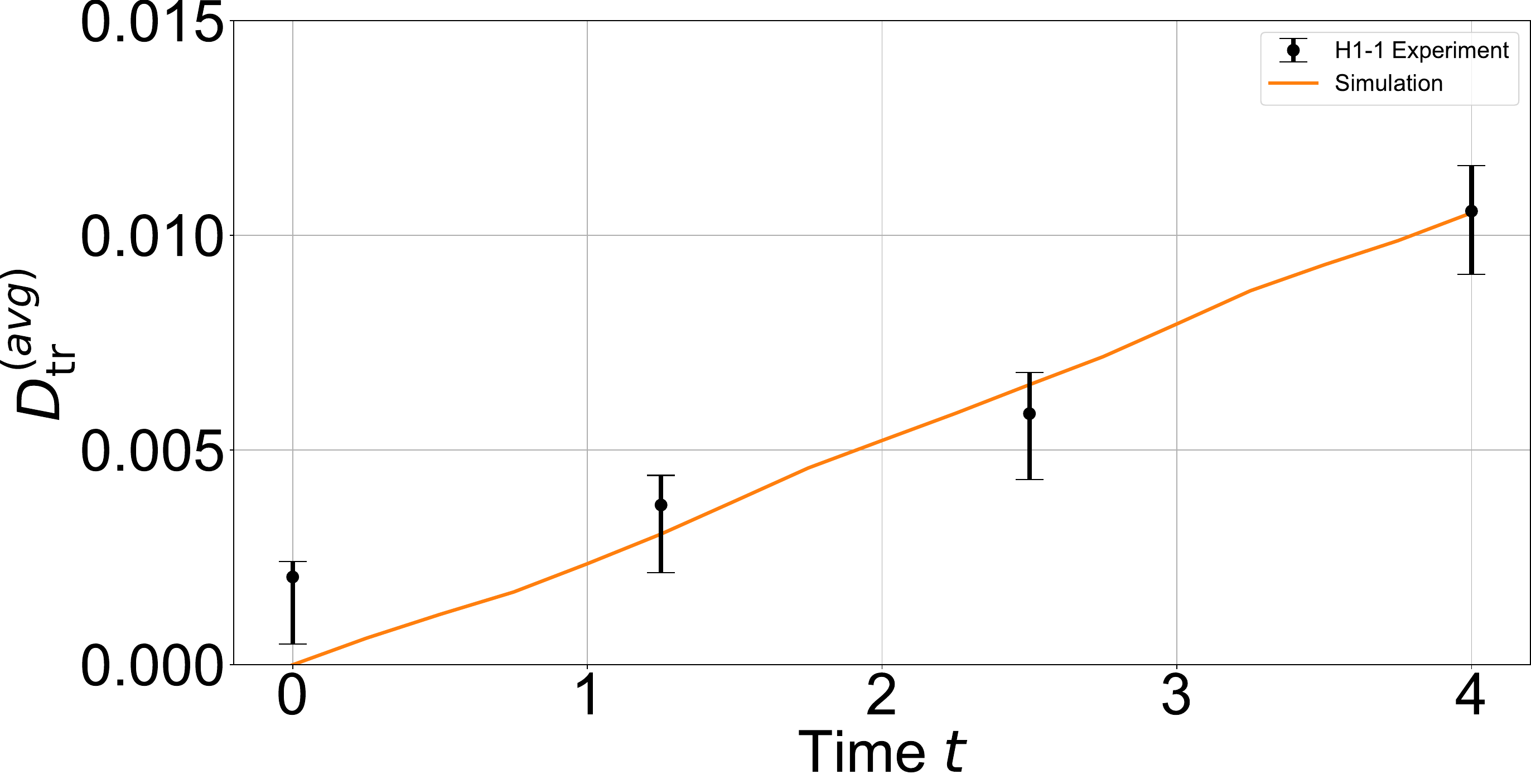}
    \caption{The observable error versus time $t$ as measured on the H1-1 quantum computer (black dots) at a fixed system size $N=20$ and Trotter step $\tau=0.25$, for early times $t \leq 4$. The observable error is quantified using the trace distance error between the site-averaged 1-RDM for a noisy quantum simulation with Trotter step $\tau=0.25$ and a noiseless simulation at the same Trotter step. The solid orange line is a noisy classical simulation, with the noise modeled as a two-qubit depolarizing channel after every 2Q gate using the gate error from Fig.~\ref{fig:linear_angle_scaling}.}
    \label{fig:error_vs_t}
\end{figure}

In Fig.~\ref{fig:error_vs_t}, we show the experimentally measured site-averaged trace distance as a function of time, using a Trotter step size $\tau=0.25$ and system size $N=20$. Each time point (and basis measurement) is repeated 8000 times. The experimental results appear consistent with the predicted linear growth. They match closely with numerical simulation using a depolarizing error model (solid orange line), which clearly displays linear-in-time behavior. Details of numerical simulations are discussed in Appendix~\ref{sec:numerical_methods}.

\textbf{Gate errors result in system-size independent local observable errors.} At late times, we expect local observable errors during thermalizing dynamics to not depend on system size $O(1)$ for a fixed time and Trotter step (see Eq.~(\ref{eq:error_eth})). By comparison, for a non-thermalizing circuit, we would expect linear in system size $O(N)$ growth of errors. 

\begin{figure}[htbp!]
    \centering
    \includegraphics[width=0.45\textwidth]{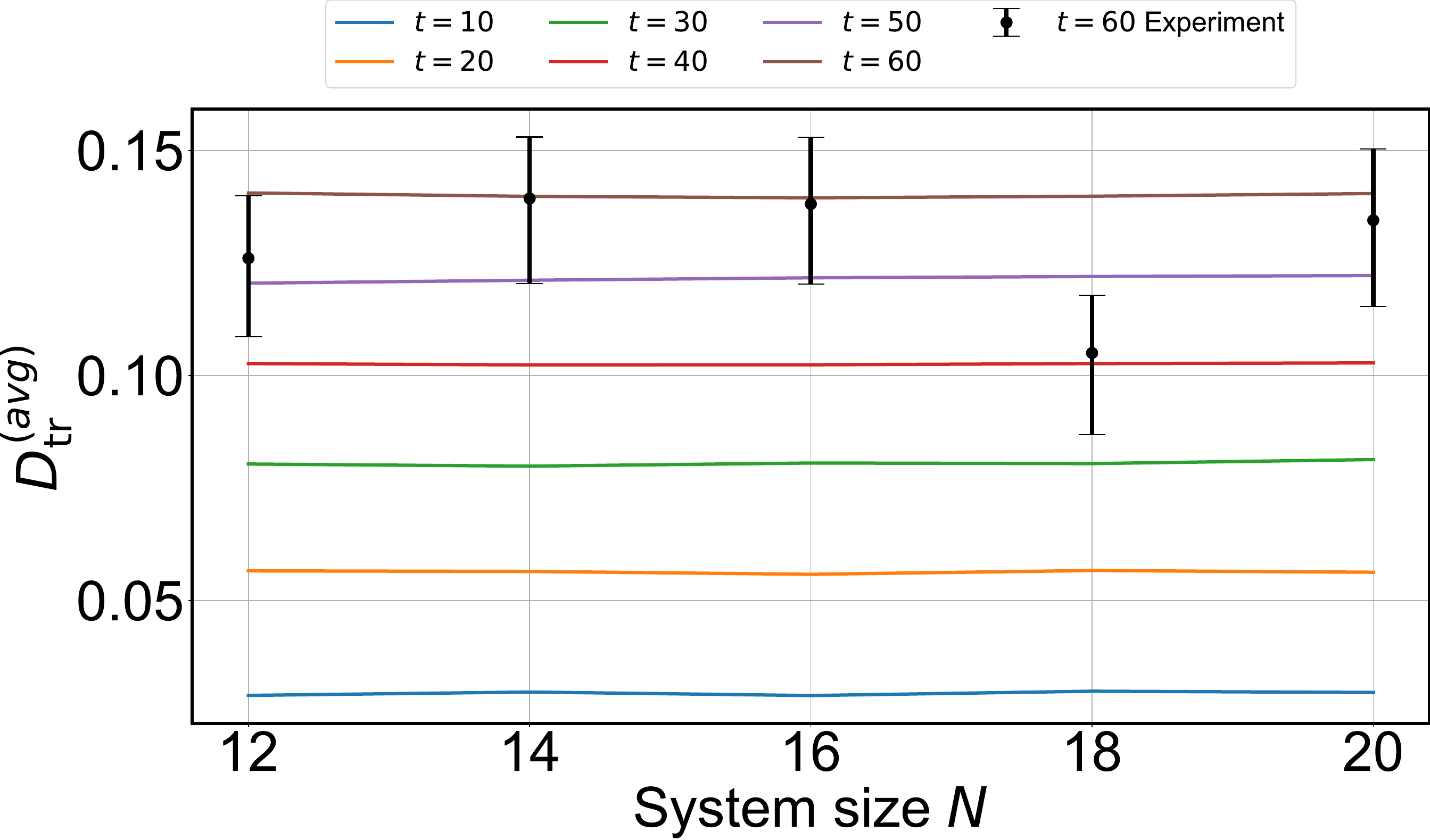}
    \caption{The observable error versus system size $N$ as measured on the H1-1 quantum computer (black dots) at fixed Trotter step $\tau=0.2$ and late time $t=60$. The observable error is quantified with the trace distance error  between the site-averaged 1-RDM for a noisy quantum simulation with Trotter step $\tau=0.2$ and for a noiseless simulation at the same Trotter step. The solid lines are obtained from noisy classical simulation at different fixed times $t$, with the noise modeled as a two-qubit depolarizing channel after every 2Q gate.}
    \label{fig:error_vs_N}
\end{figure}

In Fig.~\ref{fig:error_vs_N}, we show the experimentally measured site-averaged trace distance as a function of system size $N$, using a Trotter step size of $\tau=0.2$ and late-time value of $t=60$ (a depth $2D=2t/\tau=600$ circuit). Each system size point (and basis measurement) is repeated 200 times. Results from numerical simulations using a depolarizing error model performed at times $t=10,20,30,40,50,60$ are shown as solid lines. Numerical results show clear system size independence for the times shown, up to potentially small corrections in $N$. Experimental results agree with the numerics up to error bars. Importantly, these deep circuits have a measurable observable signal despite having many two-qubit gates (up to about $6000$) and vanishing state fidelity, highlighting the robustness of observables in near-thermal dynamics. Note that the site-averaged 1-RDM in this figure is averaged over sites $2,\ldots,N-1$ rather than all sites; this is done to remove a bias caused by a leakage detection gadget we use (discussed below). 

In Quantinuum's trapped ion quantum computers, memory error is accumulated during a quantum circuit as ions are shuttled through-out the trap. In these circuits, to avoid accidentally introducing system-size dependent memory errors due to system-size-dependent ion transport, we force the ions for different $N$ to take the same transport path in each circuit. We do this by using ``dummy'' $U_{ZZ}(0)$ gates. A dummy $U_{ZZ}(0)$ gate, due to rules specified in the H-Series compiler, causes ions to be transported into the appropriate locations to execute the gate but does not cause any lasers to be applied, and thereby does not introduce any gate error. In these experiments, each circuit is executed using all $N_0=20$ qubits on H1-1, but the 2Q gates involving qubits $N+1,\ldots, N_0$ are replaced with the $U_{ZZ}(0)$ dummy gates. 

To mitigate coherent memory errors, such as coherent $e^{-i\theta_j Z_j}$ on each qubit $j$, we implemented a heuristic form of dynamical decoupling (DD). In a Trotter step, there are two layers of 2Q gates. We insert $X$ pulses before and after the second layer. Since pairs of $X$ gates commute with the 2Q gates and $X^2=I$, they do not logically affect the circuit; however, they do toggle the sign of the phases accumulated during the ion transport of the gates, reducing the total accumulated coherent memory error.

In these circuits, we also mitigate leakage errors, which can occur during 2Q gates when a spontaneous emission event in an ion causes quantum information to leave the computational subspace. These circuits in particular, due to their large depth, can have significant leakage. In the H-Series devices: once a qubit has leaked it usually remains leaked for the remainder of the circuit; all 1Q and 2Q gates involving that qubit have no effect, i.e., act as identity; and all measurements of that qubit are registered as a $\ket{1}$. Therefore, leakage can potentially have a large impact on the dynamics of the system. This is particularly true in a 1D chain where a leakage event in the center of the chain essentially breaks it into two smaller chains for the remainder of the dynamics. Since we are primarily interested in the effect of incoherent gate errors, we use a leakage detection gadget \cite{Stricker2020,Chertkov2022,moses2023} at the end of our circuit and post-select on circuits that have no detected leakage. For the circuits in Fig.~\ref{fig:error_vs_N}, for $N=12,14,16,18,20$ the percentage of shots that detected leakage in the first $N$ qubits is $51.5\%, 65.5\%, 64.0\%,  69.0\%,  72.5\%$, respectively. Since the leakage detection gadget requires an ancilla qubit, we reset the first qubit in each circuit after it is measured to use as an ancilla to perform leakage detection on another qubit. Once the other qubit is leakage detected and measured, it is reset and used as an ancilla as well. Repeating this procedure leads to a $O(\log N)$ depth circuit that detects leakage on all but the first qubit.

\textbf{The competition between gate and Trotter errors is well described by a simple model.} At early times and for a linear-in-angle gate error model such as observed on H1-1, we expect observable errors during thermalizing dynamics to show a simple dependence on $t$ and $\tau$ described in Eq.~(\ref{eq:error_model}).

\begin{figure}[htbp!]
    \centering
    \includegraphics[width=0.45\textwidth]{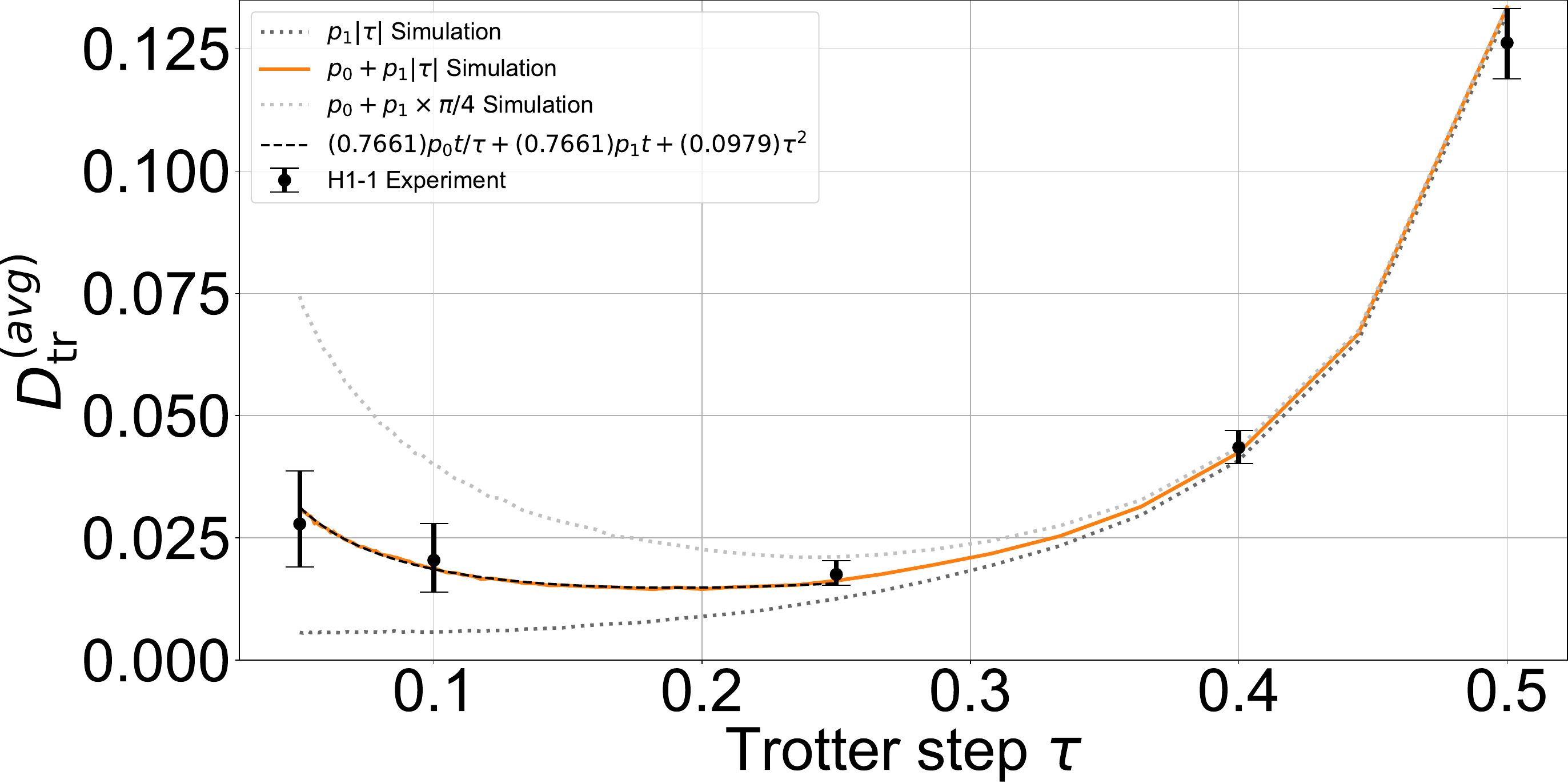}
    \caption{The observable error versus Trotter step $\tau$ as measured on the H1-1 quantum computer (black dots) at a fixed system size $N=20$ and at early time $t=4$. The observable error is quantified with the trace distance error between the site-averaged 1-RDM of a noisy quantum simulation at Trotter step $\tau$ and a noiseless simulation at a small Trotter step $\tau_0=0.04$ (chosen small enough to have essentially negligible Trotter error). In the noisy classical simulations, the noise is modeled as a two-qubit depolarizing channel with error rate $p_1|\tau|$ (dark gray dotted line), $p_0 + p_1|\tau|$ (orange solid line; See Fig.~\ref{fig:linear_angle_scaling}), or $p_0 + p_1 \pi/4$ (light gray dotted line) applied after every 2Q gate. The black dashed line is a fit of Eq.~(\ref{eq:error_model}) to the numerical simulation data with the $p_0 + p_1|\tau|$ error model, using $\tau \leq 0.25$ data.}
    \label{fig:error_vs_tau}
\end{figure}

In Fig.~\ref{fig:error_vs_tau}, we show the experimentally measured site-averaged trace distance as a function of Trotter step size $\tau$ for a fixed early time $t=4$ and system size $N=20$. In this figure, the trace distance error captures \emph{both} gate and Trotter errors; this is because here we are comparing a noisy quantum experiment at finite Trotter step with a noiseless classical simulation at small Trotter step ($\tau=0.04$) that has negligible Trotter error. Each data point (and basis measurement) is repeated between 500 and 8000 times. For comparison, we also included noisy numerical simulations with depolarizing 2Q gate errors with rates that depend differently on the angle $\tau$. The dotted light gray curve shows the expected result if the 2Q gate at angle $\tau$ performed as well as the maximum angle $\tau=\pi/4$ gate, i.e., did not improve as the angle decreased. The dotted dark gray curve shows the expected result if the 2Q gate at angle $\tau$ decreased exactly to zero at $\tau=0$, i.e., if the constant offset $p_0$ vanished. The solid orange curve shows the expected result for the 2Q gate at angle $\tau$, based on Eq.~(\ref{eq:linear_angle_gate_error}) obtained from the H1-1 benchmarking data in Fig~\ref{fig:linear_angle_scaling}. The black dashed line is a fit of Eq.~(\ref{eq:error_model}) to the \emph{numerical simulation} data, using $\tau \leq 0.25$. The fitted constants $S=0.7661$ and $C=0.0979$ describe the impact of gate errors and Trotter errors, respectively, on the observable, which also depends on the particular Hamiltonian, and energy density considered. The numerical simulations are described well at small $t$ and $\tau$ by the model and the H1-1 experimental results match closely with the simulations. We see that the experimentally measured observable errors are significantly improved by the enhanced performance of H1-1 at smaller $\tau$, but could be improved even further if the constant offset $p_0$ could be reduced. This highlights the importance of targeting the performance of the native arbitrary-angle $U_{ZZ}(\tau)$ gate at small angle $\tau \approx 0$.

Importantly, we can observe in the data the clear competition between gate errors, which dominate as $\sim 1/\tau$ at small Trotter step, and Trotter errors, which dominate as $\sim \tau^2$ at intermediate Trotter steps. This competition leads to different optimal Trotter step at different times, such as $\tau_{opt} \approx 0.2$ for this model at $t=4$.

\section{The random product state ensemble} \label{sec:rpe}

\begin{figure*}[htbp!]
    \centering
    \includegraphics[width=0.95\textwidth]{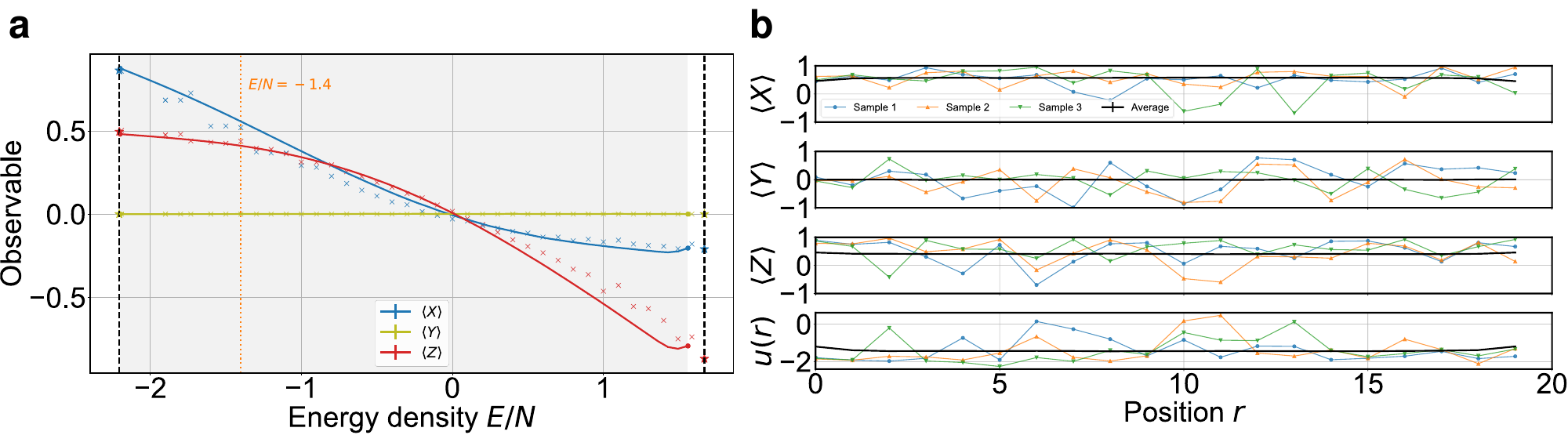}
    \caption{\textbf{a} The spatially-averaged $X,Y,Z$ observables for the random product state ensemble at different energy densities, for an $N=20$ site chain. The highlighted gray region spans the range of energies that product states occupy. The vertical dashed black lines indicate the lowest and highest energy densities, obtained with DMRG. The circles indicate the observables values for the mean-field ground state and anti-ground state product states. The stars indicate the observables values for the exact ground and anti-ground states. The crosses indicate the values of these observables for energy eigenstates at the specified energy densities, obtained with shift-invert exact diagonalization of an $N=14$ site chain. \textbf{b} The spatial profiles of the $X,Y,Z$ observables and the local energy density $u(r)$ for three random product states sampled from the RPE at $E/N \approx -1.4$ and $N=20$, as well as the average over many samples.}
    \label{fig:rpe_observables}
\end{figure*}

A key tool that we utilize in this work is the random product state ensemble (RPE). The RPE is an ensemble of random product states 
\begin{align*}
    \ket{\psi}=\prod_j (\cos(\theta_j/2) \ket{0}_j + \sin(\theta_j/2) e^{i\phi_j} \ket{1}_j)
\end{align*}
that all have the same energy $\langle \psi|H|\psi\rangle=E$. Generically, there are infinitely many such product states, which can be labeled either by the angles $\theta_j, \phi_j$ or the Bloch vectors (or spins) $\vec{\sigma}_j=(\langle X\rangle_j, \langle Y\rangle_j, \langle Z \rangle_j)=(\cos \phi_j \sin \theta_j, \sin \phi_j \sin \theta_j, \cos \theta_j)$ on each site $j$. We define the probability distribution of the RPE as follows:
\begin{align}
\lim_{\varepsilon \rightarrow 0} \left\{\ket{\psi}=\prod_j U_j \ket{0}_j \middle|\textrm{Haar }U_j,\, |\langle \psi|H|\psi\rangle - E| < \varepsilon \right\}. \label{eq:RPE}
\end{align}
Essentially, the RPE is the ensemble of Haar random product states with fixed energy $E$. In principle, one can obtain samples from this distribution by rejection sampling: one starts with a reference product state $\ket{0\cdots 0}$, applies Haar random $SU(2)$ unitaries on each site $j$, and accepts the generated product state as a valid sample if its energy is within $\varepsilon$ of the target energy $E$, then takes the limit $\varepsilon\rightarrow 0$. However, even with finite $\varepsilon$ the rejection sampling approach would have a vanishingly small acceptance rate and would be infeasible to implement except for small systems. Nonetheless, as we discuss in Appendix~\ref{sec:rpe_mcmc}, it is possible to use Markov chain Monte Carlo (MCMC) to efficiently sample product states from the RPE.

In this work, we make repeated use of the RPE mixed state, a probabilistic mixture of RPE product states 
\begin{align}
\rho_{\textrm{RPE}} = \mathbb{E}_{\ket{\psi} \sim P_{\textrm{RPE}}} \ket{\psi}\bra{\psi} \approx \frac{1}{M}\sum_{s=1}^M \ket{\psi_s}\bra{\psi_s}, \label{eq:rho_rpe}
\end{align}
where $\ket{\psi_s}$ are $M$ samples of random product states with energy $E$ obtained from MCMC sampling.

The RPE can be interpreted as a classical microcanonical ensemble. The product states in the RPE are constant energy states of a classical spin Hamiltonian obtained by replacing the vector of Pauli operators $(X_j,Y_j,Z_j)$ with a classical spin vector $\vec{\sigma}_j$. Likely due to this connection, the RPE qualitatively agrees well with the \emph{quantum} microcanonical ensemble. In Fig.~\ref{fig:rpe_observables}\textbf{a}, we show the average $X,Y,Z$ magnetizations in the RPE mixed state Eq.~(\ref{eq:rho_rpe}) versus energy density for an $N=20$ site chain using $M=11,200$ samples. Overlaid on these curves as crosses are the expectation values of these observables for individual energy eigenstates obtained from exact diagonalization of an $N=14$ site system. The eigenstate expectation values vary smoothly with energy density, as expected for systems satisfying ETH. Empirically, we find that the eigenstate values match surprisingly closely with the RPE values. While energy eigenstates from the quantum microcanonical ensemble have zero energy variance $\sigma_H^2=0$, product states from the RPE have extensive energy variance $\sigma_H^2\propto N$. For both ensembles, the fluctuations of energy density $\sigma_H/N$ goes to 0 in the thermodynamic limit $N\rightarrow \infty$. However, we do not expect these ensembles to converge to the same values, since this would imply that most finite energy (or temperature) observables of quantum Hamiltonians can be computed efficiently by sampling from the RPE, i.e., that the classical and quantum microcanonical ensembles are equivalent. Rather, the constraint that sampled states from the RPE are \emph{product} states adds non-trivial correlations to observables that do not exist in the quantum microcanonical ensemble. Note that the RPE at $E=0$, as well as unconstrained random product states \cite{chichen2024}, can be used to evaluate infinite-temperature observables.

\begin{figure}[htbp!]
    \centering
    \includegraphics[width=0.45\textwidth]{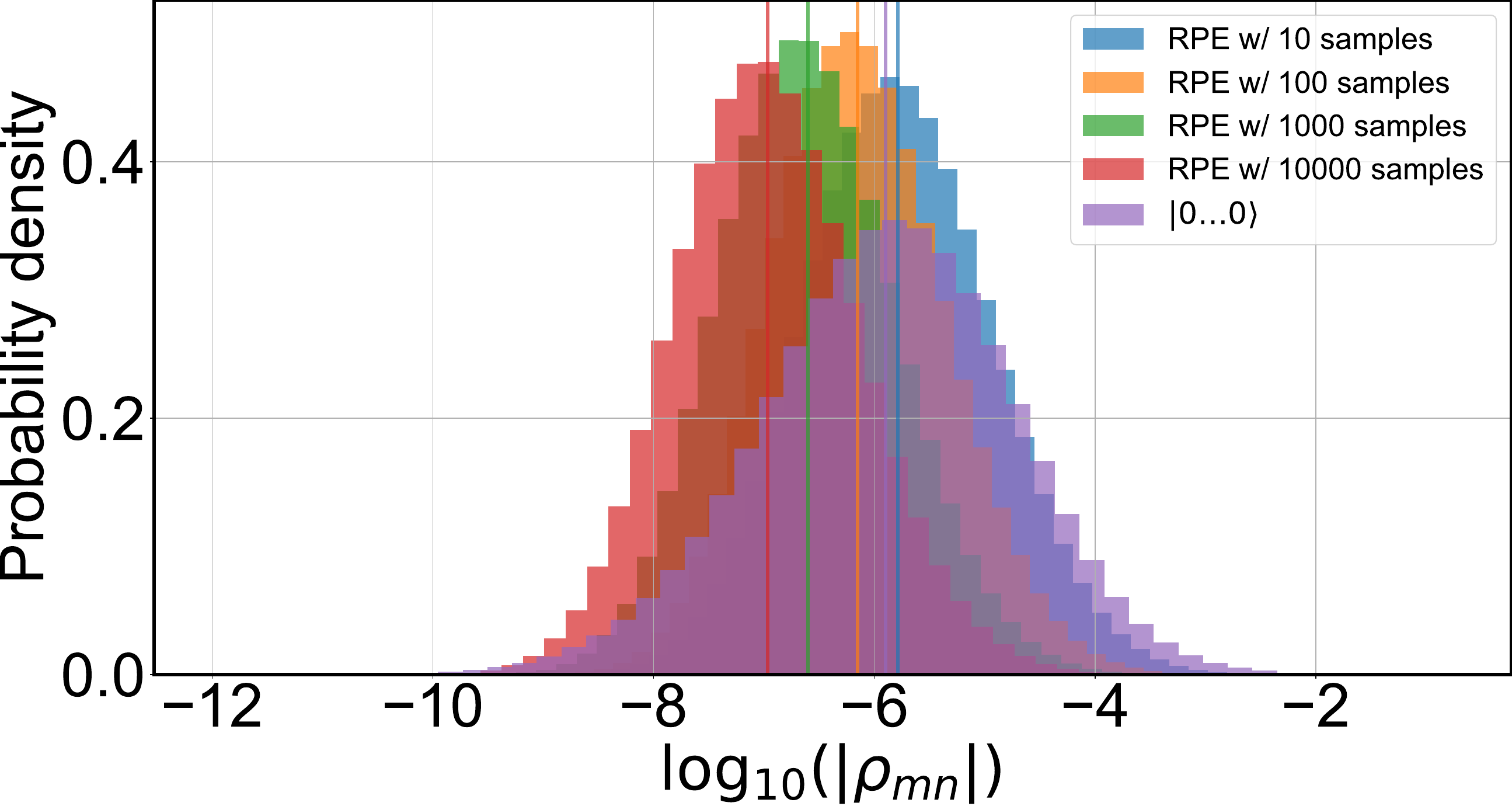}
    \caption{Histogram of the log-scaled off-diagonals of the density matrix in the energy eigenstate basis for the $\ket{0\cdots 0}\bra{0\cdots 0}$ state and the RPE mixed state at $E/N = -1.4$ and $N=12$ using different numbers of samples. The average log-scaled off-diagonal for each distribution is marked with a vertical line.}
    \label{fig:0state_rpe_offdiags}
\end{figure}

\begin{figure}[htbp!]
    \centering
    \includegraphics[width=0.45\textwidth]{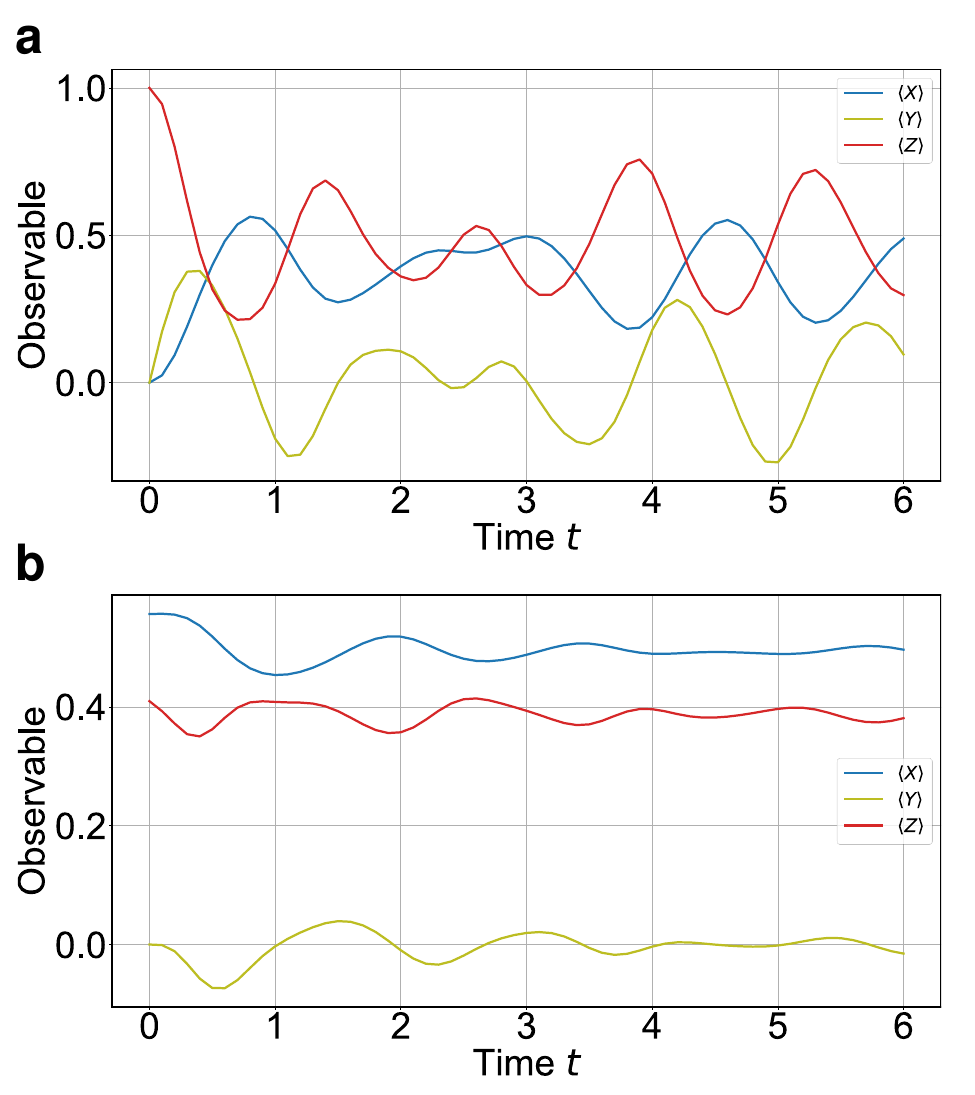}
    \caption{The average $X,Y,Z$ magnetizations versus time $t$ starting from \textbf{a} $\ket{0\cdots 0}$ and \textbf{b} the random product state ensemble mixed state (Eq.~(\ref{eq:rho_rpe})) with the same energy density $E/N=-1.4$. These results are from noiseless numerical simulations with system size $N=50$ and Trotter step $\tau=0.1$.}
    \label{fig:0state_rpe_observables_vs_t}
\end{figure}

Importantly, the RPE mixed state is closer to the diagonal ensemble than a single product state. To compute thermal observables from time evolution, one usually starts in a single product state $\ket{\psi}$, time evolves, and averages the dynamics in time. In the energy eigenstate basis $\ket{n}$ of the Hamiltonian, $\ket{\psi}=\sum_n c_n\ket{n}$ and the density matrix has non-trivial off-diagonal entries: $\rho=\ket{\psi}\bra{\psi}=\sum_{mn}c_m c_n^* |m\rangle\langle n|$. Upon time-averaging for time $t$, the off-diagonals are suppressed as $\sim 1/t$ and the density matrix approaches the diagonal ensemble $\rho_D=\sum_{n}|c_n|^2 |n\rangle\langle n|$. Due to ETH and the fact that product states have extensive energy variance, the diagonal ensemble approaches the quantum microcanonical ensemble at energy density $\langle H\rangle/N$ in the thermodynamic limit, in the sense that local observables match. Figure~\ref{fig:0state_rpe_offdiags} shows a histogram of the off-diagonals of the RPE mixed state when averaging over different numbers of product states, compared with a single product state at the same energy density. It shows that random product state averaging suppresses the off-diagonals. This suggests that state-averaging can be combined with time-averaging to speed up the approach to the diagonal ensemble. Even without time-averaging, we find that the time-evolved RPE mixed state quickly thermalizes. Figure~\ref{fig:0state_rpe_observables_vs_t} shows how local observables vary in time for a single product state compared with the RPE mixed state. For a single product state, non-trivial off-diagonals (coherences) in the initial state lead to long persistent oscillations in local observables. For the RPE mixed state, local observables decay (with oscillations) much more quickly to their thermal values.

The RPE also has connections with mean-field theory. In Fig.~\ref{fig:rpe_observables}, the shaded region corresponds to the range of energies spanned by product states. At the left (right) edge of the region is the lowest (highest) energy product state, also referred to as the mean-field (anti-) ground state. As one approaches in energy the boundaries of this region, the RPE mixed state becomes less and less mixed as it approaches the mean-field states (which is generally unique or with a small number of degenerate states), as discussed in Appendix~\ref{sec:rpe_stats}. This is an important consideration when using the RPE, as the RPE samples become more and more correlated as one approaches the boundaries, leading to less benefits from sampling. The black vertical dashed lines correspond to the exact ground and anti-ground states. The small separation between the mean-field ground state and the true ground state reflects the mean-field-like nature of the mixed-field Ising model and might not hold for other models. Nonetheless, product states always span a constant fraction of the Hilbert space and so a large range of energy densities can be explored with the RPE. 

\section{The effect of a single gate error} \label{sec:single_error}

\begin{figure}[htbp!]
    \centering
    \includegraphics[width=0.45\textwidth]{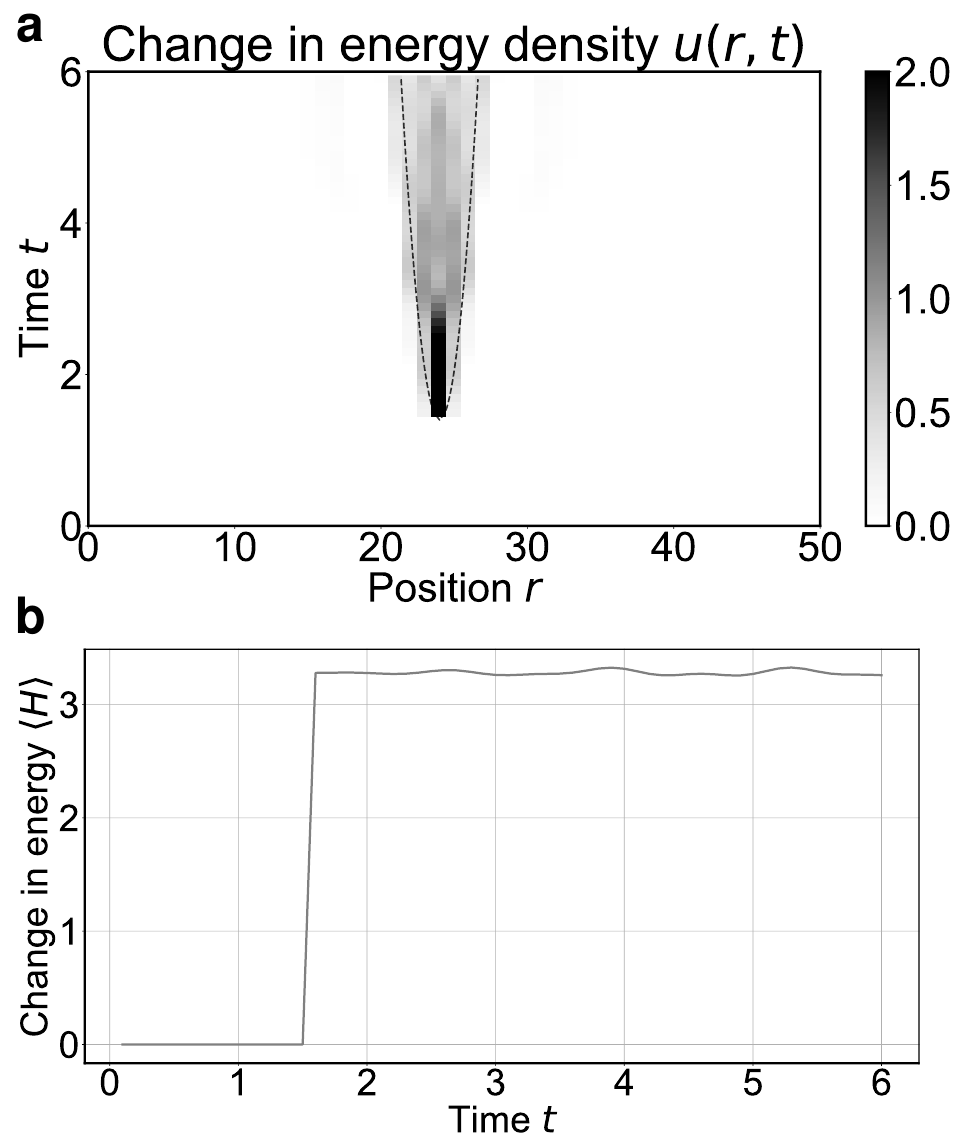}
    \caption{\textbf{a} The change in local energy density $u(r,t)=\langle \psi(t)|h_r|\psi(t)\rangle$ during time evolution starting from the $\ket{\psi(0)}=\ket{0\cdots 0}$ initial state, with a $Y$ error inserted at the center site at time $t_0=1.5$. For comparison, the dashed lines correspond to a diffusive envelope of width $\propto \sqrt{t}$. \textbf{b} The change in total energy $\langle H \rangle$ versus time $t$.}
    \label{fig:0state_energy_diffs_vs_t}
\end{figure}

We begin our analysis of gate errors by examining how a single Pauli error affects the dynamics of the Trotterized Hamiltonian simulation circuit Eq.~(\ref{eq:U}) in the regime when both $\tau$ and $t$ are small and energy is to a good approximation conserved. 

First, we examine how the error affects the energy of the system and how that energy change spreads in space. To assess this, we define the local energy density operator for our Hamiltonian Eq.~(\ref{eq:H}) as
\begin{align}
h_r = \begin{cases}
-\frac{1}{2}X_r X_{r+1} - g Z_r - h X_r& r = 1 \\
-\frac{1}{2}\left(X_{r-1} X_{r} + X_r X_{r+1} \right) - g Z_r - h X_r& 1 < r < N \\
-\frac{1}{2}X_{r-1} X_{r} - g Z_r - h X_r & r = N \\
\end{cases}, \label{eq:h_r}
\end{align}
which satisfies $H = \sum_{r=1}^N h_r$ \footnote{Note that this definition is not unique and other similar energy density operators can be defined, such as ones centered over bonds instead of sites.}. Figure~\ref{fig:0state_energy_diffs_vs_t}\textbf{a} shows how a single-site $Y$ error applied to the $\ket{0\cdots 0}$ product state at time $t_0=1.5$ changes the local energy density operator's expectation value in space and time. Figure~\ref{fig:0state_energy_diffs_vs_t}\textbf{b} shows how the total change in energy across the entire system changes with time. After the insertion of the $Y$ operator, which does not commute with the Hamiltonian $H$, the total energy abruptly increases by $\sim 3$ energy units and then remains unchanged for the duration of the evolution. This confirms our expectation that energy is approximately conserved at small Trotter steps and that an important effect of incoherent gate errors is that they drive the system's energy to zero. Because energy is approximately conserved and there are no other conserved quantities in this system, we expect the local energy density change to spread diffusively so that its front grows as $\sim \sqrt{t}$ \cite{Thomas2023}. The numerical results in Fig.~\ref{fig:0state_energy_diffs_vs_t}\textbf{a} are consistent with the expected diffusive $\sim \sqrt{t}$ scaling, which is marked by a black dashed line.

\begin{figure}[htbp!]
    \centering
    \includegraphics[width=0.45\textwidth]{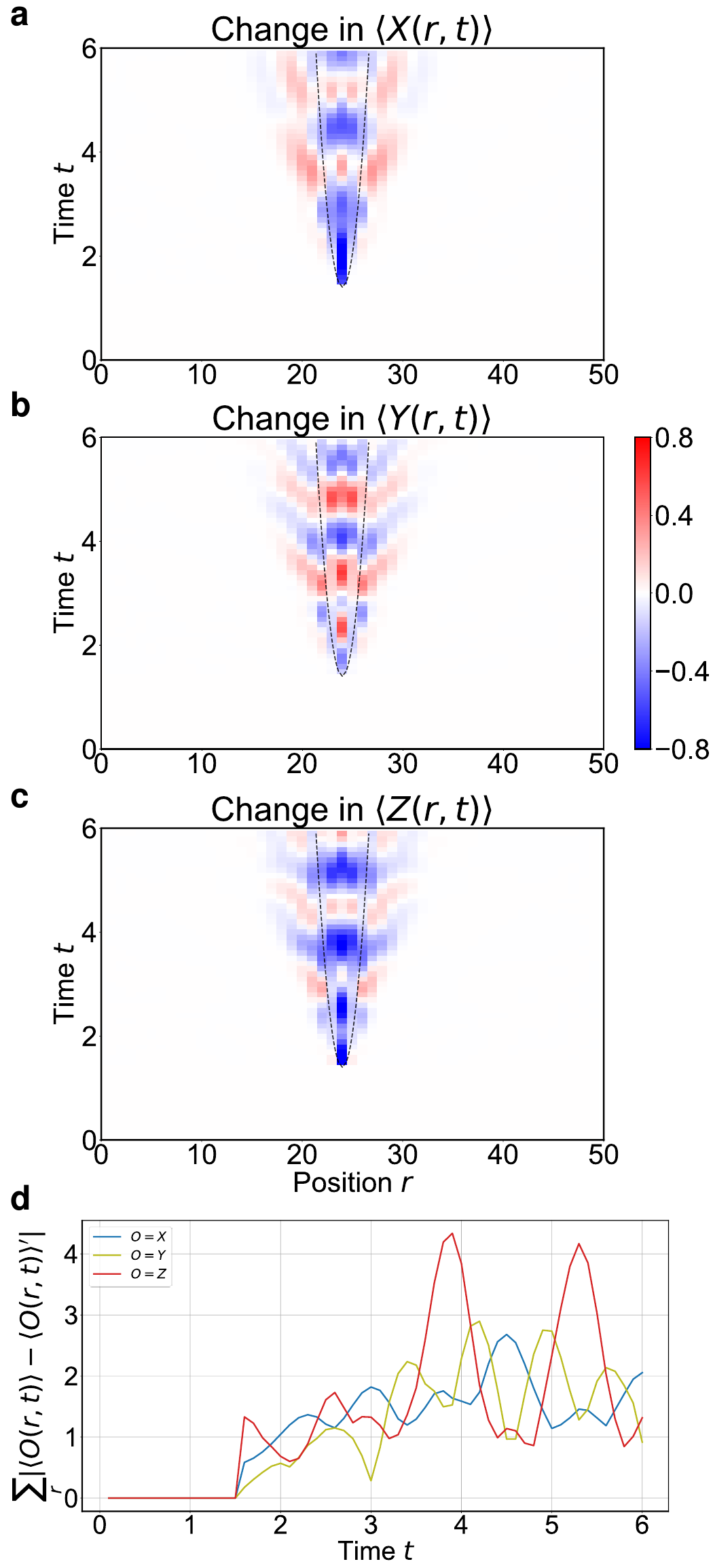}
    \caption{The change in single-site observables \textbf{a} $X$, \textbf{b} $Y$, \textbf{c} $Z$ during time evolution starting from the $\ket{\psi(0)}=\ket{0\cdots 0}$ initial state, with a $Y$ error inserted at the center site at time $t_0=1.5$. For comparison, the dashed lines correspond to a diffusive envelope of width $\propto \sqrt{t}$. \textbf{d} The total spatially-integrated change in each observable versus time $t$.}
    \label{fig:0state_observable_diffs_vs_t}
\end{figure}

Next, we consider how a single error affects local observables. Figure~\ref{fig:0state_observable_diffs_vs_t}\textbf{a}-\textbf{c} shows the change in $X,Y,$ and $Z$ observables as a function of time and space due to the same single $Y$ error discussed above. The change in these observables are clearly correlated with the change in local energy density and its $\sim\sqrt{t}$ diffusive window. However, these changes appear to have non-trivial oscillations and structure outside of the diffusive window. The total change in these observables integrated over space is shown in Fig.~\ref{fig:0state_observable_diffs_vs_t}\textbf{d}. Unlike the total energy, the total changes in observables are \emph{not} conserved with time, though are likely bounded in time, and oscillate in complicated ways among the local operators.

To simplify our analysis going forward, we mainly focus on the local trace distance error $D_{\textrm{tr}}(r)$, Eq.~(\ref{eq:tr_dist}), which holistically captures the error of all local observables. For single-qubit Paulis $P_r$, this quantity bounds the observable error $|\tr(\rho_{\textrm{ideal}}(r) P_r) - \tr(\rho_{\textrm{error}}(r) P_r)| \leq 2D_{\textrm{tr}}(r)$ \cite{nielsen2010} and so is a useful measure of observable error that encompasses errors in all possible measurement bases \footnote{Note that for 1-RDMs $\rho(r)=\frac{1}{2}(I+p_X X+ p_Y Y + p_Z Z)$, the $\ell_1$ norm distance between density matrices is equivalent to the $\ell_2$ norm distance between Bloch vectors: $||\rho_{\textrm{ideal}}(r) - \rho_{\textrm{error}}(r)||_1 =||\vec{p}_{\textrm{ideal}} - \vec{p}_{\textrm{error}}||_2$ \cite{nielsen2010}.}.

As we see from Fig.~\ref{fig:0state_energy_diffs_vs_t}~and~\ref{fig:0state_observable_diffs_vs_t}, while there is a correlation between changes in energy density and local observables in generic dynamics under an ETH Hamiltonian, it is obfuscated by the non-generic properties of the initial state. In an attempt to isolate the dynamical properties of a \emph{typical} product state, we find it helpful to study random product states with the same energy drawn from the RPE. By averaging over the RPE, we find that non-generic oscillatory behavior is averaged out and the correlation between local energy density and trace distance error becomes quantitatively sharper. We believe that this is due to the fact that the RPE mixed state at the same energy density has significantly reduced coherences between energy eigenstates compared to a single particular product state.

\begin{figure*}[htbp!]
    \centering
    \includegraphics[width=0.95\textwidth]{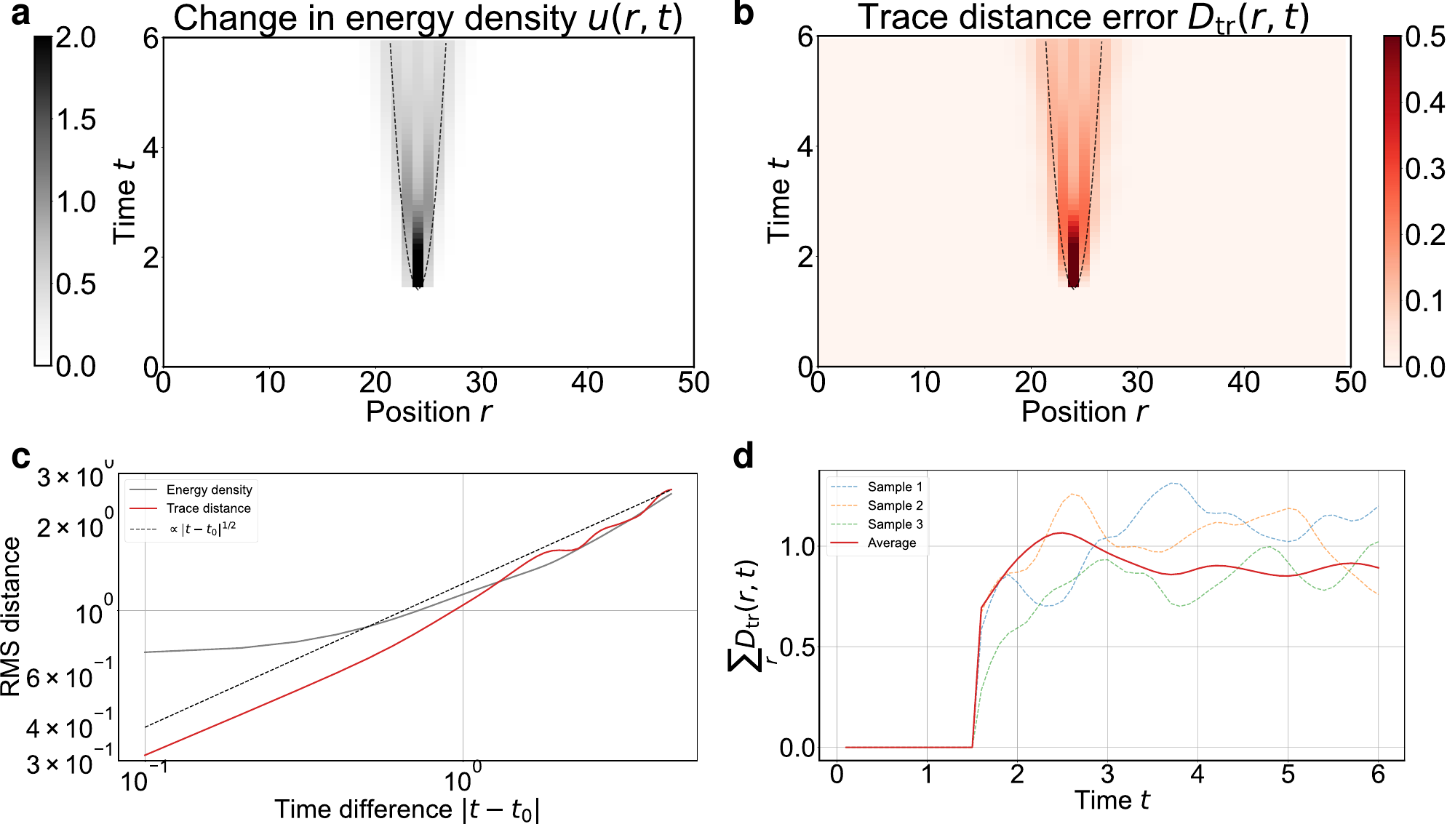}
    \caption{The change in \textbf{a} local energy density and \textbf{b} trace distance error during time-evolution starting from the random product state ensemble (RPE) mixed state at energy density $E/N=-1.4$, with $N=50$ sites and a $Y$ error inserted at the center site at time $t_0=1.5$. \textbf{c} The root-mean-square distance from the center site versus time, using the energy density and trace distance error profiles at each time as normalized probability distributions over space. \textbf{d} The total trace distance error versus time for the RPE (solid line) and individual random product states sampled from the RPE (dashed lines).}
    \label{fig:rpe_singleerror}
\end{figure*}

Upon the insertion of a single Pauli error, the RPE behaves qualitatively differently than a single product state, presumably due to its reduced coherences. In Fig.~\ref{fig:rpe_singleerror}\textbf{a}-\textbf{b}, we see for the RPE mixed state that the energy density change and trace distance error for a single error insertion are closely correlated. In Fig.~\ref{fig:rpe_singleerror}\textbf{c}, we show how the root-mean-square size $r_{\textrm{rms}} \equiv \left(\frac{\sum_r f(r) (r-N/2)^2}{\sum_{r'} f(r')} \right)^{1/2}$ of the normalized energy density change ($f(r)=\Delta u(r)$) and trace distance error ($f(r)=D_{\textrm{tr}}(r)$) profiles varied with time after the insertion of a $Y$ error at the center site $N/2$ at time $t_0=1.5$. We find that $r_{\textrm{rms}} \sim |t-t_0|^{1/2}$ displays diffusive scaling for both quantities. The trace distance error does not show additional oscillation and structure outside of the diffusive profile, which were seen in Fig.~\ref{fig:0state_observable_diffs_vs_t}\textbf{a}-\textbf{c} for the $\ket{0\cdots 0}$ state. Moreover, for the RPE, we find that the total trace distance error, shown in Fig.~\ref{fig:rpe_singleerror}\textbf{d}, appears nearly conserved in time, with only minor temporal oscillations that appear to decay in amplitude quickly with time. However, for individual samples from the RPE (marked as dashed lines), there are significant oscillations in the trace distance that are not being damped within the time scales of our simulations. Such large oscillations were also seen for the $\ket{0\cdots 0}$ state. Note that the trace-distance error $D_{\textrm{tr}}(r)$ is a non-linear function of the 1-RDM $\rho(r)$ at site $r$, which means that the average of the trace-distance error for RPE samples is not the same as the trace-distance error for the sample-averaged RPE mixed-state: $\frac{1}{M} \sum_{s=1}^M ||\rho_s(r)-\rho_s'(r)||_1 \neq  ||\frac{1}{M} \sum_{s=1}^M(\rho_s(r)-\rho_s'(r))||_1$. 

\section{The effect of many gate errors} \label{sec:many_errors}

Next, we explore how the presence of gate errors after every two-qubit gate in the quantum circuit of Eq.~(\ref{eq:U}) affects the Trotterized dynamics. In the H-Series quantum computers, two-qubit gate errors are the dominant error source. We perform numerical simulations with a simple two-qubit depolarizing error model following each two-qubit gate, with the gate error of the form $p(\tau)=p_0 + p_1 \tau$ measured in the H-Series hardware (see Fig.~\ref{fig:linear_angle_scaling}). The incoherent nature of the error channel is important since it allows us to interpret each noisy gate as independently and randomly inserting energy at some rate. Coherent gate errors can effectively add new interactions terms to the Hamiltonian being simulated and therefore can have complicated state and model-dependent effects. For this reason, and because coherent errors to some degree can be mitigated experimentally through careful calibrations and techniques such as Pauli twirling \cite{hashim2021} and dynamical decoupling \cite{viola1999}, we focus our analysis on incoherent errors.  

\begin{figure}[htbp!]
    \centering
    \includegraphics[width=0.45\textwidth]{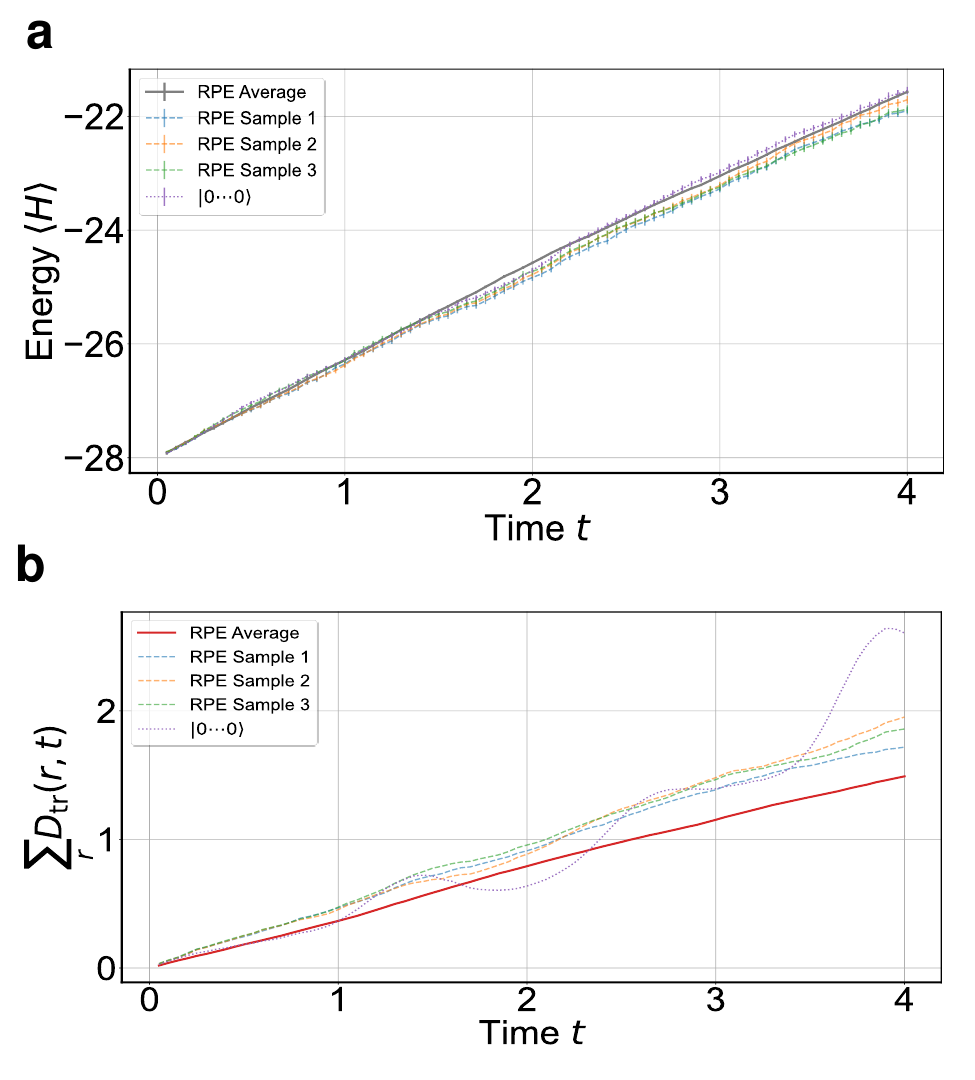}
    \caption{\textbf{a} The energy of an $N=20$ site spin chain versus time $t$ when two-qubit gates in the circuit undergo depolarizing noise. \textbf{b} The spatially-integrated trace distance error versus time. In both plots, the evolution starting from the $E/N=-1.4$ random product state ensemble (RPE) mixed state is indicated by a solid line, while the evolution starting from individually sampled product states from the RPE (and the $|0\cdots 0\rangle$ state) are shown with dashed (dotted) lines. The Trotter step is $\tau=0.05$ and the average gate infidelity is $1.1\times 10^{-3}$ (process infidelity $1.37 \times 10^{-3}$).} 
    \label{fig:rpe_many_errors_vs_t}
\end{figure}

To isolate the effects of gate errors, we again perform numerical simulations for short times at a small Trotter step, where Trotter errors are negligible and the (noiseless) dynamics are a good approximation of continuous time evolution. In this regime, we observe that gate errors cause the expected linear-in-time growth of energy, as can be seen in Fig.~\ref{fig:rpe_many_errors_vs_t}\textbf{a} for the RPE mixed state, as well as for RPE samples and the $\ket{0\cdots 0}$ state. In these simulations, the Trotter step is $\tau=0.05$ and the depolarizing noise on each 2Q gate corresponds to a process infidelity of $1.37 \times 10^{-3}$ which approximates the experimentally measured performance of the \emph{max-angle} $e^{-i\pi Z\otimes Z/4}$ two-qubit gate on the H1-1 quantum computer at the time of experimental data acquisition (this error rate is much larger than the actual H1-1 error rate at $\tau=0.05$, but is chosen here for illustrative purposes). It is important to note that by the end of this simulation, the energy density only changes by a small amount, $\Delta E/N \approx 0.3$. For large changes in energy density, which would occur at later times or for larger gate error rates, we would generally expect the growth of energy with time to become non-linear as the state approaches $E/N=0$ (infinite temperature). Here we are working in the perturbative energy density regime $\Delta E/N \ll 1$ where it is reasonable to expect that each gate error causes approximately the same change in energy and so the total change in energy is proportional to the total number of 2Q gates in the circuit $N_{2Q}=Nt/\tau$ and so grows linearly.

In Fig.~\ref{fig:rpe_many_errors_vs_t}\textbf{b}, we see that the total trace-distance error for the RPE mixed state also appears to grow linearly in time, consistent with our findings in the previous sections that total trace distance error is approximately constant for single errors. However, we see that the behavior is qualitatively different for specific product states, such as the $\ket{0\cdots 0}$ state (dotted lines) and individual RPE samples (dashed lines), which show significant oscillations in time. Nonetheless, despite the oscillations, it does seem that the trace distance error is growing approximately linear in time, particularly for ``typical'' product states drawn from the RPE.

\section{The effect of Trotter errors} \label{sec:trotter_errors}

Here we numerically explore the impact of Trotter errors on local observables. We are particularly interested in how the error scales with Trotter step $\tau$ and time $t$ for short times.

\subsection{Scaling with Trotter step}

\begin{figure}
    \includegraphics[width=0.45\textwidth]{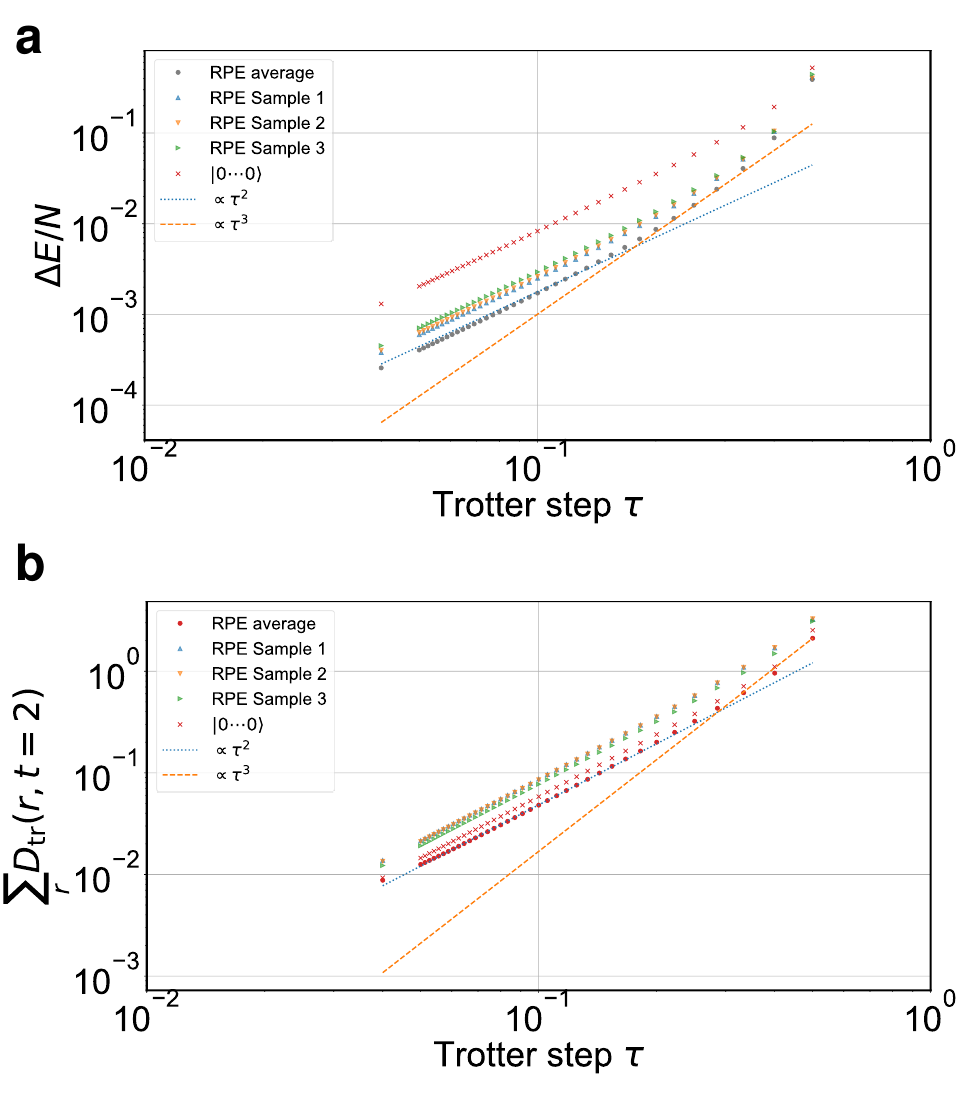}
    \caption{The \textbf{a} change in energy density and \textbf{b} total trace distance error versus Trotter step for the RPE mixed state, individual samples from the RPE, and the $|0\cdots 0\rangle$ state, for an $N=20$ site chain at time $t=2$. There are no gate errors in these simulations and $\propto \tau^2$ and $\propto \tau^3$ curves are shown for reference.}
    \label{fig:fixedstates_trotter_error_vs_tau}
\end{figure}
Here we numerically verify our expectation that Trotter errors at a fixed time scale as $O(\tau^2)$ for the second-order Trotter decomposition. One way to expect this scaling for observable errors is by considering the lowest-order in $\tau$ approximation to the Floquet Hamiltonian, Eq.~(\ref{eq:HF}). From time-dependent perturbation theory (see Ref.~\onlinecite{heyl2019} and Appendix~\ref{sec:tdpt}), the $O(\tau^2)$ perturbation to the Hamiltonian leads to an $O(\tau^2)$ perturbation to local observables.

In Fig.~\ref{fig:fixedstates_trotter_error_vs_tau}, we show the change in energy density and in local observables (as measured by the total trace distance error) for different initial states arising from Trotter errors. The results are from noiseless numerical simulations. For all states considered --- individual random product states, the RPE mixed state, and the $\ket{0\cdots 0}$ state --- both errors in energy and observables show $\sim \tau^2$ scaling at small Trotter step for a fixed time $t=2$. At large Trotter step $\gtrsim 0.3$, the Trotter errors exhibit a different scaling due to higher-order error terms.

\subsection{Scaling with time} \label{sec:trotter_scaling_with_time}

Next, we examine how Trotter errors behave in time for thermalizing dynamics. From simple bounds on Trotter errors, we would expect that a single Trotter layer would have error $O(\tau^3)$ and that the total error for $D$ layers would be $O(\tau^3 D)=O(\tau^2 t)$, leading to linear-in-time growth of errors. However, empirically we find to a good approximation that Trotter errors on local observables for thermalizing dynamics can appear constant (or bounded) in time for a range of times. Similar behavior was observed in Ref.~\onlinecite{heyl2019}.

\begin{figure}[htbp!]
    \centering
    \includegraphics[width=0.45\textwidth]{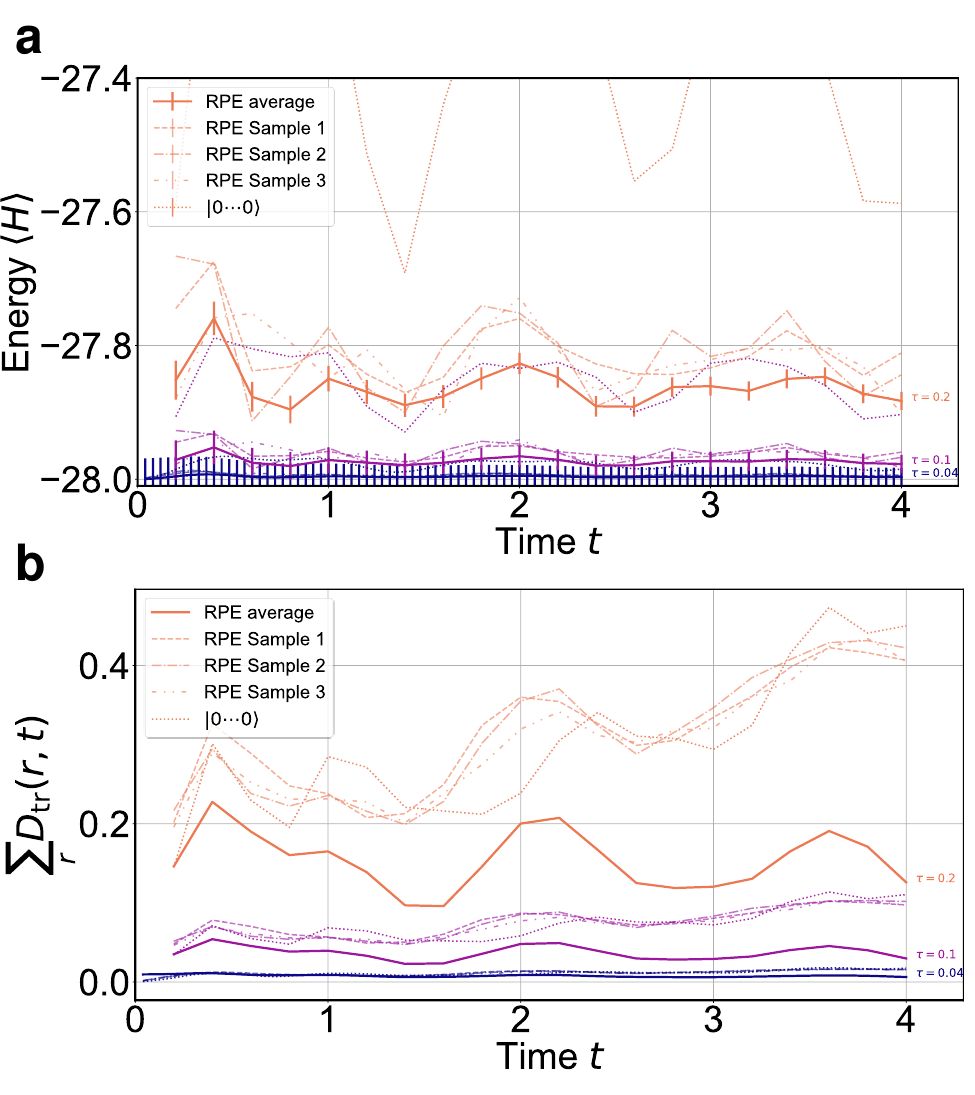}
    \caption{\textbf{a} Energy and \textbf{b} total trace distance error versus time $t$ starting from an RPE mixed state (solid lines) and individual product states sampled from the RPE and the $\ket{0\cdots 0}$ state (dashed and dotted lines), for fixed $\tau=0.04,0.1,0.2$ (different colors). These results are from numerical simulations with no gate errors.} 
    \label{fig:rpe_trottererrors_vs_t}
\end{figure}
Fig.~\ref{fig:rpe_trottererrors_vs_t} shows how Trotter errors affect the energy and local observables over time for different initial states. In Fig.~\ref{fig:rpe_trottererrors_vs_t}\textbf{a}, we see that for all initial states considered energy oscillates with time and does not appreciably grow, at least for short times. It appears that the energy of the time-evolved state quickly (in one or a few Trotter steps) jumps to a value different from the initial state and remains close to that value for a significant amount of time. Interestingly, this value appears different for the $\ket{0\cdots 0}$ state compared to random product states drawn from the RPE, perhaps because the $\ket{0 \cdots 0}$ state is not typical of product states at its energy density. Figure~\ref{fig:rpe_trottererrors_vs_t}\textbf{b} shows the observable Trotter error versus time, which displays qualitatively different behavior than the energy. While the RPE mixed state's observable error does not appreciably grow with time on this timescale (up to oscillations), the observable error of individual product states drawn from the RPE and the $\ket{0 \cdots 0}$ state appear to grow roughly linearly in time (up to oscillations).

The observed behavior can be explained using time-dependent perturbation theory, or the method presented in Ref.~\onlinecite{chen2024}. In time-dependent perturbation theory, one treats the lowest-order correction to the ideal Hamiltonian appearing in the Floquet Hamiltonian Eq.~(\ref{eq:HF}) as a perturbation,
\begin{align}
\tau^2 V\equiv H_F-H =\frac{\tau^2}{24}[H_1+2H_2, [H_1, H_2]],
\end{align}
and then computes the observable perturbatively in $\tau^2 V$. In Appendix~\ref{sec:tdpt}, we use first-order time-dependent perturbation theory to arrive at the observable error
\begin{align}
\label{eq:trotter_prediction}
\langle \Delta \mathcal{O}(t)\rangle =&-i\tau^2 t\sum_{n}V_{nn}{\rm tr}\Big(\mathcal{O}\big[\ket{n}\!\bra{n},\rho_H(t)\big]\Big)\\
-\tau^2\!\sum_{m\neq n}&\frac{V_{mn}\big(1-e^{-i t(E_m-E_n)}\big)}{E_m-E_n}{\rm tr}\Big(\mathcal{O}\big[\ket{m}\!\bra{n},\rho_H(t)\big]\Big),\nonumber
\end{align}
where $\ket{n}$ are energy eigenstates with energy $E_n$ and we assume throughout this discussion that $H$ is non-degenerate. 

From \eref{eq:trotter_prediction}, we can see how thermalization suppresses the growth of Trotter errors with time. In thermal equilibrium, $\rho_{D}=\sum_n p_n |n\rangle\langle n|$ is a diagonal ensemble, causing the first $\tau^2 t$ term in \eref{eq:trotter_prediction} to vanish \cite{chen2024} (since $[\ket{n}\!\bra{n},\rho_{D}]=0$) and suggesting that Trotter errors on local observables may not grow with time in thermal equilibrium.

\begin{figure}[htbp!]
    \centering
    \includegraphics[width=0.45\textwidth]{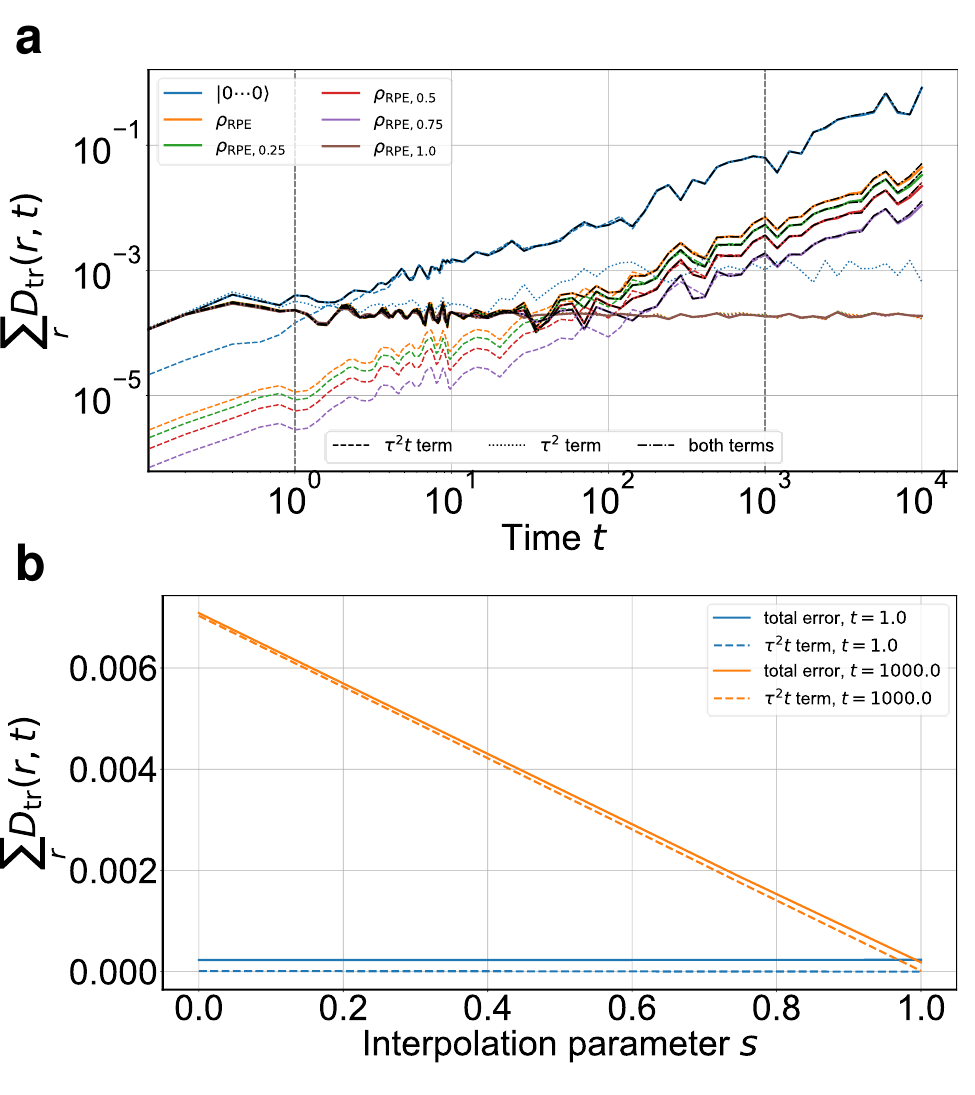}
    \caption{\textbf{a} The total trace distance error versus time with fixed Trotter step $\tau=0.01$, computed using exact diagonalization for an $N=12$ site chain and including no gate errors. The different colored curves correspond to different initial states, with $\rho_{\textrm{RPE},s} \equiv (1-s) \rho_{\textrm{RPE}} + s \rho_{\textrm{RPE},D}$ being an interpolation between the random product state ensemble (RPE) mixed state $\rho_{\textrm{RPE}}$ at $s=0$ and the RPE diagonal ensemble $\rho_{\textrm{RPE},D}$ at $s=1$. The colored dashed and dotted lines correspond to the trace distance error computed using the terms in Eq.~(\ref{eq:trotter_prediction}) with $\tau^2t$ and $\tau^2$ prefactors, respectively. The black dotted-dashed lines correspond to the error computed using the sum of the two terms. \textbf{b} The total trace distance error versus the interpolation parameter $p$ for fixed Trotter step $\tau=0.01$ at times $t=1$ and $t=1000$ (marked with vertical black dashed lines in \textbf{a}). The colored dashed lines correspond to the trace distance error contribution coming from the $\tau^2 t$ term in Eq.~(\ref{eq:trotter_prediction}).} 
    \label{fig:rpe_interpolation}
\end{figure}
Fig.~\ref{fig:rpe_interpolation} shows numerically how approaching the diagonal ensemble suppresses the growth of Trotter errors. In Fig.~\ref{fig:rpe_interpolation}\textbf{a}, we show the observable error versus time for the $\ket{0\cdots 0}$ state, the RPE mixed state $\rho_{\textrm{RPE}}$, and interpolations of the RPE mixed state and its diagonal ensemble: $\rho_{\textrm{RPE},s} \equiv (1-s) \rho_{\textrm{RPE}} + s \rho_{\textrm{RPE},D}$, where $\rho_{\textrm{RPE},D}\equiv\sum_n \langle n |\rho_{\textrm{RPE}}|n\rangle \ket{n}\bra{n}$. The dashed 
and dotted lines are obtained from numerically evaluating the $\tau^2 t$ and $\tau^2$ terms in Eq.~(\ref{eq:trotter_prediction}). The observable error computed using the sum of the two terms (black dotted-dashed lines) matches closely with the exact numerical result (solid lines), confirming that the perturbative treatment leading to Eq.~(\ref{eq:trotter_prediction}) is accurate for $\tau=0.01$ over the entire timescale explored. At late times, we see linear-in-time growth that matches well with the $\tau^2 t$ term, while at early times we see an oscillatory behavior that remains bounded. When the linear term begins to dominate is determined by how close the initial state is to the diagonal ensemble. The $\ket{0 \cdots 0}$ state, being the furthest from the diagonal ensemble, in $t\sim 1$ time is already behaving linearly. The RPE mixed state, since it is closer to the diagonal ensemble, suppresses the linear growth until $t\sim 10$. Furthermore, the interpolated RPE mixed states further suppress the linear growth, with the $\rho_{\textrm{RPE},D}$ state perfectly suppressing the linear growth. Figure~\ref{fig:rpe_interpolation}\textbf{b} shows how the observable error changes for the interpolated RPE states at early and late times (marked with black vertical dashed lines in Fig.~\ref{fig:rpe_interpolation}\textbf{a}). At early times, the $\tau^2 t$ term is already small for the RPE mixed state, so being closer to the diagonal ensemble has a negligible effect. At late times, the $\tau^2 t$ term is dominant, so approaching the diagonal ensemble suppresses the error, making the remaining error comparable to its value at early times.

Another effect determined from Eq.~(\ref{eq:trotter_prediction}), is that Trotter errors do not grow with time for observables that commute with the Hamiltonian. If $[\mathcal{O},H]=0$, $\mathcal{O}$ and $H$ can be simultaneously diagonalized so that energy eigenstates $\ket{n}$ are also eigenstates of $\mathcal{O}$. In this case, $\sum_n V_{nn} \tr(\mathcal{O}[\ket{n}\!\bra{n},\rho_H(t)])=\sum_n V_{nn} \tr([\mathcal{O},\ket{n}\!\bra{n}]\rho_H(t))=0$ and the $\tau^2 t$ term vanishes. This behavior explains why the Trotter error on the energy $\langle H \rangle$ (Fig.~\ref{fig:rpe_trottererrors_vs_t}\textbf{a}) does not grow with time for any initial state.

\section{Tools for estimating the accuracy of Trotterized dynamics} \label{sec:tools}

Due to the limited availability of quantum computing resources, for the purposes of designing experiments it is helpful to use analytical estimates or efficient classical simulations to predict the performance of a quantum circuit on a real device. Here we discuss different efficient heuristic methods for estimating the accuracy of Trotterized Hamiltonian simulation.

\subsection{Classical simulation with fitting}

A simple estimation protocol is to classically simulate \emph{for short times} the Trotterized circuit with an RPE initial state and a gate error model. Using the simulation, one can estimate the observable error and fit it to Eq.~(\ref{eq:error_model}). Then, the fitted model can be used to predict the performance of the circuit at later times or other Trotter steps.

\begin{figure}[htbp!]
    \centering
    \includegraphics[width=0.45\textwidth]{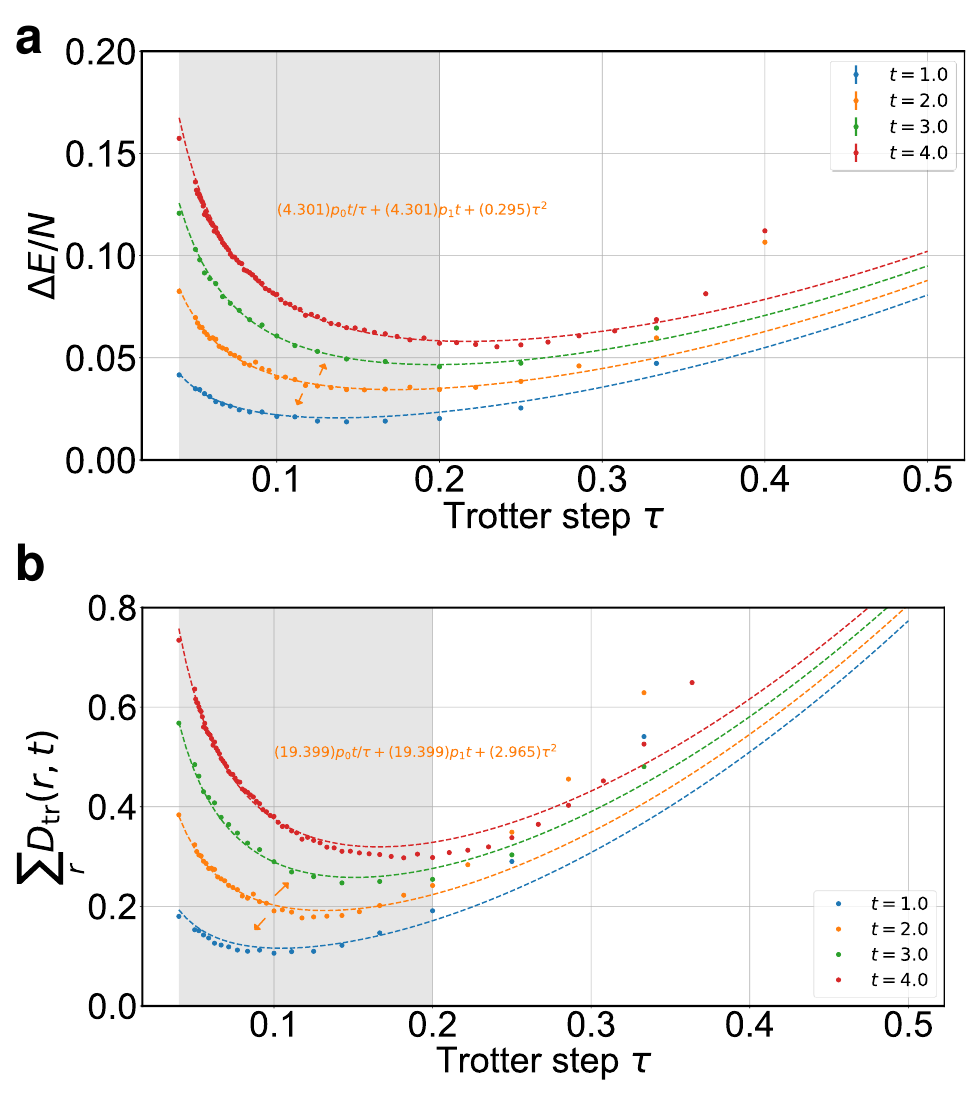}
    \caption{\textbf{a} Change in energy density and \textbf{b} total trace distance error versus Trotter step $\tau$ starting from an RPE mixed state at different times $t$. We fit the $t=2$ curve to the form $Sp_0 t/\tau + S p_1 t + C \tau^2$ with $p_0=3.5\times 10^{-4}$ and $p_1=9.6\times 10^{-4}$ using points with $0 \leq \tau \leq 0.2$ (shaded region) to obtain the fit parameters $S,C$ and then plot the fitted curve for other $t$ values.} 
    \label{fig:rpe_errorscaling_vs_tau_fits}
\end{figure}
Fig.~\ref{fig:rpe_errorscaling_vs_tau_fits} demonstrates this procedure. In this figure, numerical dynamics simulations starting from the RPE are shown. The dashed lines are a fit of Eq.~(\ref{eq:error_model}) obtained from the $t=2$ simulation data. The fit obtained at one time can be used to obtain decently accurate estimates of observable errors at other times. This agreement is heuristic and will not be perfect. 

\begin{figure*}[htbp!]
    \centering
    \includegraphics[width=0.95\textwidth]{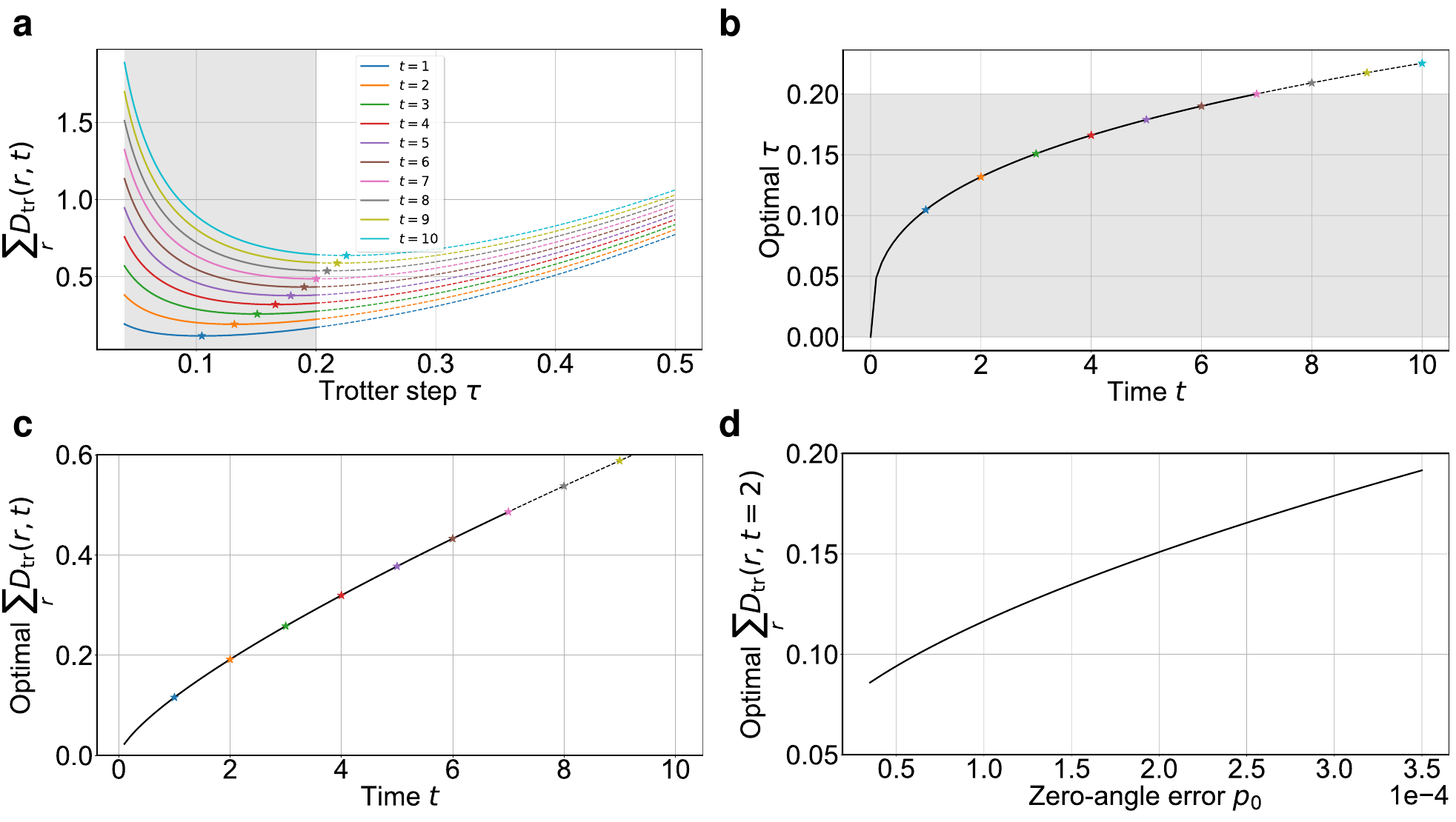}
    \caption{\textbf{a} The simple model Eq.~(\ref{eq:error_model}) for total trace distance error versus Trotter step for different times, using the fits obtained in Fig.~\ref{fig:rpe_errorscaling_vs_tau_fits}\textbf{b}. The optimal Trotter steps that minimize the trace distance error for each time are marked with  stars. The model is only valid for small $\tau$, so the shaded region (solid lines) indicates where the model is most reliable. \textbf{b} The optimal Trotter steps versus time. \textbf{c} The trace distance error versus time at the optimal Trotter steps shown in \textbf{b}. \textbf{d} The trace distance error at the optimal Trotter steps versus the zero-angle two-qubit gate infidelity $p_0$, while keeping the max-angle $p(\tau=\pi/4)=p_0 + p_1 \times \pi/4$ gate infidelity fixed to $1.1\times 10^{-3}$.} 
    \label{fig:rpe_tau_fits_analysis}
\end{figure*}

As shown in Fig.~\ref{fig:rpe_tau_fits_analysis}, using the fitted model, one can determine the optimal Trotter step that minimizes the observable error of the Trotterized dynamics. Assuming that Eq.~(\ref{eq:error_model}) holds, the optimal Trotter step is
\begin{align}
    \tau_{optimal} = \left(\frac{Sp_0 t}{2C}\right)^{1/3}
\end{align}
for a second-order Trotter decomposition, as shown in Fig.~\ref{fig:rpe_tau_fits_analysis}\textbf{b}. The values of the observable error at the optimal Trotter steps are
\begin{align}
    3C^{1/3}\left(\frac{Sp_0 t}{2}\right)^{2/3} + p_1 S t
\end{align}
and are shown in Fig.~\ref{fig:rpe_tau_fits_analysis}\textbf{c}. Note that Eq.~(\ref{eq:error_model}) only captures the lowest-order Trotter errors, so will be inaccurate when the Trotter errors become non-linear. The shaded areas with $\tau \leq 0.2$ roughly indicate where the model will hold; the dashed lines indicate where the model is likely inaccurate. Finally, Fig.~\ref{fig:rpe_tau_fits_analysis}\textbf{d} shows the $\propto p_0^{2/3}$ scaling of the optimal observable error with zero-angle gate error $p_0$, highlighting the importance of the performance of the arbitrary-angle $U_{ZZ}(\tau)$ gate at small angles.

The main benefit of the simulate-and-fit protocol is that it is classically efficient even in spatial dimension greater than one, by using tensor network techniques such as PEPS. It allows for rapid estimation of optimal Trotter steps, quantitative estimates for simulation accuracy, and even provides estimates useful for guiding hardware improvements.

The main drawback of the protocol is that it is heuristic and only applies to a limited regime of times and Trotter steps. The simple model in Eq.~(\ref{eq:error_model}) assumes a constant Trotter error, neglecting non-trivial oscillations in time (see Sec.~\ref{sec:trotter_errors}). Other discrepancies could arise from oscillations due to non-equilibrium properties of the initial state, though that should be significantly suppressed for the RPE. The method is also limited to producing short-time predictions, since Eq.~(\ref{eq:error_model}) is only expected to hold for $Spt \ll 1$. This might not be too problematic, since outside of the linear regime observable errors will be so large that a quantum simulation on real hardware might not provide useful results (without error mitigation).

\subsection{The RPE as an approximate thermal ensemble}

An alternative protocol is to use the RPE to predict the behavior of the noisy Trotterized circuit. This protocol makes the coarse approximations that (1) the time-evolved state during Trotterized dynamics is at all times in thermal equilibrium and that (2) thermal equilibrium is well-approximated by the RPE. By modeling the effect of gate errors as changing the energy of the RPE, we can then estimate how observables will change with time.

\begin{figure}[htbp!]
    \centering
    \includegraphics[width=0.45\textwidth]{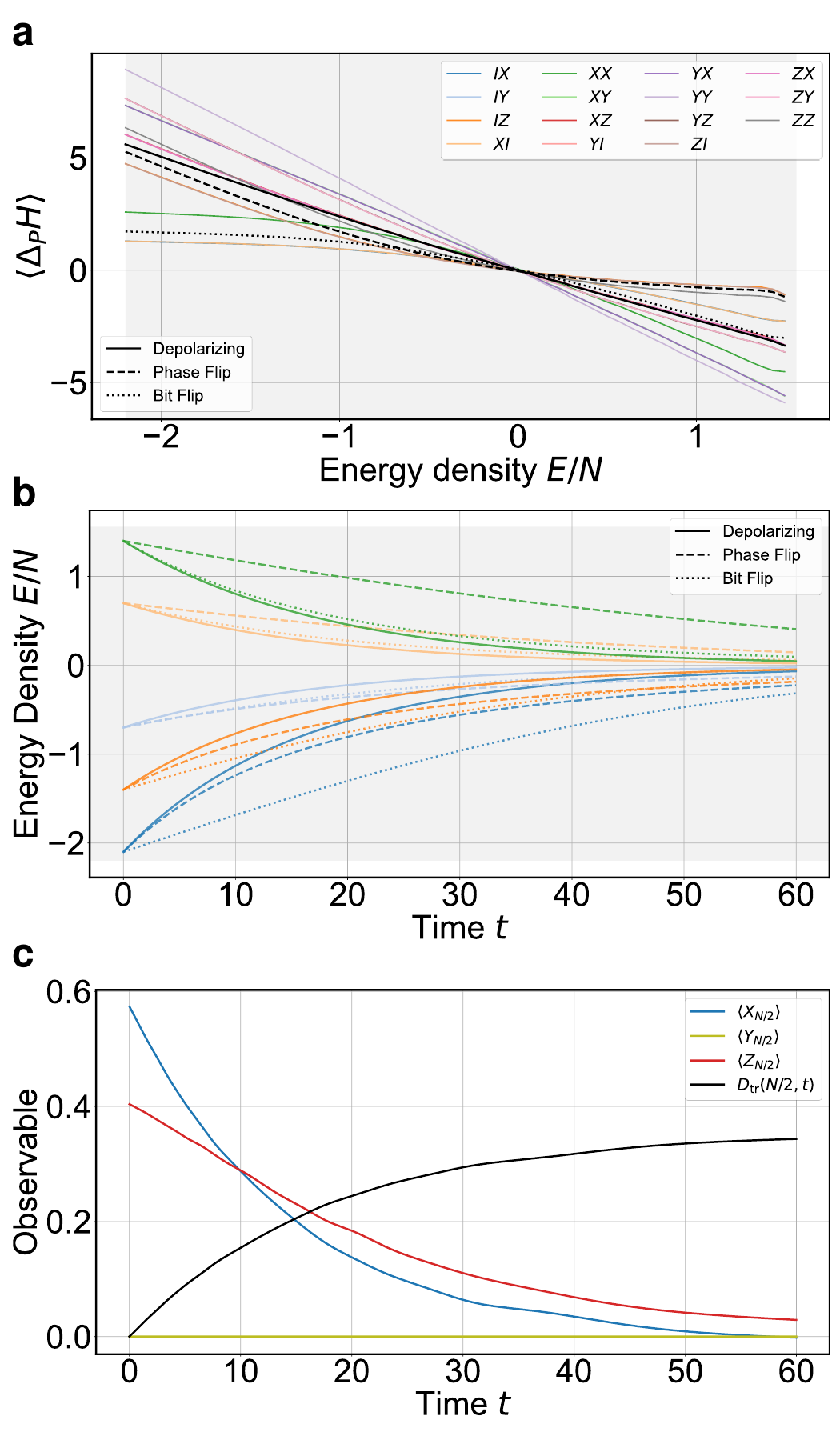}
    \caption{\textbf{a} The change in energy of the RPE mixed state versus energy density $E/N$ due to the application of a two-qubit Pauli $P$ on the center two sites of the $N=20$ spin chain. Black solid, dashed, and dotted lines indicate the average change in energy for a depolarizing, phase flip, and bit flip error, respectively. \textbf{b} The change in energy density versus time $t$ for different initial energy densities assuming the RPE ensemble behavior in \textbf{a}, for an $N=20$ chain with $\tau=0.02$ and error per two-qubit gate of $p=5\times 10^{-4}.$ The solid, dashed, and dotted lines correspond to a depolarizing error model (with $p/15$ probability for each non-identity two-qubit Pauli error $P$), a phase flip error model (with $p/3$ probability for $ZI,IZ,ZZ$ Paulis), and a bit flip error model (with $p/3$ probability for $XI,IX,XX$ Paulis). \textbf{c} Single-qubit observables and trace distance error at the center site of the spin chain versus time, starting from the RPE state at energy density $E/N=-1.4$ and assuming the depolarizing model behavior in \textbf{b}.}
    \label{fig:rpe_energy_change_predictions}
\end{figure}

Suppose that following a gate in our circuit, a single Pauli $P$ is applied to our state $\rho$. This causes a change in energy
\begin{align}
\Delta_P E &= \tr (P\rho P H) - \tr (\rho H) \nonumber \\
&= \tr(\rho(PHP - H)) \equiv \tr(\rho \Delta_P H) \label{eq:deltaE}
\end{align}
that depends on the expectation value of the operator $\Delta_P H \equiv PHP - H$ in the current state $\rho$. For example, for single-qubit Paulis and the mixed-field Ising model in Eq.~(\ref{eq:H}),
\begin{align}
\Delta_{X_r} H &= 2g Z_r, \nonumber \\
\Delta_{Y_r} H &= 2 \left( X_{r-1} X_r + X_r X_{r+1} + h X_r + g Z_r \right) \nonumber, \\
\Delta_{Z_r} H &= 2 \left( X_{r-1} X_r + X_r X_{r+1} + h X_r \right) \nonumber
\end{align}
for $1<r<N$. In general, the $\Delta_P H$ operators are the sum of Pauli terms in the Hamiltonian that anticommute with $P$. Note that for the mixed-field Ising model, $Y$ errors are special since they anti-commute with all terms in the Hamiltonian. Therefore, $Y$ errors cause the largest change in energy among the single-qubit Paulis \footnote{In fact, $\Delta_{Y_r} H$ is proportional to a slightly modified version of the energy density operator defined in Eq.~(\ref{eq:h_r})}. Figure~\ref{fig:rpe_energy_change_predictions}\textbf{a} shows how all possible two-qubit Paulis acting on the center sites change the energy of the system for the RPE mixed state at a given energy density. 

In our simple model, we assume that after a Pauli error is applied to the RPE at energy $E$, it is approximately an RPE at a new energy $E'=E+\langle \Delta_P H\rangle|_E$. For this model, we obtain the results in Fig.~\ref{fig:rpe_energy_change_predictions}\textbf{b}, which show how energy decays with time for initial states at different energy densities. We see that the energy response depends on the types of errors (purely bit-flip or purely phase-flip errors, for example) applied and on the energy density considered. In all cases, the energy loss appears well approximated by an exponential decay. Figure~\ref{fig:rpe_energy_change_predictions}\textbf{c} shows the local observables obtained for this model for an initial state starting at energy density $E/N=-1.4$ and undergoing depolarizing errors. It also shows the trace distance error, obtained by comparing the density matrix of the center site at $t=0$. We see that at early times the observables and trace distance error grow linearly with time, but eventually become non-linear.

The main benefit of this protocol is that it is classically efficient, requiring only classical Markov chain Monte Carlo sampling. It also provides a way of estimating long-time observable errors in a regime where they are non-linear and not well-described by Eq.~(\ref{eq:error_model}).

The drawback of this protocol is it is heuristic and approximate, so is not guaranteed to be accurate. It also assumes that the dynamics are in the continuous-time regime when Trotter steps are small, so does not capture effects of Trotter errors.

An improvement that can be made to this procedure that does capture Trotter errors is to use a \emph{Floquet RPE}. The Floquet RPE is the RPE for the Floquet Hamiltonian described in Eq.~(\ref{eq:HF}), which captures the lowest order in $\tau$ effects due to Trotterization. Using the Floquet Hamiltonian, one can apply the same protocol as described above and produce similar results as shown in Fig.~\ref{fig:rpe_energy_change_predictions} that also include Trotter effects.

\section{Discussion and Outlook} \label{sec:discussion}

We find that thermalizing Trotterized Hamiltonian simulation is particularly robust to gate errors in quantum hardware.
We argue that a simple model can well describe how local observables near thermal equilibrium are affected by gate errors and Trotter errors as a function of time, Trotter step, and system size.
By time-evolving an initial state built from random product states at a given energy, we numerically and experimentally validate the model.
Ultimately, we find that near thermal equilibrium observable errors significantly outperform naive estimates from gate counting: at early times errors grow linearly in time rather than quadratically and at late times errors are constant rather than linear in system size.
We also demonstrate how the improved performance of quantum gates at smaller gate angles can improve the performance of 
Trotterized Hamiltonian simulations. This makes the performance of quantum gates at small gate angle a particularly valuable benchmark for performing useful Hamiltonian simulation.

We expect that many types of quantum circuits used for Hamiltonian simulation or state preparation are less robust to gate errors than Trotterized dynamics circuits. In Trotterized circuits, before and after a gate error the dynamics is approximately energy-conserving continuous-time dynamics. This makes the error have a local effect and, since the Hamiltonian is local, raises the system's energy only by a constant amount. However, in more complicated quantum algorithms for Hamiltonian simulation, such as quantum signal processing \cite{low2017}, energy conservation is not apparent at the two-qubit gate level. While energy will be conserved after each repeated block of the algorithm, within each block, which has many 2Q gates, energy is not conserved and a single 2Q gate error could spread and lead to non-local errors and thereby potentially $O(N)$ changes to energy.
Similarly, circuits for preparing states at a given energy density could have complicated structure and no notion of energy conservation. For example, this might be the case in variational quantum circuits with geometries that differ from the underlying system Hamiltonian. This logic suggests that adiabatic state preparation using Trotterized Hamiltonian simulation is likely also robust to gate errors, a result observed in Ref.~\onlinecite{schiffer2024}.

Given that quantum dynamics simulations near thermal equilibrium are more robust to errors than other types of quantum circuits, it also begs the question whether \emph{classical} simulations of such systems are easier as well. Energy conservation and long-time hydrodynamic behavior have been incorporated into classical algorithms \cite{white2018,rakovsky2022}, though often such methods are heuristic and have accuracies that are difficult to assess. Also, while we expect that performing quantum simulations over more initial states is harder than performing simulations with a single initial state, it would be worth examining if averaging over states in the random product state ensemble leads to easier classical simulations.

The RPE was helpful in our study in a number of ways. By being closer to a thermal state, the RPE reduced the amount of time evolution needed to thermalize and reduced coherent oscillations in observables and observable errors. It would be useful to explore the utility of the RPE for other thermalizing Hamiltonians with higher spatial dimensions, larger gaps between mean-field and exact ground states, non-local interactions, or thermal phase transitions at non-zero energy density.

Moreover, it would be helpful to generalize the RPE. Some potential extensions could include ensembles of entangled states with fixed energy.
We expect that entangled state ensembles would (1) produce mixed states that are closer approximations to a diagonal ensemble than the RPE mixed state, (2) span a larger range of energy densities (i.e., be able to reach states closer to the ground state), and (3) lead to a more efficient sampling of Hilbert space.
For example, these entangled state ensembles could use random matrix product states (MPS) or other tensor networks; states generated by random quantum circuits, such as Haar random 2Q gate circuits; random valence bond states; or random fermionic states.
In designing such ensembles, it is important for the energy variance to grow at most as $\langle H^2\rangle - \langle H\rangle^2 = O(N)$ so the ensemble is peaked at an energy density in the thermodynamic limit. For local Hamiltonians, this occurs for states with finite correlation lengths, such as MPS with fixed bond dimension or circuits with constant depth.
Another direction for extending the RPE is finding ensembles defined by the Floquet unitary used in the Trotter decomposition rather than a low-order approximation to the Floquet Hamiltonian, which would more accurately capture the nonlinear large Trotter step behavior.
An hurdle in developing new ensembles will be finding efficient classical algorithms for sampling these states analogous to the Markov chain sampling algorithm presented for the RPE.

Our analysis focused on gate errors that could be modeled as stochastic Pauli errors. It is important to understand the impact of other types of errors, such as coherent and leakage errors.

Finally, it might be possible to design improved error mitigation strategies using the techniques in this work. For example, given the simple behavior of observable errors with respect to time for the RPE, zero noise extrapolation or some modified form of it could work well for extracting thermal observables from RPE-evolved dynamics.

\begin{acknowledgments}
We thank the entire Quantinuum team for numerous contributions that enabled this work. We are grateful to Etienne Granet and Henrik Dreyer for providing comments on our manuscript. We thank Karl Mayer for providing the arbitrary angle gate benchmarking data. We also thank Conor Mc Keever, Reza Haghshenas, Matthew DeCross, Michael Schecter, Ivan Deutsch, Anupam Mitra, Tameem Albash, Philip Daniel Blocher, Isaac Kim, Sarang Gopalakrishnan, and Andrew Potter for many helpful discussions.
\end{acknowledgments}

\bibliography{refs}

\newpage
\appendix

\section{Exponential decay of energy in XY model with depolarizing gate errors} \label{sec:xymodel}

Here we show how energy decays exponentially in time for Trotterized time evolution in the XY model when two-qubit gates are subject to depolarizing noise.

Consider an XY model
\begin{align}
H = \sum_{\langle ij \rangle} X_i X_j + Y_i Y_j \equiv \sum_{\langle ij \rangle} h_{ij}
\end{align}
in a square lattice geometry with periodic boundary conditions. Suppose that we implement a 2nd order Trotterized dynamics as in Eq.~(\ref{eq:UTrotter}) with $H_1 = \sum_{\langle ij \rangle} X_i X_j$ and $H_2 = \sum_{\langle ij \rangle} Y_i Y_j$ and time-evolve a translationally-invariant initial state $\rho(0)$.

Suppose that the $e^{-i\tau X_iX_j}$ and $e^{-i\tau Y_iY_j}$ two-qubit gates in the Trotterized circuit are followed by a two-qubit depolarizing error channel $\mathcal{E}_{ij}$ on qubits $i$ and $j$. The quantum channel acting on a many-body state $\rho$ is
\begin{align}
\mathcal{E}_{ij}(\rho) = (1-\lambda)\rho + \lambda \frac{I_{ij}}{4} \otimes \tr_{ij}(\rho),
\end{align}
where $\tr_{ij}$ is a partial trace over qubits on sites $i,j$, $I_{ij}$ is the identity matrix on sites $i,j$, and $\lambda$ is the depolarizing error parameter satisfying $0 \leq \lambda \leq 16/15$. For a two-qubit gate, $\lambda$ corresponds to an average gate infidelity of $p=3\lambda/4$.

After each noisy Trotter layer in the circuit, when the Trotter step $\tau$ is small, the time-evolved state $\rho(t)$ will be to a good approximation translationally invariant. Because of the translational invariance of the Hamiltonian and approximate translational invariance of the time-evolved state, we know that the energy is the same in each $ij$ bond 
\begin{align}
E(t)&\equiv \tr ( H \rho(t)) = \sum_{\langle ij\rangle} \tr (  h_{ij}\rho(t)) \nonumber \\
&= \frac{zN}{2} \tr (  h_{ij}\rho(t)) \equiv \frac{zN}{2} E_{ij}(t). \label{eq:xy_energy}
\end{align}
Here $z$ is the coordination number of the lattice ($z=4$ for square) and $E_{ij}(t)$ is the energy of the $ij$ bond at time $t$.

First, we examine how a single depolarizing channel affects the energy in the system. Importantly, for any Pauli $P$ with non-identity support on sites $i$ and/or $j$, 
\begin{align}
\tr(P I_{ij}/4 \otimes \tr_{ij}(\rho))=\overbrace{\tr_{ij}(P_{ij})}^{=0} \times \tr_{\overline{ij}}(P_{\overline{ij}}\tr_{ij}(\rho))=0,
\end{align}
because non-identity Paulis are traceless. Here $P_{ij}, P_{\overline{ij}}$ are the Pauli $P$ restricted to sites $ij$ and its complement, respectively. Example Paulis include $X_i X_j$ or $Y_j Y_k$ with $k\neq i,j$. Using this fact, we know that
\begin{align}
&\tr(h_{kl}\frac{I_{ij}}{4}\otimes \tr_{ij}(\rho)) \nonumber \\
&\quad = \begin{cases}
0 &\langle k,l\rangle \in N(i,j) \\
\tr_{\overline{ij}}(h_{kl}\tr_{ij}(\rho))=\tr(h_{kl}\rho) &\langle k,l\rangle \notin N(i,j)
\end{cases}, \label{eq:hkl_zero}
\end{align}
where $N(i,j)$ is the set of all edges in the graph defining the lattice whose vertices include $i$ and/or $j$, $\tr_{\overline{ij}}$ is a partial trace on all sites other than $i,j$. Note that Eq.~(\ref{eq:hkl_zero}) is true for the XY model because $h_{ij}$ is made up of only 2-qubit Paulis on $ij$ and would not hold if the Hamiltonian, for example, contained 1-qubit Paulis on those sites. From Eq.~(\ref{eq:hkl_zero}), we can see that the bond expectation value after a single depolarizing channel becomes
\begin{align}
&\tr(h_{kl}\mathcal{E}_{ij}(\rho)) \nonumber \\
&\, = \begin{cases}
(1-\lambda)\tr(h_{kl}\rho) + \lambda \cdot 0 = (1-\lambda)\tr(h_{kl}\rho) & \langle k,l\rangle \in N(i,j) \\
(1-\lambda)\tr(h_{kl}\rho) + \lambda \cdot \tr(h_{kl}\rho) =\tr(h_{kl}\rho) & \langle k,l\rangle \notin N(i,j)
\end{cases}. \label{eq:hkl_single_channel}
\end{align}
From Eq.~(\ref{eq:hkl_single_channel}), we then see that the energy after a single depolarizing channel is 
\begin{align}
E'_{ij} &= (1-\lambda)\sum_{\langle kl\rangle \in N(i,j)} \tr(h_{kl}\rho) + \sum_{\langle kl\rangle \notin N(i,j)} \tr(h_{kl}\rho) \nonumber \\
&= E -\lambda \sum_{\langle kl\rangle \in N(i,j)} \tr(h_{kl}\rho) \nonumber \\
&= E-\lambda |N(i,j)| E_{ij} \label{eq:xy_energy_single_channel}
\end{align}
where $|N(i,j)|=2z-1$ is the number of edges in the graph whose vertices include sites $i$ and/or $j$. Note that after the depolarizing error, energy is approximately conserved in the Trotterized circuit. Therefore, even though the state will evolve and become more complicated over time, we are still able to determine its energy by considering the energy of the state immediately following the depolarizing channel, which is simple.

Next, we consider how energy changes after an entire Trotter layer of gates with depolarizing errors. To first order in $\lambda$, we can think of the change in energy for the entire Trotter layer as being a sum of the change in energies from each single depolarizing channel in the layer. For the second-order Trotter decomposition considered, there are $N_{\textrm{gate}}=\frac{3zN}{2}$ gates in a Trotter layer, where the depolarizing channel can act. (In principle, gates between Trotter layers can be combined to reduce this number, but for simplicity we do not consider that case here.) We then expect that the change in energy after a Trotter layer is
\begin{align}
E(t+\tau) - E(t) &= N_{\textrm{gate}} (E_{ij}'-E) \nonumber \\
&= \left(\frac{3zN}{2}\right)\left(-\lambda (2z-1)E_{ij}\right) \nonumber \\
&= -3(2z-1)\lambda E, \label{eq:energy_change}
\end{align}
which is proportional to the total energy $E$. We used Eq.~(\ref{eq:xy_energy_single_channel}) in the first line and Eq.~(\ref{eq:xy_energy}) in the third line.

For small $\tau$, we can then approximate the energy dynamics with a simple differential equation
\begin{align}
\frac{d E}{dt} \approx \frac{E(t+\tau) - E(t)}{\tau} = -\gamma E,
\end{align}
whose solution is an exponential decay
\begin{align}
E(t) = E(0) e^{-\gamma t} \label{eq:exp_decay}
\end{align}
with decay rate
\begin{align}
\gamma=3(2z-1)\frac{\lambda}{\tau}=4(2z-1)\frac{p}{\tau}.
\end{align}

\begin{figure}[htbp!]
    \centering
    \includegraphics[width=0.45\textwidth]{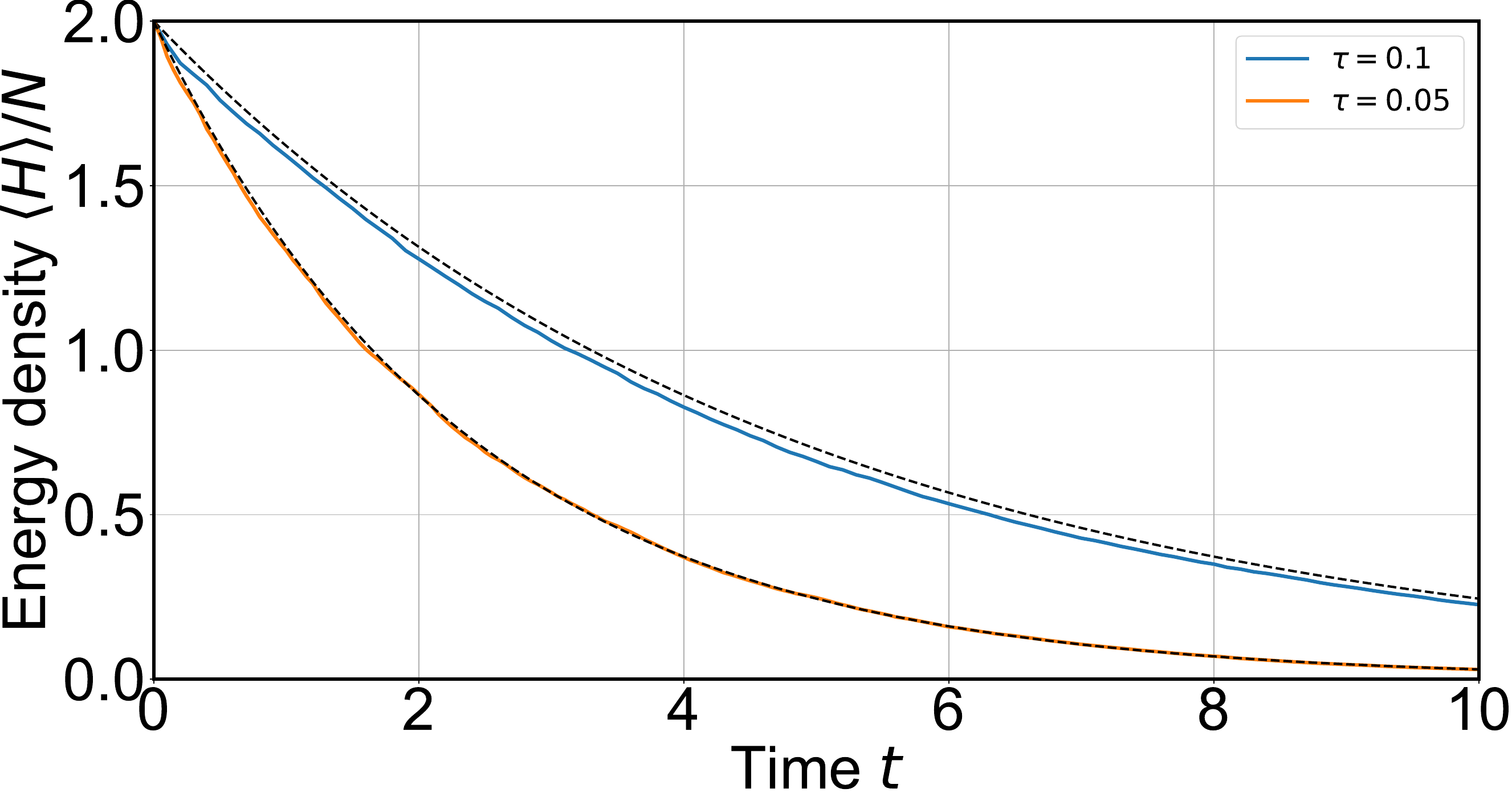}
    \caption{Energy density versus time for the XY model with depolarizing noise. The results are for a noisy numerical statevector simulation on a $4 \times 4$ square lattice with periodic boundaries, using a 2nd order Trotter decomposition with Trotter step $\tau$ and initial state $\ket{\psi}=\ket{+\cdots +}$. The 2Q gates experience depolarizing noise with $\lambda=10^{-3}$. The black dashed lines are Eq.~(\ref{eq:exp_decay}).} 
    \label{fig:xy_2d_depolarizing_energy_vs_t}
\end{figure}

We verify this prediction numerically. We perform a noisy statevector simulation of the XY model, with 1000 shots, on a $4 \times 4$ 2D square lattice with periodic boundaries. We use the second-order Trotter decomposition to time-evolve the translationally invariant initial state $\ket{\psi}=\ket{+\cdots +}$, which has energy density $E/N=2$. Figure~\ref{fig:xy_2d_depolarizing_energy_vs_t} shows that the numerical results agree well with Eq.~(\ref{eq:exp_decay}), which are marked with black dashed lines, when using the 2D square lattice value for $\gamma=\frac{21\lambda}{\tau}$. Note that the agreement becomes better as $\tau$ is decreased.

We expect that this essentially perfect exponential decay will occur in the Trotterized dynamics of a large class of models undergoing depolarizing gate errors. In particular, the same arguments will apply for translationally-invariant spin models (in any dimension) with only two-body Pauli terms, such as Heisenberg or XXZ models. If a Hamiltonian has, for example, single-body Pauli terms, then the change in energy after a single Trotter layer (Eq.~(\ref{eq:energy_change})) will not be exactly proportional to the energy, but will also include additional perturbations. 

\section{Symmetries of the mixed-field Ising model} \label{sec:mfi_symmetries}

The mixed-field Ising model possesses a few non-trivial symmetries worth noting that can impact the expectation values of certain observables. The model is time-reversal symmetric: $K H K^{-1} = H$ where $K$ is the complex conjugation operator that acts in the $Z$-basis. This symmetry implies that all energy eigenstates $\ket{n}$ of $H$ are real in the $Z$-basis, and consequently that $\langle n | P_Y |n\rangle=0$ for any imaginary Pauli $P_Y$ with an odd number of single-site $Y$ operators. With open boundary conditions, this model is reflection-symmetric: $R H R^{-1} = H$ where $R$ reflects spins at sites $j$ around the center sites $c$: $j-c \rightarrow c-j$. Since $R^2=I$, this symmetry splits the spectrum of $H$ into two sectors, with reflection symmetric $R=+1$ and anti-symmetric $R=-1$ eigenstates. These symmetries minimally impact the thermalizing nature of this model, though do impact the values of certain observables (such as $Y$). Note that adding an additional $Y$-field to Eq.~(\ref{eq:H}) does not remove the time-reversal symmetry, but actually modifies the time-reversal symmetry operator to $K'=KU$ where $U=\prod_j e^{-i\theta X_j}$ is a uniform rotation about the Ising axis. To remove the time-reversal symmetry, one needs to introduce additional interactions, which increases the number of 2Q gates needed to implement the model as a Trotter circuit.

\section{Time-dependent perturbation theory derivation of Trotter error time scaling} \label{sec:tdpt}

To leading order in the Trotter step $\tau$, the error on observables can be calculated in time-dependent perturbation theory by computing the Floquet Hamiltonian to leading non-trivial order in the Trotter step $\tau$,
\begin{align}
H_{F} = H + \tau^2\overbrace{\frac{1}{24}[H_1+2H_2,[H_1,H_2]]}^{V}+O(\tau^{4}),
\end{align}
ignoring the $O(\tau^4)$ terms, and treating $\tau^2 V$ as a perturbation to $H$. The full time-dependence of a generic observable $\mathcal{O}$ can then be computed as
\begin{align}
\mathcal{O}(t)=\mathscr{U}^{\dagger}(t)\mathcal{O}_{H}(t)\mathscr{U}(t),
\end{align}
where $\mathcal{O}_H(t)$ is the exact (without Trotter error) evolution under $H$ and the interaction-picture time evolution operator $\mathscr{U}(t)$ is given by
\begin{align}
\mathscr{U}(t)&\equiv \mathcal{T}\exp\bigg(-i\tau^2\int_{0}^{t}dt' V_{H}(t')\bigg)\\
&=1-i\tau^2\int_0^t d t' V_H(t')+\dots
\end{align}
To lowest order in $\tau$, the difference between the exact and Trotterized evolutions of $\mathcal{O}$, $\Delta \mathcal{O}(t)\equiv \mathcal{O}(t)-\mathcal{O}_H(t)$, is then given by
\begin{align}
\label{eq:delta_O}
\Delta \mathcal{O}(t)=-i\tau^2\int_{0}^{t}dt'[\mathcal{O}_H(t),V_{H}(t')].
\end{align}
We can decompose $V$ into pieces that are purely diagonal or off-diagonal in the energy eigenbasis of $H$ as $V=V_{\parallel}+V_{\perp}$, where
\begin{align}
V_{\parallel}&=\sum_{n}V_{nn}\ket{n}\bra{n},\\
V_{\perp}&=\sum_{m\neq n}V_{mn}\ket{m}\bra{n}.
\end{align}
Here $V_{mn}=\bra{m}V\ket{n}$, where $\ket{m},\ket{n}$ are the eigenstates of $H$. Inserting this decomposition into \eref{eq:delta_O}, taking its expectation value, and carrying out the time integral (assuming no degeneracies) we ultimately arrive at the following expression for the expectation value $\langle \Delta \mathcal{O}(t)\rangle$,
\begin{align}
&\langle \Delta \mathcal{O}(t)\rangle =-i\tau^2 t \sum_{n}V_{nn}{\rm tr}\Big(\mathcal{O}\big[\ket{n}\!\bra{n},\rho_H(t)\big]\Big)\nonumber\\
-&\tau^2\!\sum_{m\neq n}\frac{V_{mn}\big(1-e^{-i t(E_m-E_n)}\big)}{E_m-E_n}{\rm tr}\Big(\mathcal{O}\big[\ket{m}\!\bra{n},\rho_H(t)\big]\Big).
\end{align}
Here, the interaction picture time evolution of the density matrix (i.e., the exact Schrodinger-picture time evolution in the absence of Trotter error) is given by $\rho_{H}(t)= e^{-i H t}\rho e^{i H t}$.

\section{Details of numerical methods} \label{sec:numerical_methods}

We utilize multiple numerical methods to study the dynamics of a one-dimensional spin chain. Depending on the specific problem being studied, one method could be much more efficient than the other.

\subsection{Matrix product state time-evolution}

Most of the calculations in this work involve matrix product states (MPS) calculations, which are performed using the ITensor library \cite{itensor} written in \texttt{Julia} \cite{julia}. Due to the one-dimensional nature of the systems studied, MPS methods can accurately and efficiently capture the quantum state's evolution even for large system sizes (we look at $N \leq 50$). However, due to the linear growth of entanglement entropy with time, the runtime of these methods scales exponentially with time. This limits their use to studying short-time dynamics (we look at $t \leq 6$). We perform two types of MPS dynamics calculations, purely unitary evolution of the quantum circuit in Eq.~(\ref{eq:U}) and dissipative non-unitary evolution of the circuit subject to depolarizing two-qubit gate noise.

\subsubsection{Unitary evolution} \label{sec:methods_mps_unitary}

Our unitary MPS evolution essentially amounts to the time-evolved block decimation (TEBD) method \cite{vidal2004}. In our calculations, we represent our time-evolved state as an MPS and apply each unitary gate in the Trotter circuit Eq.~(\ref{eq:U}) one at a time to the state. The application of each two-qubit unitary can increase the bond-dimension of the MPS. To control the size of the MPS and thereby the cost of the algorithm, we perform singular value decomposition (SVD) and adaptively truncate the bond dimension of the MPS so that the truncated squared singular values are less than $10^{-10}$ and the bond dimension is less than $D=1024$. This ensures that the bond dimension is more than large enough so that tensor network truncation effects are negligible when measuring local observables. We compute local observables using standard contraction techniques that are efficient due to the canonical form of the MPS \cite{paeckel2019,cirac2021}.

This simulation method was used in Figs.~\ref{fig:0state_rpe_observables_vs_t}, \ref{fig:0state_energy_diffs_vs_t}, \ref{fig:0state_observable_diffs_vs_t}, \ref{fig:rpe_singleerror}, \ref{fig:fixedstates_trotter_error_vs_tau}, and \ref{fig:rpe_trottererrors_vs_t}.

\subsubsection{Dissipative evolution} \label{sec:methods_mps_dissipative}

In our dissipative MPS calculations, we utilize MPS to represent trajectories in a quantum trajectories simulations \cite{weimer2021}. A quantum channel $\mathcal{E}$ can be represented as $\mathcal{E}(\rho)=\sum_{j=1}^k K_j \rho K_j^\dagger$ where the Kraus operators $K_j$ satisfy $\sum_j K_j^\dagger K_j=I$. When acting on a pure state $\rho=\ket{\psi}\bra{\psi}$, the channel has the effect of creating a probabilistic mixture of states 
\begin{align}
    \mathcal{E}(\ket{\psi}\bra{\psi}) = \sum_j p_j \ket{\psi_j}\bra{\psi_j} \label{eq:channel_trajectories}
\end{align}
where $\ket{\psi_j}=K_j \ket{\psi} / \sqrt{p_j}$ are normalized states ($\langle \psi_j|\psi_j\rangle=1$) and $p_j = \bra{\psi} K^\dagger_j K_j \ket{\psi}$ are normalized probabilities ($\sum_j p_j=1$). The $\ket{\psi_j}$ pure states are referred to as \emph{quantum trajectories}. When evaluating an expression involving a quantum channel, rather than directly computing each term and each trajectory, one can sample a trajectory with probability $p_j$ and evaluate the term for that individual trajectory. When averaged over many trajectories, the average will approximate the total sum. The trajectory sampling approach is useful when there are many channels $N_{c}$ since the exponentially many $k^{N_c}$ terms can be approximated with $M \ll k^{N_c}$ samples. In our dissipative simulations, we have a two-qubit depolarizing channel after every two-qubit gate in our circuit so that $k=16$ and $N_c = DN$. We use MPS to represent the trajectories and parallelize the independent trajectory calculations across the cores of a high-performance computing cluster.

Using the quantum trajectories method, we evolve $N=20$ site chains up to time $t=4$, with adaptive bond dimensions chosen for the MPS so that the SVD truncation error is below $10^{-10}$ and the maximum bond dimension is $512$. We average over $M=11,200$ trajectories when computing observables. Due to the $M \times $ sampling overhead, we perform smaller-scale calculations (smaller $N$ and $t$) than for the purely unitary evolution described in Sec.~\ref{sec:methods_mps_unitary}. For the $\ket{0\cdots 0}$ initial state calculations, each trajectory has the same initial state, but different sampled Kraus operators $K_j$ after each two-qubit gate. For the RPE initial state calculations, each trajectory has a \emph{different} random product state initial state and different sampled Kraus operators. Empirically, we find that the MPS bond dimension, number of trajectories, and number of RPE samples are chosen large enough so that the effects of truncation error and statistical sampling error are negligible compared to the effects of gate and Trotter errors.

This simulation method was used in Figs.~\ref{fig:error_vs_t}, \ref{fig:error_vs_N}, \ref{fig:error_vs_tau}, \ref{fig:rpe_many_errors_vs_t}, and \ref{fig:rpe_errorscaling_vs_tau_fits}.

\subsection{Statevector simulation}

For the unitary and dissipative circuit dynamics simulations used in Figs.~\ref{fig:qeast}\textbf{c}~and~\ref{fig:xy_2d_depolarizing_energy_vs_t}, we used the Qiskit Aer statevector simulator \cite{javadiabhari2024} written in \texttt{Python}. 

\subsection{Exact diagonalization}

In order to simulate dynamics to long times or to resolve small observable errors not easily resolvable with sampling methods, we also use the exact diagonalization (ED) method. The runtime of these methods do not depend on time but scale exponentially in system size. Therefore, we use exact diagonalization to obtain long-time ($t\leq 10000$) small-system-size ($N \leq 12$) results. We use ED both to simulate unitary and dissipative evolution. 

\subsubsection{Unitary evolution} \label{sec:methods_ed_unitary}

For unitary evolution, our goal is to time-evolve a density matrix $\rho$ by a unitary $U$. The initial state density matrix can be a single product state $\rho=|0\cdots 0\rangle\langle 0 \cdots 0|$ or a sum over many product states $\rho=\frac{1}{M}\sum_{s=1}^M \ket{\psi_s}\bra{\psi_s}$ and the unitary operator can either be the exact continuous-time unitary $e^{-itH}$ or the approximate Trotterized unitary $U_{\textrm{Trotter}}^D$. In either case, we implement the evolution by first representing $U$ as a $2^N \times 2^N$ matrix and diagonalizing $U$ numerically. 

To illustrate, if evolving by $U=e^{-itH}$, we first diagonalize $U=\sum_n e^{-it E_n} \ket{n}\bra{n}$ where $\ket{n}$ are the $2^N$ eigenstates of $U$ (and also of $H$ with energy $E_n$, assuming no degeneracy). We also represent $\rho$ as a matrix in the eigenstate basis: $\rho=\sum_{mn} \rho_{mn} \ket{m}\bra{n}$. Then, the time-evolved density matrix is
\begin{align}
\rho(t) = \sum_{mn} \rho_{mn} e^{-it(E_m - E_n)} \ket{m}\bra{n}.
\end{align}
To diagonalize $U$ takes  $O((2^N)^3)$ runtime and to compute $\rho(t)$ takes  $O((2^N)^2)$ runtime, regardless of the amount of time-evolution $t$. To compute observables expectation values, one represents an observable $\mathcal{O}$ as a matrix in the same basis $\mathcal{O}=\sum_{mn} \mathcal{O}_{mn}\ket{m}\bra{n}$ and evaluates $\langle O(t) \rangle = \tr(\rho(t)\mathcal{O})=\sum_{mn} \rho_{mn} e^{-it(E_m - E_n)} \mathcal{O}_{nm}$, which also takes  $O((2^N)^2)$ runtime.

This simulation method was used in Fig.~\ref{fig:rpe_interpolation}.

\subsubsection{Spectral properties} \label{sec:methods_ed_spectral}

In addition to simulating dynamics, we use ED to look at the properties of energy eigenstates in the mixed-field Ising Hamiltonian Eq.~(\ref{eq:H}). In Fig.~\ref{fig:rpe_observables}, we present expectation values of observables for energy eigenstates of the mixed-field Ising Hamiltonian $H$ for eigenstates at specific energy densities $E/N$ for $N=14$ site chains. To compute these eigenstates, we do not fully diagonalize the Hamiltonian, but instead use the shift-invert Lanczos algorithm, as implemented in the \texttt{eigsh} function in the \texttt{scipy} package \cite{scipy} written in \texttt{Python}. This algorithm more efficiently finds eigenvectors with eigenvalues near a specified value than full diagonalization. In Fig.~\ref{fig:qeast}\textbf{a}, we do a similar analysis but using full diagonalization for the Quantum East model.

\section{Markov chain Monte Carlo sampling of random product states} \label{sec:rpe_mcmc}

Product states from the random product state ensemble (RPE) can be generated using a classical Markov chain Monte Carlo (MCMC) sampling algorithm. This algorithm essentially amounts to sampling three-dimensional spin configurations with fixed energy $E$ from a classical spin Hamiltonian, i.e., generating spin configurations from a classical microcanonical ensemble.

For concreteness, consider a quantum Hamiltonian of the form $\hat{H} = \sum_{jk\alpha \beta} J_{jk}^{\alpha\beta} \hat{\sigma}^\alpha_j \hat{\sigma}^\beta_k + \sum_{j\alpha} J_j^\alpha \hat{\sigma}^\alpha_j$. For a product state with Bloch vectors $\vec{\sigma}_j$, the energy of the product state is the energy of a spin configuration in a classical version of this Hamiltonian with the replacement $\langle \hat{\sigma}^\alpha_j\rangle \rightarrow \vec{\sigma}_j$:
\begin{align*}
E = \sum_{jk\alpha\beta} J_{jk}^{\alpha\beta} \sigma^\alpha_j \sigma^\beta_k + \sum_{j\alpha} J_j^\alpha \sigma^\alpha_j.
\end{align*}
Importantly, we can determine how modifying a single spin $j$ changes a product state's energy by considering the \emph{effective local field} on site $j$
\begin{align}
h_j^\alpha \equiv \sum_{k\beta} J_{jk}^{\alpha\beta} \sigma^\beta_k + J_j^\alpha ,
\end{align}
which depends on the spins on other sites $k \neq j$. Using this definition, we see that the energy of a product state can be simply expressed as
\begin{align*}
E = \sum_j \vec{h}_j \cdot \vec{\sigma}_j \equiv \sum_j E_j.
\end{align*}
If one keeps the $k \neq j$ spins fixed but changes spin $j$ from $\vec{\sigma}_j$ to $\vec{\sigma}_j'$, then the change in energy is $\Delta E_j \equiv \vec{h}_j \cdot (\vec{\sigma}_j' - \vec{\sigma}_j)$. Note that one can construct an effective local field in this way for \emph{any} spin Hamiltonian, not only ones of the form discussed above. Also note that the Bloch vectors $||\vec{\sigma}_j||=1$ are normalized while the effective fields $||\vec{h}_j||\neq 1$ are not, which indicates that the local energy $E_j=\vec{h}_j \cdot \vec{\sigma}_j$ of a spin falls between $-||\vec{h}_j||$ and $+||\vec{h}_j||$, with the boundary values obtained when the spin is anti-aligned ($\vec{\sigma}_j \propto -\vec{h}_j$) or aligned ($\vec{\sigma}_j \propto +\vec{h}_j$) with the effective field.

We generate product state samples from the RPE using a Metropolis MCMC algorithm. The algorithm has the following steps:
\begin{enumerate}
\item Start with a product state $\sigma$ with spins $\vec{\sigma}_j$ with energy $E$.
\item Repeat 
    \begin{enumerate}
    \item[a.] Propose a new product state $\sigma'$ with $m$ spins updated to $\vec{\sigma}_{j_1}',\ldots,\vec{\sigma}_{j_m}'$ that has the same energy $E$. The proposal probability is denoted as $T(\sigma \rightarrow \sigma')$.
    \item[b.] Accept the new state with the Metropolis acceptance probability $A(\sigma \rightarrow \sigma')=\min\left(1, \frac{P(\sigma')T(\sigma \rightarrow \sigma')}{P(\sigma)T(\sigma' \rightarrow \sigma)}\right)$.
    \item[c.] Save the current state to the list of samples.
    \end{enumerate}
\end{enumerate}
For the RPE, each Bloch vector $\vec{\sigma}_j$ is as equally likely as another so that $P(\sigma)=P(\sigma')$. In the Metropolis algorithm, one is free to choose the proposal move, which affects the acceptance rate and autocorrelation time of the Markov chain \cite{landau2021}. Below, we describe the proposal moves that we devised for generating new product states with the same energy. We specifically designed these moves so that $T(\sigma \rightarrow \sigma')=T(\sigma' \rightarrow \sigma)$ and therefore the acceptance probability is always exactly $A(\sigma \rightarrow \sigma')=1$ and no states are rejected.

\begin{figure}
    \centering
    \includegraphics[width=0.45\textwidth]{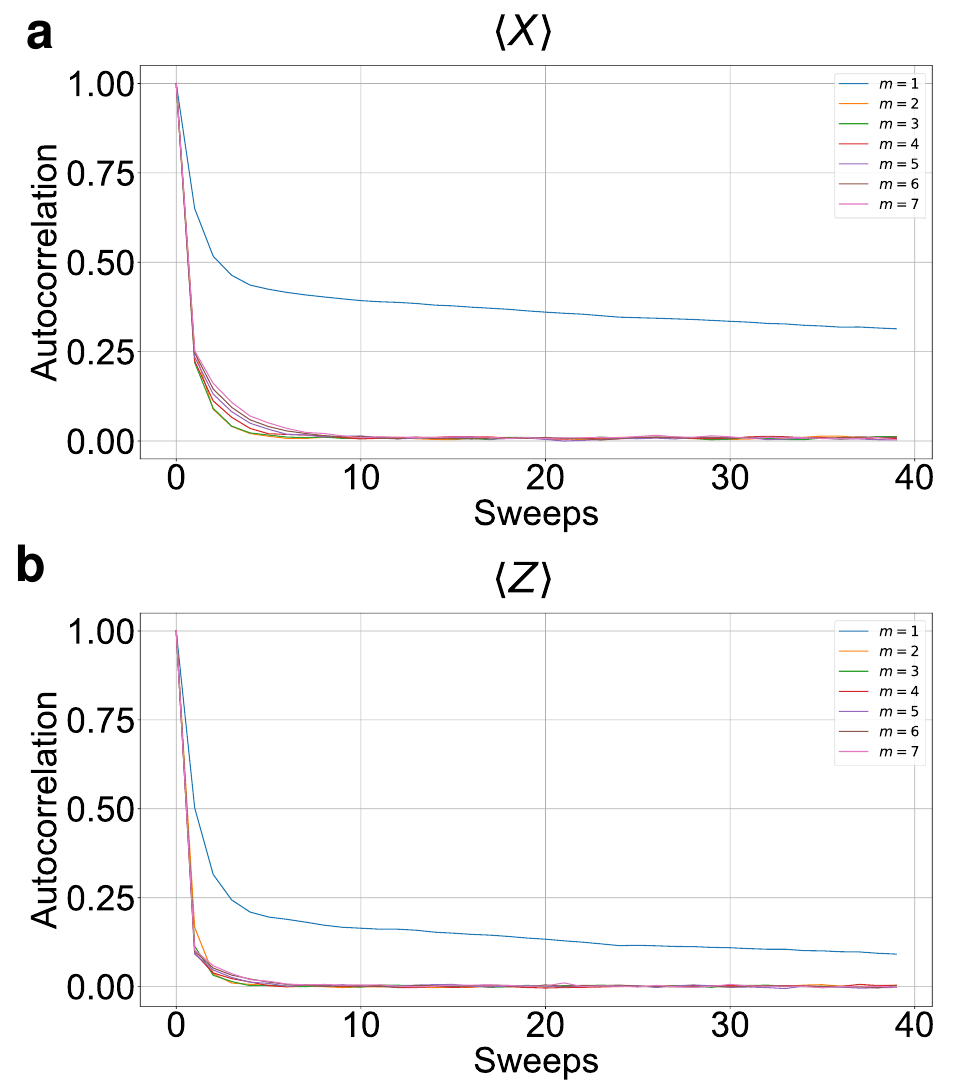}
    \caption{The autocorrelation function versus MCMC sweep for \textbf{a} the average $x$-magnetization $X=\frac{1}{N}\sum_{j=1}^N X_j$ and \textbf{b} the average $z$-magnetization $Z=\frac{1}{N}\sum_{j=1}^N Z_j$ using different $m$-site proposal moves in the Metropolis MCMC algorithm. A MCMC sweep is $N$ proposal moves.}
    \label{fig:figautocorrelation}
\end{figure}

\subsection{One-site move}

The simplest move we considered involves updating a single spin and proceeds as follows:
\begin{enumerate}
\item Choose uniformly at random a spin $j \in [1,N]$ for an $N$-site system. The local energy of spin $j$ is $E_j$.
\item Compute the unit vector $\hat{n}_{j,\parallel}\equiv \vec{h}_j / ||\vec{h}_j||$ parallel to the effective field and choose a random unit vector $\hat{n}_{j,\perp}$ that is orthogonal to $\hat{n}_{j,\parallel}$.
\item Change the spin to 
\begin{align*} 
\vec{\sigma}_j'= \left(E_j/||\vec{h}_j||\right) \hat{n}_{j,\parallel} + \left(1 - E_j^2/||\vec{h}_j||^2\right)^{1/2} \hat{n}_{j,\perp}.
\end{align*}
\end{enumerate}

Since rotating the spin $j$ about the local field $\vec{h}_j$ does not change the energy of the state, the new state produced at step 3 must have the same energy as the original state. Since the new state proposed (the chosen $\hat{n}_{j,\parallel}$ vector) does not depend on the current state, $T(\sigma \rightarrow \sigma')=T(\sigma' \rightarrow \sigma)$. While this move works, it produces a Markov chain with a large autocorrelation time [see Fig.~\ref{fig:figautocorrelation}], likely due to the slow diffusive spreading of energy induced by the single-site Markov chain dynamics.

\subsection{Two-site move}

To reduce the autocorrelation time of the Markov chain, we consider a more complicated two-site move that produces larger changes to the product state in a single proposal. Importantly, this update allows energy to distribute non-locally in space and thereby helps avoid the slow diffusive spreading caused by the local move. The two-site move has the steps:
\begin{enumerate}
\item Choose uniformly at random two spins $j_1,j_2 \in [1,N]$ for an $N$-site system such that $j_1$ and $j_2$ are not neighboring spins according to the Hamiltonian (i.e., $J_{j_1,j_2}=0$). The local energies of the spins are $E_{j_1}, E_{j_2}$.
\item Compute the unit vectors $\hat{n}_{j_{1/2},\parallel}\equiv \vec{h}_{j_{1/2}} / ||\vec{h}_{j_{1/2}}||$ parallel to the effective fields and choose random unit vectors $\hat{n}_{j_{1/2},\perp}$ that are orthogonal to $\hat{n}_{j_{1/2},\parallel}$.
\item Pick an energy change $\Delta E$ uniformly at random in the interval 
\begin{align*}
    \Delta E &\geq \min(-||\vec{h}_{j_1}||-E_{j_1}, ||\vec{h}_{j_2}||-E_{j_2}) \\
    \Delta E &\leq \max(||\vec{h}_{j_1}||-E_{j_1}, -||\vec{h}_{j_2}||-E_{j_2}).
\end{align*} 
Set the new energies for the two spins to $E_{j_1}'=E_{j_1} + \Delta E$ and $E_{j_2}' = E_{j_2} - \Delta E$.
\item Change the spins to 
\begin{align*} 
\vec{\sigma}_{j_{1/2}}'= &\left(E_{j_{1/2}}'/||\vec{h}_{j_{1/2}}||\right) \hat{n}_{j_{1/2},\parallel} \nonumber \\
&\quad+ \left(1 - E_{j_{1/2}}'^2/||\vec{h}_{j_{1/2}}||^2\right)^{1/2} \hat{n}_{j_{1/2},\perp}.
\end{align*}
\end{enumerate}

In addition to rotating each spin about its local field, in the two-site move we also redistribute energy $\Delta E$ between the two spins by changing how much each spin points along its local field. Since the range of possible $\Delta E$ is the same for the proposed state as the current state, $T(\sigma \rightarrow \sigma')=T(\sigma' \rightarrow \sigma)$. Also, it is important that the two chosen spins are not neighboring so that each of their local fields do not depend on the other spin. This move significantly decreases the autocorrelation time of the Markov chain [see Fig.~\ref{fig:figautocorrelation}] compared to the one-site move.

\subsection{$m$-site move}

Finally, we generalize to an $m$-site move with $m \geq 2$ that can produce even more non-local changes to the state in a single proposal. The $m$-site move has the steps:
\begin{enumerate}
\item Choose uniformly at random $m$ spins $j_1,j_2,\ldots,j_m \in [1,N]$ for an $N$-site system such that none of the $j_1,\ldots,j_m$ spins are neighboring according to the Hamiltonian (i.e., $J_{j_a,j_b}=0 \, \forall a \neq b$) (This is not possible if $m$ is too large. For example, for a 1D chain $m \leq N/3$ must hold for this to always be possible.). The local energies of the spins are $E_{j_1}, \ldots, E_{j_m}$.
\item Compute the unit vectors $\hat{n}_{j_{a},\parallel}\equiv \vec{h}_{j_a} / ||\vec{h}_{j_a}||$ parallel to the effective fields and choose random unit vectors $\hat{n}_{j_a,\perp}$ that are orthogonal to $\hat{n}_{j_a,\parallel}$, for $a=1,\ldots,m$.
\item Choose a uniformly random $m$-dimensional unit vector $\hat{r}$ that is orthogonal to the all-ones vector $(1,\ldots,1)$. Define the energy change vector as $\Delta E_{j_a} = (\Delta E) \hat{r}_a$.
\item Determine the minimum $\Delta E_{\textrm{min}}$ and maximum $\Delta E_{\textrm{max}}$ allowed energy changes along the $\hat{r}$ direction ($\Delta E_{\textrm{min}}$ can be negative). Choose $\Delta E$ uniformly at random in this range $\Delta E_{\textrm{min}} \leq \Delta E \leq \Delta E_{\textrm{max}}$. Set $E_{j_a}' = E_{j_a} + \Delta E_{j_a}$.
\item Change the spins to 
\begin{align*} 
\vec{\sigma}_{j_a}'= &\left(E_{j_a}'/||\vec{h}_{j_a}||\right) \hat{n}_{j_a,\parallel} \nonumber \\
&\quad+ \left(1 - E_{j_a}'^2/||\vec{h}_{j_a}||^2\right)^{1/2} \hat{n}_{j_a,\perp}.
\end{align*}
\end{enumerate}

In step 3, the unit vector $\hat{r}$ specifies the direction in ``energy difference space'' to move. The vector is constrained so that $\sum_a \hat{r}_a = 0$ which ensures that the total energy change in the proposal is zero: $\sum_{j_a} \Delta E_{j_a} = (\Delta E)\sum_a \hat{r}_a = 0$. Step 4 can be done efficiently, and involves considering the possible boundaries in local energy space of each site: the allowed energy changes form a hyperrectangle defined by  $-||\vec{h}_j|| \leq E_j + \Delta E_j \leq +||\vec{h}_j||$. Note that in the energy change part of the update described in steps 3 and 4, the probability of picking a particular $\hat{r}$ and $\Delta E$ is the same before and after the proposal, ensuring that $T(\sigma \rightarrow \sigma')=T(\sigma' \rightarrow \sigma)$. 

In Fig.~\ref{fig:figautocorrelation}, we compare the autocorrelation times for Markov chains using the $m$-site move with different $m$. We find that for our specific 1D Hamiltonian, observables, and energy, $m=3,4$ site moves appear to work the best, though they do not offer significant improvement over the $m=2$ site move. It seems that the choice of the random $\hat{r}$ could have a significant impact on the algorithm performance at large $m$. We choose it so that it is distributed uniformly on the surface of the $m-1$ dimensional sphere that exists in the subspace orthogonal to the $(1,\ldots,1)$ vector. However, when $m$ is large, $\hat{r}$ chosen this way are likely to point along a ``thin'' direction of the energy change hyperrectangle, making the allowed values of $\Delta E$ small. It would be interesting to explore alternative energy proposals that encourage larger energy distributions per move.

\subsection{$m$-site move with energy window}

We also develop a variant of the MCMC algorithm that allows for sampling states in an energy window $[E-\varepsilon,E+\varepsilon]$ with target energy $E$. The algorithm is quite similar to the one described in the previous section, but with a few minor changes. During this sampling, the current energy $E_{curr}$ is logged. 

Step 3 is modified, so that the vector $\hat{r}$ is a random $m$-dimensional vector that can have overlap with $(1,\ldots,1)$, which causes a change in energy. In particular, we choose to sample the entries of $\hat{r}$ from a Gaussian distribution with diagonal covariance matrix and unit standard deviation on the $(m-1)$ vectors orthogonal to $(1,\ldots,1)$ and $\varepsilon$ standard deviation on the $(1,\ldots,1)$ vector; we then normalize the vector. This helps bias the update move towards updates that do not change the energy. This bias prevents the autocorrelation time from diverging as $1/\varepsilon$ as $\varepsilon \rightarrow 0$ and makes the energy window algorithm exactly agree with the original algorithm when $\varepsilon=0.$

Also, step 4 is modified with an additional constraint that the energy change keeps the energy in the energy window so that $E- \varepsilon \leq E_{curr} + \Delta E \leq E+\varepsilon$.

\subsection{Validation of algorithm}

\begin{figure}
    \includegraphics[width=0.45\textwidth]{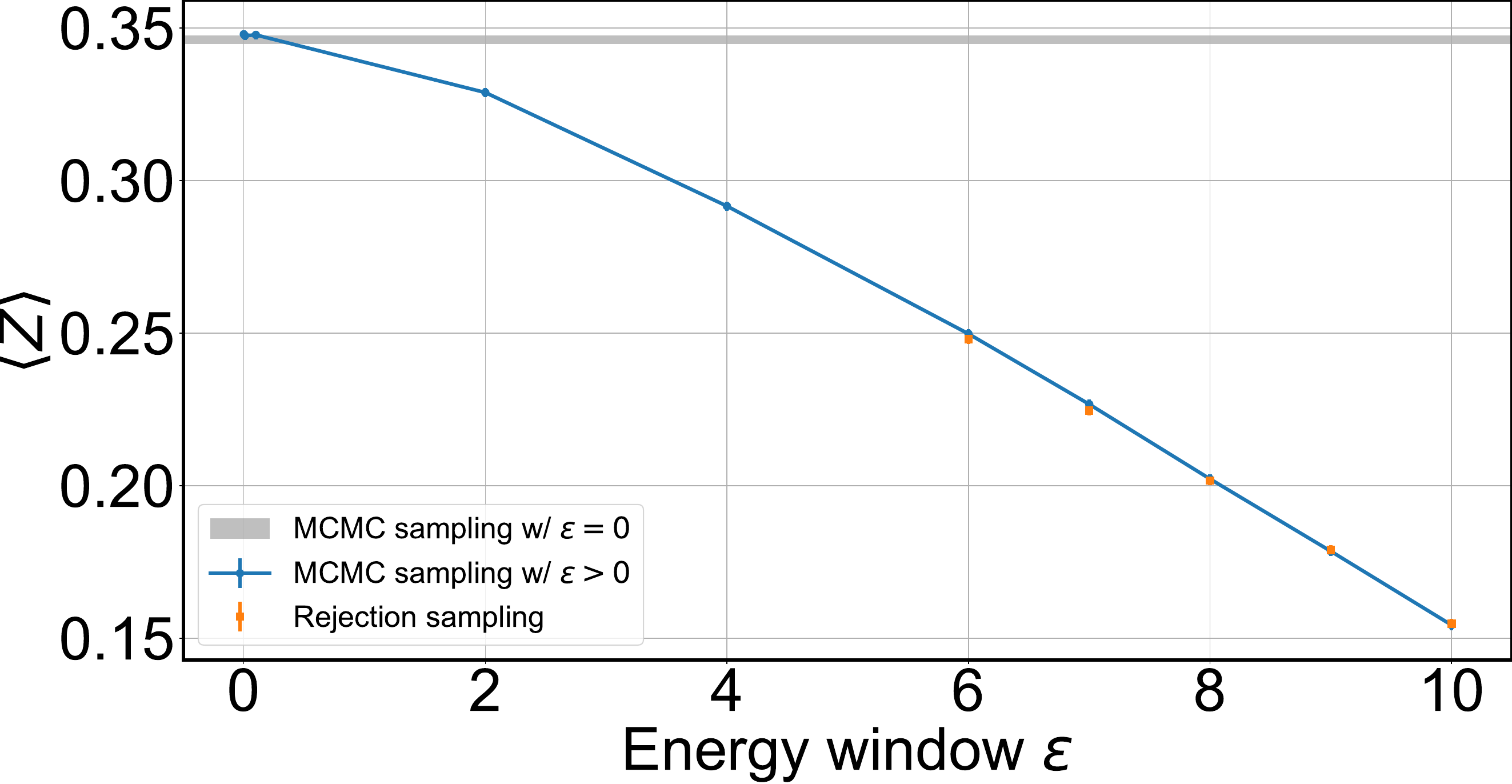}
    \caption{The average $z$-magnetization versus energy window size $\varepsilon$ for random product states sampled from the RPE with a finite energy window, for an $N=12$ site chain. The shaded line is obtained by MCMC sampling the RPE with $\varepsilon=0$. The blue points are obtained by MCMC sampling algorithm the RPE with $\varepsilon>0$. The orange points are obtained from rejection sampling.}
    \label{fig:rpe_sampling_validation}
\end{figure}

To validate that our MCMC sampling algorithms are sampling the intended distribution, we compare them against the rejection sampling approach described in Sec.~\ref{sec:rpe}. Figure~\ref{fig:rpe_sampling_validation} shows a comparison of the MCMC sampling algorithms with and without an energy window against rejection sampling. The target energy corresponds to the $\ket{0\cdots 0}$ state energy and the system size is $N=12$. We indeed see close agreement between the two MCMC sampling algorithms and between the energy window MCMC sampling algorithm and rejection sampling, demonstrating a clear consistency.

\section{Properties of the RPE} \label{sec:rpe_stats}

\begin{figure}
    \includegraphics[width=0.45\textwidth]{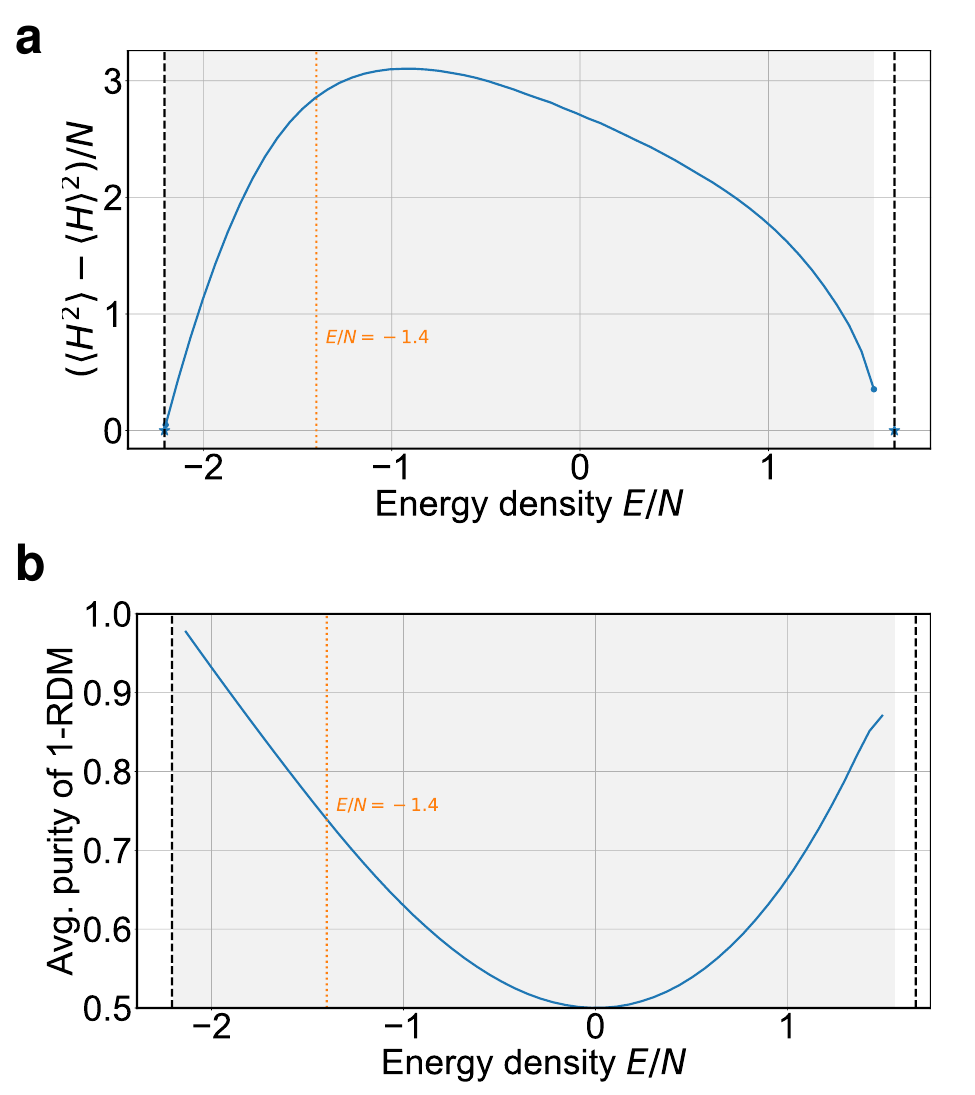}
    \caption{\textbf{a} The energy variance per site versus the energy density for the RPE with $N=20$ sites. \textbf{b} The spatial average of the purities $\frac{1}{N}\sum_{j=1}^N\tr (\rho_j^2)$ of the 1-RDMs $\rho_j$ on each site $j$ versus energy density for the RPE.}
    \label{fig:rpe_energy_variance_and_purity_vs_energy_density}
\end{figure}

Fig.~\ref{fig:rpe_energy_variance_and_purity_vs_energy_density}\textbf{a} shows the energy variance per site as a function of energy density for the RPE. The borders of the shaded region correspond to the energy densities of the mean-field ground state and anti-ground state (i.e., the lowest and highest energy product states for the Hamiltonian). For the ferromagnetic mixed-field Ising model in Eq.~(\ref{eq:H}), the mean-field ground state is unique. Therefore, as the RPE energy approaches the mean-field ground state energy $\ket{\psi_{\textrm{MF}}}$, the RPE itself becomes closer and closer to the single mean-field ground state product state $\rho_{\textrm{RPE}} \approx \ket{\psi_{\textrm{MF}}} \bra{\psi_{\textrm{MF}}}$, and so the variance approaches its minimal value and purity of reduced density matrices (see Fig.~\ref{fig:rpe_energy_variance_and_purity_vs_energy_density}\textbf{b}) approaches 1. Interestingly, the antiferromagnetic mixed-field Ising model has a doubly-degenerate mean-field ground state (in the thermodynamic limit), which one can also see in the RPE results. As one approaches the highest energy product states of the ferromagnetic model (the lowest energy states of the antiferromagnetic model), one sees that the RPE's energy variance stays non-zero and purity of 1-RDMs does not approach 1. The RPE in this case is approaching an equal mixture of the two mean-field anti-ground states: $\rho_{\textrm{RPE}} \approx \frac{1}{2}\left(\ket{\psi_{\textrm{AMF,1}}} \bra{\psi_{\textrm{AMF,1}}} + \ket{\psi_{\textrm{AMF,2}}} \bra{\psi_{\textrm{AMF,2}}}\right)$.

\end{document}